\documentclass[11pt]{article}
\usepackage[a4paper,margin=2.6cm]{geometry}
\usepackage{amsmath,mathtools,amssymb,amsthm,mathrsfs,graphicx,enumitem}
\usepackage{newtxtext}
\usepackage{tocloft}
\usepackage{xcolor}

\usepackage[hidelinks]{hyperref}

\numberwithin{equation}{section}
\allowdisplaybreaks[1]
\theoremstyle{plain}
\newtheorem{theorem}{Theorem}[section]
\newtheorem{proposition}[theorem]{Proposition}
\newtheorem{lemma}[theorem]{Lemma}
\newtheorem{rhp}[theorem]{RH Problem}
\theoremstyle{definition}

\newtheorem{assumption}{Assumption}
\newtheorem{remark}[theorem]{Remark}

\newlist{conds}{itemize}{1}
\setlist[conds]{leftmargin=3.6em,labelsep=0.6em,itemsep=2pt,topsep=3pt}
\newcommand{\cond}[2]{\item[\textbf{\textit{(#1)}}]\textbf{\textit{#2}}}

\newcommand{\rmi}{\mathrm{i}}
\newcommand{\rmd}{\mathrm{d}}

\newcommand{\Mc}{M}
\newcommand{\Mg}{M^{(glo)}}
\newcommand{\Ml}{M^{(loc)}}
\newcommand{\Nc}{N}
\newcommand{\Ng}{N^{(glo)}}
\newcommand{\Nl}{N^{(loc)}}
\newcommand{\hEj}{\hat E_{j_1}}
\newcommand{\Ej}{E_{j_1}}
\newcommand{\ve}{\varepsilon}
\newcommand{\bphi}{\boldsymbol\phi}
\newcommand{\bdelta}{\boldsymbol\delta}
\newcommand{\psm}[1]{\begin{psmallmatrix}#1\end{psmallmatrix}}

\begin{document}

\begin{center}
{\Large\bfseries Long-time asymptotics of the defocusing complex\\[3pt] short pulse equation with a finite-genus\\[3pt] algebro-geometric background}\\[14pt]
{\large Wenhao Liu$^a$, Yufeng Zhang$^a$\footnote{\footnotesize
 Corresponding author. {\sl Email address}: zyfusf@163.com}, Zhenbo Wang$^a$, Zhijun Qiao$^b$}\\[6pt]
{\small $^a$ School of Mathematics, China University of Mining and Technology, 1 Daxue Road,}\\
{\small Xuzhou, Jiangsu 221116, PR China}\\
{\small $^b$ School of Mathematical and Statistical Sciences, University of Texas Rio Grande Valley,}\\
{\small Edinburg, TX 78539, USA}
\end{center}

\vspace{6pt}
\begin{abstract}
We study the finite-genus algebro-geometric solutions of the complex short pulse equation and the long-time asymptotics of such solutions in the defocusing regime. The finite-genus Baker–Akhiezer function is formulated as an explicitly solvable Riemann–Hilbert problem on the complex spectral plane, which yields theta-function representations of the solutions together with reconstruction formulae for the field and the reciprocal coordinate at the spectral point $\lambda=0$. We then consider the Cauchy problem for the defocusing complex short pulse  equation with initial data that coincide with a finite-genus algebro-geometric solution outside a compact set. By means of the nonlinear steepest descent method, we obtain the asymptotics in bounded transition windows around the critical speeds and on fixed compact sets of speeds away from them, namely two families of transition regions, a Zakharov-Manakov region, and a fast-decay region. The leading term of the asymptotics is a phase-shifted finite-genus solution, while the corrections are of order $t^{-1/3}$ and are described by the Painlev\'e XXXIV transcendent in the transition regions, of order $t^{-1/2}$ and are described by parabolic cylinder functions in the Zakharov-Manakov region, and are exponentially small in the fast-decay region.


\end{abstract}

\tableofcontents

\section{Introduction}\label{sec1}
We consider the complex short pulse (CSP) equation
\begin{equation}\label{csp}
q_{xt}+q+\frac{\sigma}{2}\big(|q|^2q_x\big)_x=0,\quad (x,t)\in\mathbb R\times\mathbb R^+,\quad \sigma=\pm1,
\end{equation}
where $q=q(x,t)$ is complex valued. We focus on the defocusing case $\sigma=-1$.

The short pulse equation arises as an integrable reduction of the negative Wadati–Konno–Ichikawa (WKI) hierarchy. Equation \eqref{csp} is obtained from the two-component short pulse system
\begin{equation*}
u_{x\varsigma}=u+\tfrac12(uvu_x)_x,\quad v_{x\varsigma}=v+\tfrac12(uvv_x)_x,
\end{equation*}
via the reduction $v=\sigma\bar u$ followed by the time reversal $q(x,t)=u(x,-t)$, that is $\varsigma=-t$. The two-component system itself corresponds to the $m=-3$ flow of the hierarchy, written after a constant normalization of the flow and with the sign convention for the second field used here. Early work of Qiao developed commutator representations of the WKI hierarchy, its Neumann-constrained Hamiltonian systems, and parametric solutions from a Bargmann constraint \cite{Qiao92,Qiao93,Qiao95JMP}. Qiao's 1995 study of stationary systems explicitly constructs the inverse WKI recursion and its lower-order branch \cite{Qiao95PLA}. The two-component member appears as an isospectral negative flow in his monograph \cite{Qiao02}, and the Lenard scheme developed by Qiao, Cao and Strampp recovers the same system together with its Lax pair \cite{QCH03}.

The real diagonal reduction $v=u$  yields the SP equation
\begin{equation}
u_{xt}=u+\frac16(u^3)_{xx},
\end{equation}
derived by Sch\"afer and Wayne from Maxwell's equations \cite{SW04}, which approximates ultra-short pulses more accurately than the nonlinear Schr\"odinger (NLS) equation \cite{CJSW05}. The SP equation is integrable with a WKI-type Lax pair \cite{SS05,WKI79}; subsequent work has covered its hierarchy \cite{Bru05}, loop solitons, breathers and periodic solutions \cite{Mat07,Mat08}, and global well-posedness and wave breaking \cite{PS10,LPS09}. Feng \cite{Feng15} formulated \eqref{csp} as a model for ultra-short pulse propagation in nonlinear media, constructed its soliton solutions, and proposed a coupled model of two complex envelopes; a focusing–defocusing extension was subsequently studied in \cite{YZ18}.

For the scalar CSP equation \eqref{csp} we work with its $2\times2$ WKI spectral problem, in which the spectral parameter multiplies the potential in the spatial part while the temporal part has a pole at $\lambda=0$ \cite{WKI79}. The analysis therefore employs a reciprocal coordinate $y$, as in the treatment of the Camassa–Holm equation \cite{BS06,CGI06,CH93}; both the field and the physical coordinate $x$ are then recovered from the first Taylor coefficient at $\lambda=0$  of the solution to the Riemann–Hilbert (RH) problem. In the defocusing case, the reciprocal density takes the form $p=\sqrt{1-|q_x|^2}$ in physical coordinates, and the transformation remains regular whenever $|q_x|<1$.

The RH approach \cite{Zhou89} to equations with WKI-type Lax pairs and decaying data is by now well developed. For the SP equation, the RH problem was formulated in \cite{BSZ17}, following the treatment of the short-wave model of the Camassa–Holm equation \cite{BSZ11}, and the long-time asymptotics were derived in \cite{Xu18}. For the CSP equation, the RH formalism and the long-time asymptotics were developed in \cite{PTF20,XF20}, the modified CSP equation was treated in \cite{CGW22}, and the coupled modified CSP equation with decaying data was addressed in our previous work \cite{LGL24}. Soliton resolution and the asymptotic stability of solitons for the SP and CSP equations were established in \cite{LTYF22,YF21}, and the periodic SP equation was studied in \cite{KST26}. All of these results, however, concern the zero background; perturbations of nonconstant quasi-periodic CSP backgrounds remain untreated.

Finite-genus solutions form a central class of exact solutions of integrable systems \cite{BBEIM94,Dub81,GHMT08,NMPZ84}. Asymptotic analysis is more convenient in a matrix formulation on the complex spectral plane. Kotlyarov and Shepelsky introduced a planar unimodular Baker-Akhiezer function for the NLS equation \cite{KS17}, and it was adapted to the WKI equation in \cite{LYZ26}. The finite-genus background then enters the nonlinear steepest descent method \cite{DZ93} as a global parametrix.

The nonlinear steepest descent method \cite{DIZ93,DZ93,DZ03,ZM76} is the standard tool for long-time asymptotics of integrable equations. It has been applied, among others, to the derivative nonlinear Schr\"odinger, Camassa-Holm and spin-1 Gross-Pitaevskii equations \cite{BM17,BJM18,BKS09,CL20,Geng21,Len17,Liu18}. Nonzero and step-like backgrounds were treated for the KdV equation \cite{EGKT13}, the focusing NLS equation \cite{BKS11} and the periodic Toda lattice \cite{KT09}, and Painlev\'e asymptotics in transition regions for the defocusing NLS equation with nonzero background were obtained in \cite{WF23}. Fan et al.\ \cite{FLYZ26} obtained the long-time asymptotics of the defocusing NLS equation with a finite-genus background, with Painlev\'e XXXIV corrections in the transition regions; related results followed for the focusing NLS and mKdV equations \cite{MF26a,MF26b} and for the defocusing WKI equation \cite{LYZ26}. The Painlev\'e XXXIV model problem originates from critical edge behavior in random matrix theory \cite{FIKN06,IKO08,IKO09}.

In this paper we develop the planar finite-genus formalism for \eqref{csp} and study the long-time asymptotics of the Cauchy problem for the defocusing ($\sigma=-1$) CSP equation with initial data
\begin{equation}\label{ic}
q(x,0)=q_0(x)\sim q^{(AG)}(x,0),\quad x\to\pm\infty,
\end{equation}
where $q^{(AG)}(x,t)=q^{(AG)}(x,t;\mathbf E,\hat{\mathbf E},\bphi)$ is a finite-genus algebro-geometric solution associated with the hyperelliptic contour
\begin{equation}\label{contour1}
\begin{aligned}
&w^2(\lambda)=\prod_{j=0}^n(\lambda-E_j)(\lambda-\hat E_j),\quad E_0<\hat E_0<E_1<\cdots<E_n<\hat E_n,\quad 0\notin\bigcup_{j=0}^n[E_j,\hat E_j],\\
&\mathbf E=(E_0,\dots,E_n), \quad \hat{\mathbf E}=(\hat E_0,\dots,\hat E_n), \quad \bphi=(\phi_1,\dots,\phi_n).
\end{aligned}
\end{equation}

Three features of the negative flow \eqref{csp} separate it from the NLS and WKI problems treated so far.
(i) The phase function $g(\lambda)$ has a simple pole at $\lambda=0$ instead of polynomial growth at $\lambda=\infty$. The function $\Xi(\lambda)$ in \eqref{Xi1} below is positive on the gaps, has a double pole at $\lambda=0$ and vanishes at $\lambda=\pm\infty$. Hence, for $\xi=y/t>0$, the stationary points appear on both sides of the origin and move in opposite directions, the critical values $\xi_j,\hat\xi_j$ are not ordered along the bands, and the whole half-plane $\xi<0$ is a fast-decay region. (ii) The field $q$ and the coordinate $x$ are reconstructed from the expansion at $\lambda=0$ without differentiation in $y$, and $x(y,t)$ acquires the $\mathcal O(1)$ shift $\rmi\delta'(0)$. (iii) The parameter of the Painlev\'e XXXIV model is determined from the endpoint values of the reflection coefficients (Proposition \ref{pro61}). In both transition regions the parameters of the model are $(\alpha,\omega)=(\frac14,0)$.

Before stating our main results, we introduce the necessary definitions and notation. Let $w(\lambda)$ be defined by \eqref{contour1} with $w(\lambda)=\lambda^{n+1}(1+\mathcal O(\lambda^{-1}))$ as $\lambda\to\infty$, and let
\begin{equation}\label{fg1}
f(\lambda)=\int_{\hat E_0}^\lambda\frac{P_f(s)}{w(s)}\rmd s,\quad g(\lambda)=\int_{\hat E_0}^\lambda\frac{c_gP_g(s)}{s^2w(s)}\rmd s,\quad P_f(\lambda)=\prod_{k=0}^n(\lambda-\lambda_k^f),\quad P_g(\lambda)=\prod_{k=0}^n(\lambda-\lambda_k^g),
\end{equation}
where the constants $\lambda_k^f,\lambda_k^g$ and $c_g$ are uniquely determined by
\begin{equation}\label{fg1norm}
\begin{cases}f(\lambda)=\lambda+\mathcal O(1),\quad g(\lambda)=\mathcal O(1), & \lambda\to\infty,\\[2pt] g(\lambda)=\dfrac1{4\lambda}+\mathcal O(1), & \lambda\to0,\\[4pt] \displaystyle\int_{E_j}^{\hat E_j}\rmd f=\int_{E_j}^{\hat E_j}\rmd g=0, & j=0,\dots,n.\end{cases}
\end{equation}
Then $\lambda_k^f,\lambda_k^g\in(E_k,\hat E_k)$, $c_gP_g(0)=-\frac14w(0)$ and $c_g<0$. Set
\begin{equation}\label{Xi1}
\Xi(\lambda)=-\frac{g'(\lambda)}{f'(\lambda)}=-\frac{c_gP_g(\lambda)}{\lambda^2P_f(\lambda)},\quad \xi_j=\Xi(E_j),\quad \hat\xi_j=\Xi(\hat E_j),\quad j=0,\dots,n,
\end{equation}
and
\begin{equation}\label{Zxi}
\mathcal Z(\xi)=\Big\{\lambda\in\mathbb R\setminus\Big(\bigcup_{j=0}^n[E_j,\hat E_j]\cup\{0\}\Big):\ \Xi(\lambda)=\xi\Big\}.
\end{equation}
\begin{assumption}\label{ass1}\leavevmode
\begin{enumerate}[label=\textup{(A\arabic*)},leftmargin=2.5em]
\item The numbers $\{\xi_j,\hat\xi_j\}_{j=0}^n$ are pairwise distinct.
\item $q_0(x)=q^{(AG)}(x,0)$ for $|x|>C$, with some $C>0$.
\item The Cauchy problem \eqref{csp}, \eqref{ic} has a smooth global solution with, for $t\ge0$, $|q_x|<1$, $|q^{(AG)}_x|<1$ and
\begin{equation*}
\int_{\mathbb R}\big(|q-q^{(AG)}|+|q_x-q^{(AG)}_x|+|q_{xx}-q^{(AG)}_{xx}|\big)(x,t)\,\rmd x<\infty.
\end{equation*}
\item $S_{11}(\lambda)\neq0$ for $\lambda\in\overline{\mathbb C^-}\setminus E$, $E=\{E_j,\hat E_j\}_{j=0}^n$, including the nonendpoint boundary values on the bands \textup{(}see \eqref{Sentries}\textup{)}.
\item Every endpoint $p\in E$ is nonresonant: $c_p\neq0$ in Proposition~\ref{pro61}. Then $|r_1(p)|=|r_2(p)|=1$.
\end{enumerate}
Theorem~\ref{thm14} is proved under these assumptions. The global well-posedness \textup{(A3)} is not addressed here.
\end{assumption}

\begin{figure}[htbp]
\centering
\includegraphics[width=0.78\textwidth]{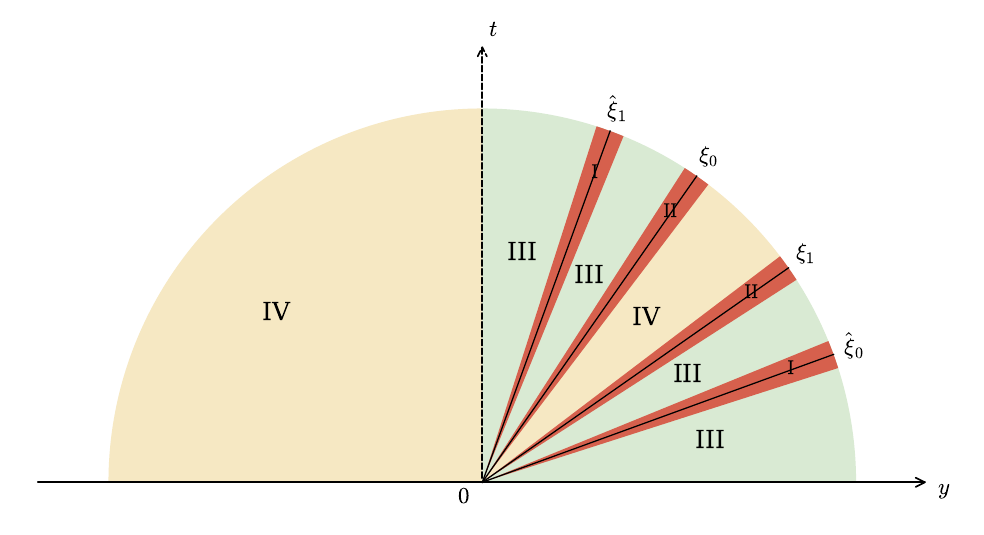}
\caption{The regions I-IV in the $(y,t)$-half-plane for $n=1$, where $\hat\xi_1<\xi_0<\xi_1<\hat\xi_0$.}\label{figreg}
\end{figure}

\begin{figure}[htbp]
\centering
\includegraphics[width=0.78\textwidth]{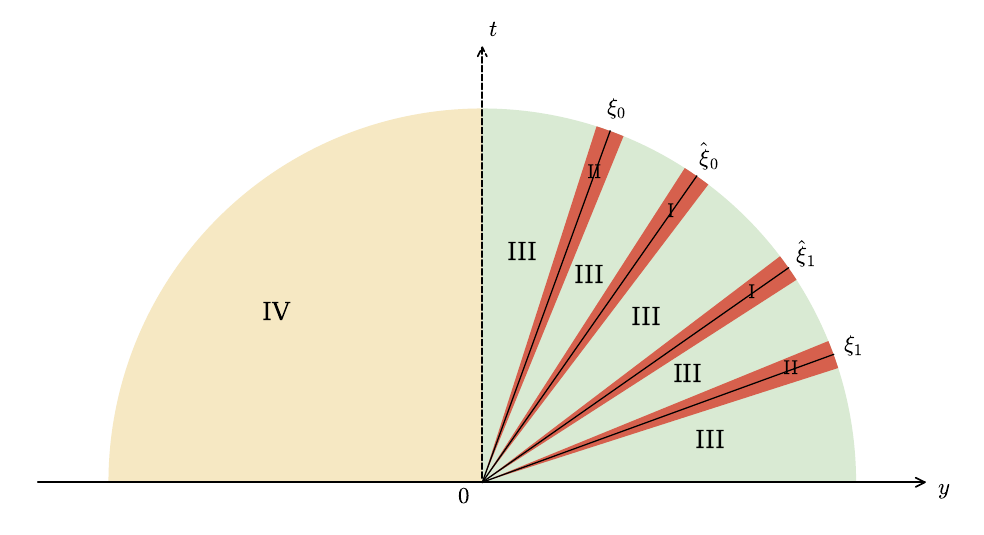}
\caption{The regions I-IV in the $(y,t)$-half-plane for $n=1$, where $\xi_0<\hat\xi_0<\hat\xi_1<\xi_1$.}\label{figregb}
\end{figure}

As shown in Figures~\ref{figreg} and \ref{figregb}, we consider the following regions of the $(y,t)$-half-plane:
\begin{itemize}[leftmargin=1.5em]
\item Transition region I: $\bigcup_{j=0}^n\big\{\xi:|\xi-\hat\xi_j|t^{2/3}\le C\big\}$.
\item Transition region II: $\bigcup_{j=0}^n\big\{\xi:|\xi-\xi_j|t^{2/3}\le C\big\}$.
\item Zakharov-Manakov region III: $\xi\in K$, where $K$ is a compact subset of $\{\xi>0:\mathcal Z(\xi)\neq\emptyset\}\setminus\{\xi_p\}_{p\in E}$.
\item Fast-decay region IV: $\xi\le-c_0$ for some $c_0>0$, or $\xi\in K$, where $K$ is a compact subset of $\{\xi>0:\mathcal Z(\xi)=\emptyset\}\setminus\{\xi_p\}_{p\in E}$.
\end{itemize}
Here $C>0$ and $c_0>0$ are fixed and $\xi=y/t$. In region I or II near $p\in E$, the Painlev\'e parameter $s$ of \eqref{zeta}, \eqref{zeta6} satisfies $|s|\le C'$ for $t$ large, where
\begin{equation*}
C'=2C\max_{0\le j\le n}\max\Big\{\big|\theta^{(1,\hat j)}\big|\,\big|\theta^{(3,\hat j)}(\hat\xi_j)\big|^{-1/3},\ \big|\theta^{(1,j)}\big|\,\big|\theta^{(3,j)}(\xi_j)\big|^{-1/3}\Big\},
\end{equation*}
with $\theta^{(1,\cdot)},\theta^{(3,\cdot)}$ as in \eqref{thetaexp} and \eqref{thetaexp6}. This follows from the asymptotic formulas for $s$ after \eqref{zetaprop} and in \eqref{zetaprop6}. For $s\ge0$ the stationary point that merges with $p$ lies in the adjacent gap, and for $s<0$ it lies in the band. The order of the critical values and the presence of region IV for $\xi>0$ depend on the positions of the bands relative to the origin: in Figure~\ref{figreg}, $\mathcal Z(\xi)=\emptyset$ for $\xi_0<\xi<\xi_1$, whereas in Figure~\ref{figregb}, $\mathcal Z(\xi)\neq\emptyset$ for all $\xi>0$, and region IV reduces to $\xi<0$. Our main result is the following.

\begin{theorem}\label{thm13}
Let $M(y,t;\lambda)$ be defined by \eqref{MA}. Then $M$ solves RH Problem \ref{rhp32}, and the finite-genus algebro-geometric solution of \eqref{csp} is given in parametric form by \eqref{qAG1}-\eqref{xAG1}.
\end{theorem}

\begin{theorem}\label{thm14}
Let $q^{(AG)}(x,t;\mathbf E,\hat{\mathbf E},\bphi)$ be the finite-genus solution of Theorem \ref{thm13}, and let Assumption~\ref{ass1} hold. Then the solution of the Cauchy problem \eqref{csp}, \eqref{ic} admits the representation
\begin{equation}\label{mainq}
q(x,t)=e^{-2\delta(0)}\Big[q^{(AG)}\big(x^{(glo)}(y,t),t;\mathbf E,\hat{\mathbf E},\bphi+\bdelta\big)+e^{2\rmi\varphi_\ve}F_{12}\Big],
\end{equation}
\begin{equation}\label{mainx}
x(y,t)=x^{(glo)}(y,t)+\rmi\delta'(0)+\rmi F_{11},
\end{equation}
where the real vector $\bdelta=(\delta_1,\dots,\delta_n)$ and the function $\delta(\lambda)$ are defined by \eqref{deltasys}, \eqref{delta} in region I, by \eqref{deltasys6}, \eqref{delta6} in region II and by \eqref{deltasys7}, \eqref{delta7} in regions III and IV, $\delta(0)\in\rmi\mathbb R$ and $\delta'(0)\in\rmi\mathbb R$ are determined by \eqref{delta0}, $x^{(glo)}$ reads \eqref{x8}, $\varphi_\ve$ is defined before \eqref{M0} \textup{(}before \eqref{N0} in region II\textup{)}, and the matrix $F=(F_{kl})_{k,l=1,2}$ has the following asymptotics as $t\to\infty$:
\begin{itemize}[leftmargin=1.5em]
\item Transition region I: For fixed $j_1\in\{0,\dots,n\}$ and $|\xi-\hat\xi_{j_1}|t^{2/3}\le C$,
\[
F=\frac{t^{-1/3}}{|\theta^{(3,\hat j_1)}(\hat\xi_{j_1})|^{2/3}\hat E_{j_1}^2}\Mg(0)^{-1}H_1(\hat E_{j_1})\begin{pmatrix}0&\rmi a(s)\\0&0\end{pmatrix}H_1(\hat E_{j_1})^{-1}\Mg(0)+\mathcal O(t^{-1/2}), 
\]
where the global parametrix $\Mg$ is defined by \eqref{Mglo}, $\theta^{(3,\hat j_1)}$, $s=s(\xi,t)$ and $H_1$ are given by \eqref{thetaexp}, \eqref{zetaprop} and \eqref{Mlocdef}, and $a(s)$ is represented by \eqref{as} in terms of the solution $v$ of the Painlev\'e XXXIV equation \eqref{P34} with
\begin{equation*}
\begin{aligned}
&v(s)=\frac14s^{-1/2}+\mathcal O(s^{-2}), \quad s\to+\infty,\\
&v(s)=-\frac s2+\mathcal O(s^{-2}), \quad s\to-\infty.
\end{aligned}
\end{equation*}
\item Transition region II: For fixed $j_1\in\{0,\dots,n\}$ and $|\xi-\xi_{j_1}|t^{2/3}\le C$,
\[
F=-\frac{t^{-1/3}}{|\theta^{(3,j_1)}(\xi_{j_1})|^{2/3}E_{j_1}^2}\Ng(0)^{-1}\tilde H_1(E_{j_1})\begin{pmatrix}0&\rmi a(s)\\0&0\end{pmatrix}\tilde H_1(E_{j_1})^{-1}\Ng(0)+\mathcal O(t^{-1/2}),
\]
where the global parametrix $\Ng$ can be found in \eqref{Nglo6}, $\theta^{(3,j_1)}$, $s$ and $\tilde H_1$ are given by \eqref{thetaexp6}, \eqref{zetaprop6} and \eqref{Nlocdef}.
\item Zakharov-Manakov region III: Uniformly for $\xi\in K$, $K$ a compact subset of $\{\xi>0:\mathcal Z(\xi)\neq\emptyset\}\setminus\{\xi_p\}_{p\in E}$,
\[
F=\sum_{\lambda_k\in\mathcal Z(\xi)}\frac{\ve_kt^{-1/2}}{2\sqrt{\kappa_k}\,\lambda_k^2}\Ng(0)^{-1}\Ng(\lambda_k)A_k\Ng(\lambda_k)^{-1}\Ng(0)+\mathcal O(t^{-1}\log t),
\]
where $\ve_k=\operatorname{sgn}\lambda_k$, and $\kappa_k$ and $A_k$ are given by \eqref{kappa7} and \eqref{beta7}.
\item Fast-decay region IV: Uniformly for $\xi\le-c_0$ and for $\xi\in K$, $K$ a compact subset of $\{\xi>0:\mathcal Z(\xi)=\emptyset\}\setminus\{\xi_p\}_{p\in E}$, $F=\mathcal O(e^{-ct})$ with some $c>0$. In particular, for $\xi\le-c_0$, $\bdelta=0$, $\delta\equiv0$ and
\[
q(x,t)=q^{(AG)}(x,t;\mathbf E,\hat{\mathbf E},\bphi)+\mathcal O(e^{-ct}).
\]
\end{itemize}
\end{theorem}

The rest of the paper is organized as follows. Section \ref{sec2} recalls the hyperelliptic contour, the Baker-Akhiezer function, the gauge transformation of the Lax pair and the reconstruction at $\lambda=0$. In Section \ref{sec3}, we construct the phase functions and solve the planar finite-genus RH problem in terms of Riemann theta functions, which proves Theorem \ref{thm13}. Section \ref{sec4} develops the Jost solutions, the scattering data and the signature table for the defocusing background. Sections \ref{sec5} and \ref{sec6} are devoted to the transition regions I and II and the associated Painlev\'e XXXIV parametrices. In Section \ref{sec7} we study the Zakharov-Manakov and fast-decay regions. Finally, in Section \ref{sec8} we evaluate the transformed RH problems at $\lambda=0$ and complete the proof of Theorem \ref{thm14}. The Painlev\'e XXXIV parametrix used in Sections \ref{sec5} and \ref{sec6} is recalled in Appendix \ref{appA}.

\section{The basic RH problem and Baker-Akhiezer function}\label{sec2}

In this section we recall some basic facts on hyperelliptic Riemann surfaces \cite{BBEIM94,Dub81,KS17} and formulate the Baker-Akhiezer function of \eqref{csp} as a RH problem. Let $p=p(x,t)=\sqrt{1+\sigma|q_x|^2}$, where $|q_x|<1$ is assumed in the defocusing case. Then
\begin{equation}\label{cons}
p_t=-\frac{\sigma}{2}\big(|q|^2p\big)_x,
\end{equation}
and the reciprocal variable $y=y(x,t)$ is defined by
\begin{equation}\label{y}
\rmd y=p\,\rmd x-\frac{\sigma}{2}|q|^2p\,\rmd t,
\end{equation}
normalized by $y=x+\int_{-\infty}^x\big(p(s,t)-1\big)\rmd s$ when $q$ decays as $x\to-\infty$.

\begin{figure}[htbp]
\centering
\includegraphics[width=0.82\textwidth]{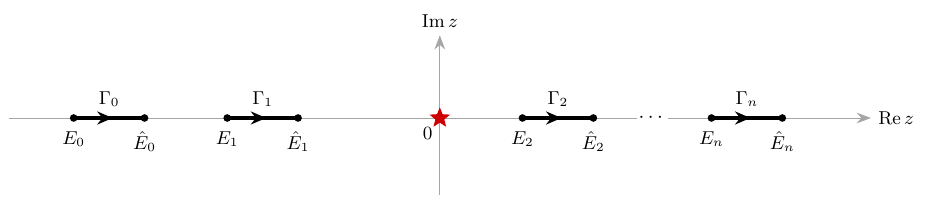}
\caption{The jump contour on the $\lambda$-plane for the defocusing CSP equation.}\label{fig3}
\end{figure}

Let $\{E_j,\hat E_j\}_{j=0}^n$ be $2n+2$ distinct points in $\mathbb C$, where $E_0<\hat E_0<E_1<\cdots<E_n<\hat E_n$ are real. Let $ X$ be the Riemann surface of genus $n$ defined by
\begin{equation}\label{contour}
 X:\ w^2(\lambda)=P(\lambda),\quad P(\lambda)=\prod_{j=0}^n(\lambda-E_j)(\lambda-\hat E_j),
\end{equation}
with cuts along the arcs $\Gamma_j=(E_j,\hat E_j)$, $j=0,1,\dots,n$, which lie on the real axis (Figure~\ref{fig3}). We assume $0\notin\bigcup_{j=0}^n\overline{\Gamma_j}$. The upper and lower sheets $ X_\pm$ are fixed by
\begin{equation}\label{sheet}
\sqrt{P(\lambda)}=\pm \lambda^{n+1}\big(1+\mathcal O(\lambda^{-1})\big),\quad \lambda=\pi( P)\to\infty,\quad  P\in X_\pm,
\end{equation}
where $\pi( P)$ is the projection of $ P=(\lambda,w)$ onto $\mathbb P^1$. Denote by $\infty^\pm$ and $0^\pm$ the preimages of $\lambda=\infty$ and $\lambda=0$ on $ X_\pm$. The function $\sqrt{P(\lambda)}$ is meromorphic on $ X$, with $2n+2$ zeros at $E_j,\hat E_j$ and poles of order $n+1$ at $\infty^\pm$.

With respect to the variable \eqref{y}, the vector-valued Baker-Akhiezer function $\Phi(y,t; P)=(\Phi_1,\Phi_2)^T$ of \eqref{csp} is defined by the following conditions \cite{BBEIM94}:
\begin{itemize}[leftmargin=1.5em]
\item $\Phi(y,t; P)$ is meromorphic on $ X\setminus\{\infty^\pm,0^\pm\}$ with pole divisor $ D=\sum_{j=1}^n P_j$, which is nonspecial with $\pi( P_j)\notin\{E_j,\hat E_j,0\}$ and $\pi( P_j)\neq\pi( P_k)$, $j\neq k$.
\item $\Phi(y,t; P)$ has the asymptotic behaviors
\begin{equation}\label{BAinf}
\Phi(y,t; P)=\begin{cases}\big[(1,0)^T+\mathcal O(\lambda^{-1})\big]e^{-\rmi \lambda y}, &  P\to\infty^-,\\[2pt] c\lambda\big[(0,1)^T+\mathcal O(\lambda^{-1})\big]e^{\rmi \lambda y}, &  P\to\infty^+,\end{cases}
\end{equation}
\begin{equation}\label{BA0}
\Phi(y,t; P)=\begin{cases}\mathcal O(1)\,e^{-\frac{\rmi t}{4\lambda}}, &  P\to0^-,\\[2pt] \mathcal O(1)\,e^{\frac{\rmi t}{4\lambda}}, &  P\to0^+,\end{cases}
\end{equation}
where $c$ is a complex constant.
\end{itemize}

The CSP equation \eqref{csp} is the compatibility condition of the Lax pair
\begin{equation}\label{lax}
\Psi_x=X(x,t;\lambda)\Psi,\quad \Psi_t=T(x,t;\lambda)\Psi,
\end{equation}
\begin{equation}\label{XT}
X=-\rmi \lambda\sigma_3+\lambda Q_x,\quad T=-\frac{\sigma}{2}|q|^2X-\frac{\rmi}{4\lambda}\sigma_3-\frac{\rmi}{2}Q\sigma_3,\quad Q=\begin{pmatrix}0&q\\-\sigma\bar q&0\end{pmatrix},
\end{equation}
where $\sigma_{1,2,3}$ are the Pauli matrices
\begin{equation*}
\sigma_{1}=\begin{pmatrix}0&1\\1&0\end{pmatrix}, \quad \sigma_{2}=\begin{pmatrix}0&-i\\i&0\end{pmatrix}, \quad \sigma_{3}=\begin{pmatrix}1&0\\0&-1\end{pmatrix}.
\end{equation*}
Next we recall the RH formalism for \eqref{csp} under zero boundary conditions \cite{PTF20,XF20}.

\begin{rhp}\label{rhp21}
The function $M^{b}(\lambda)=M^{b}(y,t;\lambda)$ satisfies the following basic RH problem.
\begin{conds}
\cond{1}{Analyticity:} $M^{b}(\lambda)$ is analytic in $\mathbb C\setminus\Gamma$, where $\Gamma$ is a piecewise smooth oriented contour.
\cond{2}{Jump condition:} $M^{b}(\lambda)$ satisfies
\begin{equation}\label{Mtjump}
M^{b}_-(\lambda)=M^{b}_+(\lambda)e^{-\rmi\tilde\theta(\lambda)\hat\sigma_3}J(\lambda),\quad \lambda\in\Gamma,
\end{equation}
where
\begin{equation*}
\tilde\theta(\lambda)=\lambda y+\dfrac{t}{4\lambda}.
\end{equation*}
\cond{3}{Asymptotic behaviors:}
\begin{equation}\label{Mtasy}
M^{b}(\lambda)=\begin{cases}I+\mathcal O(\lambda^{-1}), & \lambda\to\infty,\\[2pt] M^{b}_0(y,t)\big[I+\lambda H(x(y,t),t)+\mathcal O(\lambda^2)\big], & \lambda\to0,\end{cases}
\end{equation}
where $M^{b}_0=e^{\eta\sigma_3}G^{-1}$, $H=Q+\rmi(y-x)\sigma_3$, and
\[
G=\sqrt{\frac{1+p}{2p}}\begin{pmatrix}1&-\dfrac{\rmi q_x}{1+p}\\[6pt]-\dfrac{\rmi\sigma\bar q_x}{1+p}&1\end{pmatrix},\quad
\eta=-\sigma\int_{-\infty}^x\frac{\bar q_sq_{ss}-q_s\bar q_{ss}}{4p(p+1)}(s,t)\,\rmd s.
\]
\cond{4}{Local behavior:} $M^{b}(\lambda)$ has inverse fourth-root singularities at the endpoints of $\Gamma$.
\end{conds}
\end{rhp}

\begin{proposition}\label{prop22}
Suppose that $M^{b}(y,t;\lambda)$ satisfies RH Problem \ref{rhp21}. Then the solution $q(x,t)$ of \eqref{csp} is given by
\begin{equation}\label{recq}
q(x,t)=\lim_{\lambda\to0}\frac{\big[M^{b}(y,t;0)^{-1}M^{b}(y,t;\lambda)\big]_{12}}{\lambda},
\end{equation}
with
\begin{equation}\label{recx}
x(y,t)=y+\rmi\lim_{\lambda\to0}\frac{\big[M^{b}(y,t;0)^{-1}M^{b}(y,t;\lambda)\big]_{11}-1}{\lambda}.
\end{equation}
\end{proposition}

Let $\Gamma_j=(E_j,\hat E_j)$, $j=0,1,\dots,n$, be oriented arcs connecting $E_j$ and $\hat E_j$ (see Figure~\ref{fig3}), and $\hat\Gamma_j=(\hat E_{j-1},E_j)$, $j=1,\dots,n$. The Baker-Akhiezer function of \eqref{csp} is defined by
\begin{equation}\label{PsiAG}
\psi(y,t;\lambda)=M^{b}(y,t;\lambda)\,e^{-\rmi\left(\lambda y+\frac{t}{4\lambda}\right)\sigma_3},
\end{equation}
which satisfies the following RH Problem \ref{rhp24}.

\begin{rhp}\label{rhp24}
The function $\psi(y,t;\lambda)$ satisfies the following RH problem.
\begin{conds}[leftmargin=5.2em]
\cond{1}{Analyticity:} $\psi(y,t;\lambda)$ is analytic in $\mathbb C\setminus(\Gamma\cup\{0\})$, where
\begin{equation}\label{Gamma contour}
\Gamma=\big(\bigcup_{j=0}^n\Gamma_j\big)\cup\big(\bigcup_{j=1}^n\hat\Gamma_j\big).
\end{equation}
\cond{2}{Jump condition:} $\psi_y\psi^{-1}$ is an entire function, and $\psi_t\psi^{-1}$ is analytic in $\mathbb C\setminus\{0\}$ with at most a simple pole at $\lambda=0$. Moreover,
\[
\psi_-(y,t;\lambda)=\psi_+(y,t;\lambda)J(\lambda),\quad \lambda\in\Gamma,
\]
where
\begin{equation}\label{JcaseA}
\begin{aligned}
&J(\lambda)=\begin{pmatrix}0&\rmi e^{\rmi\phi_j}\\ \rmi e^{-\rmi\phi_j}&0\end{pmatrix},\quad \lambda\in(E_j,\hat E_j),\quad j=0,1,\dots,n,\\
&J(\lambda)=I,\quad \lambda\in\hat\Gamma_j,\ j=1,\dots,n.
\end{aligned}
\end{equation}
Here $\phi_j\in\mathbb R$ with $\phi_0=0$.
\cond{3}{Normalization:}
\[
\begin{aligned}
&\psi(y,t;\lambda)=\big(I+\mathcal O(\lambda^{-1})\big)e^{-\rmi\left(\lambda y+\frac{t}{4\lambda}\right)\sigma_3}, && \lambda\to\infty,\\
&\psi(y,t;\lambda)\,e^{\frac{\rmi t}{4\lambda}\sigma_3}=\mathcal O(1), && \lambda\to0.
\end{aligned}
\]
\cond{4}{Local behavior:} $\psi$ has inverse fourth-root singularities at $E_j,\hat E_j$, $j=0,\dots,n$.
\cond{5}{Symmetry condition:} $\psi(y,t;\lambda)$ satisfies
\[
 \psi(y,t;\lambda)=\sigma_1\overline{\psi(y,t;\bar \lambda)}\sigma_1,
\]
\end{conds}
\end{rhp}

\section{Algebro-geometric solutions of the CSP equation}\label{sec3}

In this section we construct explicit solutions of RH Problem \ref{rhp24}. they give algebro-geometric solutions of \eqref{csp} of arbitrary finite genus. The construction is based on two scalar phase functions $f(\lambda)$ and $g(\lambda)$, which are characterized as follows.
\begin{rhp}\label{rhp31}
Find scalar functions $f(\lambda)$ and $g(\lambda)$ with the following properties.
\begin{conds}
\cond{1}{Analyticity:} $f(\lambda)$ is holomorphic in $\mathbb C\setminus\Gamma$, while $g(\lambda)$ is holomorphic in $\mathbb C\setminus(\Gamma\cup\{0\})$.
\cond{2}{Jump condition:}
\begin{equation}\label{fgjump}
\begin{cases}f_+(\lambda)+f_-(\lambda)=C_j^f,\quad g_+(\lambda)+g_-(\lambda)=C_j^g, & \lambda\in\Gamma_j,\ j=0,\dots,n,\\[2pt] f_+(\lambda)-f_-(\lambda)=\hat C_j^f,\quad g_+(\lambda)-g_-(\lambda)=\hat C_j^g, & \lambda\in\hat\Gamma_j,\ j=1,\dots,n,\end{cases}
\end{equation}
where $C_j^f$ and $C_j^g$ are real constants, while $\hat C_j^f$ and $\hat C_j^g$ are purely imaginary.
\cond{3}{Asymptotic behavior:}
\begin{equation}\label{fgasy}
\begin{aligned}
&f(\lambda)=\lambda+f_0+\mathcal O(\lambda^{-1}),\quad g(\lambda)=g_0+\mathcal O(\lambda^{-1}), && \lambda\to\infty,\\
&f(\lambda)=\mathcal O(1),\quad\quad\ \ \ g(\lambda)=\frac{1}{4\lambda}+\mathcal O(1), && \lambda\to0.
\end{aligned}
\end{equation}
\cond{4}{Symmetry relations:} $\overline{f(\bar \lambda)}=f(\lambda)$, $\overline{g(\bar \lambda)}=g(\lambda)$.
\end{conds}
\end{rhp}

Following \cite{KS17}, we write $f(\lambda)$ and $g(\lambda)$ as Cauchy-type integrals:
\begin{equation}\label{f}
f(\lambda)=\begin{cases}w(\lambda)\Big(1+\dfrac{1}{2\pi\rmi}\displaystyle\int_{\Gamma_0}\frac{C_0^f}{w_+(\xi)(\xi-\lambda)}\rmd\xi\Big), & n=0,\\[12pt]
\dfrac{w(\lambda)}{2\pi\rmi}\Big(\displaystyle\sum_{j=0}^n\int_{\Gamma_j}\frac{C_j^f}{w_+(\xi)(\xi-\lambda)}\rmd\xi+\sum_{l=1}^n\int_{\hat\Gamma_l}\frac{\hat C_l^f}{w(\xi)(\xi-\lambda)}\rmd\xi\Big), & n\geq1,\end{cases}
\end{equation}
and
\begin{equation}\label{g}
g(\lambda)=\frac{w(\lambda)}{2\pi\rmi}\Big(\sum_{j=0}^n\int_{\Gamma_j}\frac{C_j^g}{w_+(\xi)(\xi-\lambda)}\rmd\xi+\sum_{l=1}^n\int_{\hat\Gamma_l}\frac{\hat C_l^g}{w(\xi)(\xi-\lambda)}\rmd\xi\Big)+\frac{w(\lambda)}{4w(0)\lambda},\quad n\geq0,
\end{equation}
with $w(\cdot)$ as in \eqref{contour}. Expanding \eqref{fgasy} as $\lambda\to\infty$ leads to the linear systems
\begin{equation}\label{sysf}
\sum_{j=0}^n\int_{\Gamma_j}\frac{\xi^mC_j^f}{w_+(\xi)}\rmd\xi+\sum_{l=1}^n\int_{\hat\Gamma_l}\frac{\xi^m\hat C_l^f}{w(\xi)}\rmd\xi=-2\pi\rmi\,\delta_{m,n-1},\quad m=0,1,\dots,n-1,
\end{equation}
\begin{equation}\label{sysg}
\sum_{j=0}^n\int_{\Gamma_j}\frac{\xi^mC_j^g}{w_+(\xi)}\rmd\xi+\sum_{l=1}^n\int_{\hat\Gamma_l}\frac{\xi^m\hat C_l^g}{w(\xi)}\rmd\xi=\frac{\pi\rmi}{2w(0)}\,\delta_{m,0},\quad m=0,1,\dots,n-1.
\end{equation}
For each of $f$ and $g$ this gives $n$ equations for the $2n+1$ jump constants $C_j$, $j=0,\dots,n$, and $\hat C_l$, $l=1,\dots,n$. We take
\begin{equation}\label{cases}
C_0^f=C_0^g=\hat C_j^f=\hat C_j^g=0, \quad j=1,2,\dots,n
\end{equation}
Define the algebro-geometric solution $\psi(y,t;\lambda)$ of \eqref{csp} in the form
\begin{equation}\label{ansatz}
\psi(y,t;\lambda)=e^{\rmi(f_0y+g_0t)\sigma_3}M(y,t;\lambda)e^{-\rmi(f(\lambda)y+g(\lambda)t)\sigma_3},
\end{equation}
where $M(y,t;\lambda)$ is constructed below.

\subsection{Algebro-geometric solution}

\begin{rhp}\label{rhp32}
Let $\Gamma_A=\bigcup_{j=0}^n(E_j,\hat E_j)$. We seek $M(y,t;\lambda)$ such that:
\begin{conds}
\cond{1}{Analyticity:} $M(\lambda)$ is holomorphic in $\mathbb C\setminus\Gamma_A$.
\cond{2}{Jump condition:} the boundary values are related by $M_-(\lambda)=M_+(\lambda)J_A(\lambda)$ with
\begin{equation}\label{JA}
J_A(\lambda)=\begin{pmatrix}0&\rmi e^{-\rmi(yC_j^f+tC_j^g-\phi_j)}\\ \rmi e^{\rmi(yC_j^f+tC_j^g-\phi_j)}&0\end{pmatrix},\quad \lambda\in\Gamma_A,
\end{equation}
with $C_0^f=C_0^g=\phi_0=0$.
\cond{3}{Normalization:}
\begin{equation*}
M(\lambda)=I+\mathcal O(\lambda^{-1}), \quad \lambda\to\infty.
\end{equation*}
Moreover, $M(\lambda)$ is holomorphic near $\lambda=0$.
\cond{4}{Local behavior:} Away from $E_j$ and $\hat E_j$, $M(\lambda)$ is bounded on $\Gamma_A$. At these points the singularity is at most of inverse fourth-root type
\end{conds}
\end{rhp}

Let
\begin{equation}\label{kappa}
\tau(\lambda)=\prod_{j=0}^n\sqrt[4]{\frac{\lambda-E_j}{\lambda-\hat E_j}},\quad \lambda\in\mathbb C\setminus\Gamma_A,
\end{equation}
with branch cuts $(E_j,\hat E_j)$ and normalized such that $\tau(\infty)=1$. In particular
\begin{equation*}
\tau_-(\lambda)=\rmi\tau_+(\lambda), \quad \lambda\in\Gamma_A,
\end{equation*}
and
\begin{equation}\label{kappaasy}
\tau(\lambda)=\begin{cases}c_j(\lambda-\hat E_j)^{-1/4}\big(1+\mathcal O(\lambda-\hat E_j)\big), & \lambda\to\hat E_j,\\[2pt] 1+\displaystyle\sum_{j=0}^n\frac{\hat E_j-E_j}{4\lambda}+\mathcal O(\lambda^{-2}), & \lambda\to\infty,\end{cases}
\end{equation}
where $c_{j}$ are nonzero constants ($j=1,2,\ldots,n$).

\begin{figure}[htbp]
\centering
\includegraphics[width=0.8\textwidth]{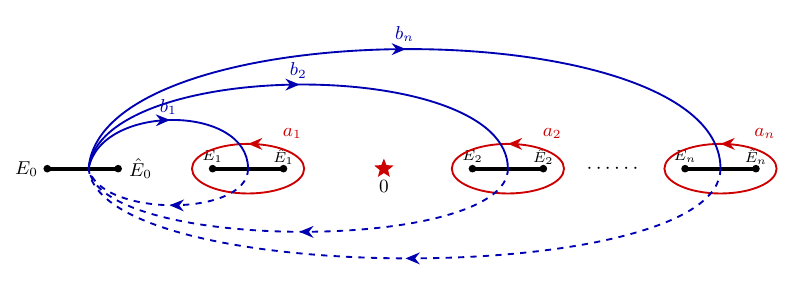}
\caption{Defocusing CSP equation: branch cuts along the real line.}\label{fig5}
\end{figure}

The branch cuts lie along the real line and are shown in Figure \ref{fig5}. Consider the integrals
\begin{equation}\label{omega}
\omega_j(\lambda)=\int_{\hat E_0}^\lambda\psi_j(s)\,\rmd s,\quad j=1,2,\dots,n,
\end{equation}
where
\[
\psi_j(\lambda)=\frac{\sum_{i=1}^nc_{ji}\lambda^{n-i}}{\sqrt{P(\lambda)}}.
\]
The coefficients $c_{jl}$ are fixed uniquely by requiring
\begin{equation}\label{normA}
\int_{\mathbf a_l}\rmd\omega_j( P)=2\int_{E_l}^{\hat E_l}\psi_{j-}(\lambda)\,\rmd \lambda=-2\int_{E_l}^{\hat E_l}\psi_{j+}(\lambda)\,\rmd \lambda=\delta_{jl},\quad j,l=1,2,\dots,n.
\end{equation}
Here $\psi_{j\pm}$ are the boundary values from $\mathbb C^\pm$, and $\mathbf a_l$ is the counterclockwise loop around $[E_l,\hat E_l]$ on the upper sheet in Figure~\ref{fig5}; with $\mathbf b_l$ as in Figure~\ref{fig5}, the cycles $\mathbf a_l,\mathbf b_l$ form a canonical basis.
Their $\mathbf b$-periods form the matrix
\begin{equation}\label{BA}
B_{jl}=\int_{\mathbf b_l}\rmd\omega_j( P)=2\sum_{k=1}^l\int_{\hat E_{k-1}}^{E_k}\psi_j(\lambda)\,\rmd \lambda,
\end{equation}
which is symmetric and has positive definite imaginary part, by the Riemann bilinear relations for the canonical basis fixed by \eqref{normA}. Since the cuts are real, $\psi_j$ is purely imaginary on the gaps and real on the bands, so $B=\rmi Y$ with $Y$ real and positive definite. The corresponding Riemann theta function is
\begin{equation}\label{theta}
\Theta(\mathbf u)=\sum_{\mathbf l\in\mathbb Z^n}\exp\big\{\pi\rmi(B\mathbf l,\mathbf l)+2\pi\rmi(\mathbf l,\mathbf u)\big\},\quad \mathbf u\in\mathbb C^n,
\end{equation}
where $(\mathbf l,\mathbf u)=l_1u_1+\cdots+l_nu_n$. It is an even function with the quasi-periodicity properties
\begin{equation}\label{thetaper}
\Theta(\mathbf u\pm\mathbf e_j)=\Theta(\mathbf u),\quad \Theta(\mathbf u\pm B\mathbf e_j)=e^{\mp2\pi\rmi u_j-\pi\rmi B_{jj}}\Theta(\mathbf u),
\end{equation}
$\mathbf e_j$ denoting the $j$-th standard basis vector of $\mathbb C^n$, so that $B\mathbf e_j$ is column $j$ of $B$.

The vectors $\mathbf e_j$ and $B\mathbf e_j$, $j=1,\dots,n$, generate a lattice $\Lambda\subset\mathbb C^n$, and $\operatorname{Jac}\{ X\}=\mathbb C^n/\Lambda$ is the Jacobian variety. The Abel map $\mathbf A: X\to\operatorname{Jac}\{ X\}$ has components
\begin{equation}\label{abel}
A_j( P)=\int_{ P_0}^{ P}\rmd\omega_j(\mathcal Q),\quad j=1,2,\dots,n,
\end{equation}
where the base point satisfies $\pi( P_0)=\hat E_0$. It extends to integral divisors $ D= P_1+\cdots+ P_m$ by
\begin{equation}\label{abelD}
\mathbf A( D)=\mathbf A( P_1)+\cdots+\mathbf A( P_m).
\end{equation}
If $ D= P_1+\cdots+ P_n$ is nonspecial, then on $ X$ cut along $\mathbf a_j,\mathbf b_j$ the zeros of $\Theta(\mathbf A( P)+\mathbf w( D))$ are exactly the $n$ points $ P_j$. Here
\begin{equation}\label{wD}
\mathbf w( D)=-\mathbf A( D)-\mathbf K,
\end{equation}
and the Riemann constant vector reads
\begin{equation}\label{K}
K_j=\frac12\sum_{l=1}^nB_{lj}-\frac j2.
\end{equation}
Restricted to the upper sheet of $X$, the Abel map has the jumps
\begin{equation}\label{abeljumpA}
\begin{aligned}
&\mathbf A_-(\lambda)-\mathbf A_+(\lambda)\equiv0\pmod{\mathbb Z^n}, && \lambda\in(\hat E_{j-1},E_j),\ j=1,\dots,n,\\
&\mathbf A_-(\lambda)+\mathbf A_+(\lambda)=0, && \lambda\in(E_0,\hat E_0),\\
&\mathbf A_-(\lambda)+\mathbf A_+(\lambda)=B\mathbf e_j, && \lambda\in(E_j,\hat E_j),\ j=1,\dots,n.
\end{aligned}
\end{equation}
Set
\begin{equation}\label{FH}
F_s(\lambda)=\frac{\Theta(\mathbf A(\lambda)+\mathbf c+\mathbf d_s)}{\Theta(\mathbf A(\lambda)+\mathbf d_s)},\quad H_s(\lambda)=\frac{\Theta(-\mathbf A(\lambda)+\mathbf c+\mathbf d_s)}{\Theta(-\mathbf A(\lambda)+\mathbf d_s)},\quad \lambda\in\mathbb C\setminus\Gamma_A,\quad s=1,2,
\end{equation}
with $\mathbf c=(c_1,\dots,c_n)^T$, $\mathbf d_1=-\mathbf w( D)$ and $\mathbf d_2=\mathbf w( D)$. Combining \eqref{thetaper} with \eqref{abeljumpA} gives
\begin{equation}\label{FHjumpA}
\begin{aligned}
&F_{s-}(\lambda)=F_{s+}(\lambda),\quad\quad\ \ H_{s-}(\lambda)=H_{s+}(\lambda), && \lambda\in(\hat E_{j-1},E_j),\quad  j=1,\dots,n,\\
&F_{s-}(\lambda)=e^{-2\pi\rmi c_j}H_{s+}(\lambda),\quad H_{s-}(\lambda)=e^{2\pi\rmi c_j}F_{s+}(\lambda), && \lambda\in(E_j,\hat E_j),\quad j=0,1,\dots,n,
\end{aligned}
\end{equation}
where we put $c_0=0$. Therefore, the matrix
\begin{equation}\label{NA}
N(\lambda)=\begin{pmatrix}F_1(\lambda)&H_1(\lambda)\\ F_2(\lambda)&H_2(\lambda)\end{pmatrix}
\end{equation}
satisfies
\begin{equation*}
N_-(\lambda)=N_+(\lambda)J_{N}(\lambda),
\end{equation*}
where
\begin{equation}\label{JAN}
J_{N}(\lambda)=\begin{cases}I, & \lambda\in(\hat E_{j-1},E_j),\ j=1,\dots,n,\\[2pt] \begin{pmatrix}0&e^{2\pi\rmi c_j}\\ e^{-2\pi\rmi c_j}&0\end{pmatrix}, & \lambda\in(E_j,\hat E_j),\ j=0,1,\dots,n.\end{cases}
\end{equation}
We then set
\begin{equation}\label{MA}
M(\lambda)=\frac12\begin{pmatrix}F_1^{-1}(\infty)&0\\0&H_2^{-1}(\infty)\end{pmatrix}\begin{pmatrix}\big(\tau(\lambda)+\tau^{-1}(\lambda)\big)F_1(\lambda)&\big(\tau(\lambda)-\tau^{-1}(\lambda)\big)H_1(\lambda)\\[2pt] \big(\tau(\lambda)-\tau^{-1}(\lambda)\big)F_2(\lambda)&\big(\tau(\lambda)+\tau^{-1}(\lambda)\big)H_2(\lambda)\end{pmatrix},
\end{equation}
the divisor $ D$ being chosen so that $\tau\pm\tau^{-1}$ vanishes at every pole of $F_s$ and $H_s$. The divisor may be taken from the zeros of
\begin{equation}\label{divpoly}
\prod_{j=0}^n(\lambda-E_j)-\prod_{j=0}^n(\lambda-\hat E_j),
\end{equation}
which has degree $n$ and changes sign across each gap $(\hat E_{j-1},E_j)$, lifted to the sheet where $\tau-\tau^{-1}=0$. For this choice,
\begin{equation*}
h=\prod_{j=0}^n(\lambda-E_j)/w=\tau^2
\end{equation*}
on the upper sheet and
\begin{equation*}
\operatorname{div}(h-1)=\mathbf D+\infty^+-\sum_{j=0}^n\hat E_j.
\end{equation*}
 Then, Abel's theorem and \eqref{abeljumpA} give $\mathbf A(\infty^+)+\mathbf A(\mathbf D)+\mathbf K\in\mathbb R^n$ modulo periods, and $B=\rmi Y$ with $Y$ positive definite then gives $\Theta>0$ there. In particular the normalization factors in \eqref{MA} do not vanish for real phases. If a root of \eqref{divpoly} is $\lambda=0$, the corresponding factor is removable there. In this section $\mathbf D$ denotes this auxiliary divisor; it need not coincide with the pole divisor of the vector function $\Phi$ in Section~\ref{sec2}. With this choice, $M(\lambda)$ satisfies the analyticity, normalization conditions and
\begin{equation}\label{MAjump}
M_-(\lambda)=M_+(\lambda)\begin{pmatrix}0&\rmi e^{2\pi\rmi c_j}\\ \rmi e^{-2\pi\rmi c_j}&0\end{pmatrix},\quad \lambda\in(E_j,\hat E_j),\quad j=0,1,\dots,n.
\end{equation}
This coincides with the jump \eqref{JA} provided
\begin{equation}\label{cj}
c_j=-\frac{yC_j^f+tC_j^g-\phi_j}{2\pi},\quad j=1,\dots,n.
\end{equation}
Accordingly, $M(y,t;\lambda)$ solves RH Problem \ref{rhp32}. Put 
\begin{equation}\label{tilde Phi}
\tilde\Phi(\lambda)=(f(\lambda)-\lambda)y+\hat g(\lambda)t,
\end{equation}
where
\begin{equation*}
\hat g(\lambda)=g(\lambda)-\frac1{4\lambda}.
\end{equation*}
Proposition \ref{prop22} then expresses the finite-genus solution of \eqref{csp} as
\begin{equation}\label{qAG1}
q^{(AG)}(x,t)=e^{2\rmi\tilde\Phi(0)}\lim_{\lambda\to0}\frac{M_{22}(y,t;0)M_{12}(y,t;\lambda)-M_{12}(y,t;0)M_{22}(y,t;\lambda)}{\lambda},
\end{equation}
with
\begin{equation}\label{xAG1}
x(y,t)=y+\tilde\Phi'(0)+\rmi\lim_{\lambda\to0}\frac{M_{22}(y,t;0)M_{11}(y,t;\lambda)-M_{12}(y,t;0)M_{21}(y,t;\lambda)-1}{\lambda},
\end{equation}
$M$ being the matrix \eqref{MA}.

\begin{remark}[Genus zero]\label{rem35}
Let $n=0$, $E_0\hat E_0>0$, and put
\begin{equation*}
m=\frac{E_0+\hat E_0}2,\quad d=E_0\hat E_0,\quad w_0=w(0)=-\operatorname{sgn}(m)\sqrt d,\quad A_0=\frac{E_0-\hat E_0}{4d}.
\end{equation*}
Then $f=w$, $g=w/(4w_0\lambda)$, $M=\frac12\psm{\tau+\tau^{-1}&\tau-\tau^{-1}\\ \tau-\tau^{-1}&\tau+\tau^{-1}}$, and \eqref{qAG1}-\eqref{xAG1} give
\begin{equation*}
q^{(AG)}=A_0e^{2\rmi(w_0y-mt/(4d))},\quad x=-\frac m{w_0}y-\frac{A_0^2}2t,
\end{equation*}
that is, $q^{(AG)}=A_0e^{\rmi(kx+\omega t)}$ with
\begin{equation*}
k=-\frac{2d}m,\quad \omega=\frac1k+\frac{A_0^2k}2,\quad |q^{(AG)}_x|=\frac{\hat E_0-E_0}{2|m|}<1.
\end{equation*}
This plane wave satisfies \eqref{csp} with $\sigma=-1$.
\end{remark}

\subsection{Admissibility and the converse reconstruction}\label{sec3c}

The following converse justifies the parametric formulae \eqref{qAG1}-\eqref{xAG1}.

\begin{proposition}\label{prop33}
Let $\psi$ solve RH Problem~\ref{rhp24} for $\sigma=-1$, with jump matrices of determinant one and the inverse fourth-root behavior at $E_j,\hat E_j$. Suppose that $\psi(y,t;\lambda)e^{\frac{\rmi t}{4\lambda}\sigma_3}$ is holomorphic at $\lambda=0$ and that, near $\lambda=0$,
\[
\psi(y,t;\lambda)e^{\frac{\rmi t}{4\lambda}\sigma_3}=A(y,t)\Big[I+\lambda\big(Q(y,t)+\rmi(y-x(y,t))\sigma_3\big)+\mathcal O(\lambda^2)\Big]e^{-\rmi\lambda y\sigma_3}.
\]
Then $x(y,t)$ is real and the parametric fields obey
\begin{equation}\label{conv}
q_{yt}=-x_yq,\quad x_t=-\frac{1}{2}|q|^2,\quad x_y^2-|q_y|^2=1 .
\end{equation}
\end{proposition}

\begin{proof}
The determinant is entire and equals one by the normalization, and at the endpoints it is single-valued of order at most $|\lambda-p|^{-1/2}$, so the singularities are removable. The logarithmic derivatives $\psi_y\psi^{-1}$ and $\psi_t\psi^{-1}$ have trivial jumps. At $\lambda=\infty$ we have
\begin{equation*}
\psi_y\psi^{-1}=-\rmi\lambda\sigma_3+\mathcal O(1), \quad \psi_t\psi^{-1}=\mathcal O(\lambda^{-1}).
\end{equation*}
At $\lambda=0$, the $y$-derivative is regular and the residue of the $t$-derivative is $-\frac{\rmi}{4}A\sigma_3A^{-1}$. By Liouville's theorem, $\Phi=A^{-1}\psi$ satisfies
\[
\Phi_y=\lambda L_y\Phi,\quad L_y=-\rmi A^{-1}\sigma_3A,
\]
and a $t$-equation whose singular part is $-\frac{\rmi}{4\lambda}\sigma_3-\frac{\rmi}{2}Q\sigma_3$. Matching the coefficients of $\lambda^{-1}$, $\lambda^0$ and $\lambda$ in the compatibility condition of the two equations gives the first relation of \eqref{conv}, and $L_y^2=-I$ gives the third. the second relation follows from the $t$-equation itself. With
\begin{equation*}
\Phi=\big(I+\lambda H+\mathcal O(\lambda^2)\big)e^{-\rmi\lambda y\sigma_3}e^{-\frac{\rmi t}{4\lambda}\sigma_3}, \quad H=Q+\rmi(y-x)\sigma_3.
\end{equation*}
the function $\Phi_t\Phi^{-1}$ has no term of order $\lambda$. The diagonal part of this coefficient is
\begin{equation*}
-\rmi x_t\sigma_3-\frac{\rmi}{2}|q|^2\sigma_3=0,
\end{equation*}
i.e.\ $x_t=-\frac12|q|^2$, with no integration constant. Besides, the first row $(a,b)$ of $A$ gives
\begin{equation*}
x_y=|a|^2+|b|^2\ge1.
\end{equation*}
With $x_y\ge1$, \eqref{conv} is \eqref{csp} written in $(y,t)$, $x(\cdot,t)$ is a diffeomorphism, and
\begin{equation*}
|q_x|^2=\frac{|q_y|^2}{x_y^2}=1-\frac1{x_y^2}<1.
\end{equation*}
\end{proof}

\section{The Jost functions}\label{sec4}

Let $q^{(AG)}$ be as in \eqref{qAG1}-\eqref{xAG1}, and let $p_0,y_0,\eta_0,G_0$ denote the quantities constructed from $q^{(AG)}$ in Section~\ref{sec2}, with the integration constants in $y_0$ and $\eta_0$ chosen so that \eqref{qAG1}-\eqref{xAG1} hold. Take $E=\{E_j,\hat E_j\}_{j=0}^n$ and
\begin{equation}\label{Psi0}
\Psi_0(x,t;\lambda)=G_0(x,t)e^{-\eta_0(x,t)\sigma_3}\psi\big(y_0(x,t),t;\lambda\big),
\end{equation}
so that $\Psi_0$ solves \eqref{lax} with $q=q^{(AG)}$. The Jost solutions $\Psi^\pm$ are defined through the Volterra equations
\begin{equation}\label{volterra}
\Psi^\pm(x,t;\lambda)=\Psi_0(x,t;\lambda)+\int_{\pm\infty}^x\Psi_0(x,t;\lambda)\Psi_0(s,t;\lambda)^{-1}\,\lambda\big(Q-Q^{(AG)}\big)_s(s,t)\,\Psi^\pm(s,t;\lambda)\,\rmd s,
\end{equation}
where
\begin{equation*}
Q^{(AG)}=\begin{pmatrix}0&q^{(AG)}\\ \overline{q^{(AG)}}&0\end{pmatrix},
\end{equation*}
By \eqref{cons}, the quantities
\begin{equation}\label{Delta}
\Delta=\int_{\mathbb R}(p-p_0)(x,t)\,\rmd x,\quad \eta_\Delta=\int_{\mathbb R}\big(\eta_x-\eta_{0,x}\big)(x,t)\,\rmd x,
\end{equation}
are independent of $t$. We therefore normalize $y$ and $\eta$ so that
\begin{equation*}
y-y_0\to0, \quad \eta-\eta_0\to0, \quad x\to-\infty,
\end{equation*}
obtaining
\begin{equation*}
y-y_0\to\Delta, \quad \eta-\eta_0\to\eta_\Delta, \quad x\to+\infty.
\end{equation*}
The two normalization constants are fixed by these conditions, which determine the gauge freedom $G\mapsto Ge^{c\sigma_3}$ of \eqref{Mtasy} with a constant $c$. The phase is determined by
\begin{equation}\label{gauge}
\eta_x=\big(G^{-1}G_x\big)_{11},\quad \eta_t=\big(G^{-1}G_t\big)_{11}+\frac{\rmi}{2}\big(G^{-1}Q\sigma_3G\big)_{11},
\end{equation}
and the compatibility condition $(\eta_x)_t=(\eta_t)_x$ holds by \eqref{csp}. the first identity is the one used in \eqref{Mtasy}, and the two lines there refer to $x\to-\infty$ and $x\to+\infty$, respectively. The shifts $\Delta$ and $\eta_\Delta$ are the boundary fluxes of $y-y_0$ and $\eta-\eta_0$. Conservation of $\Delta$ in \eqref{Delta} requires the flux of \eqref{cons} to be balanced at the two ends, and conservation of $\eta_\Delta$ requires in addition that the difference of the gauge fluxes in \eqref{gauge} vanish as $x\to\pm\infty$.

Next let
\begin{equation}\label{mu}
\mu^\pm(x,t;\lambda)=e^{\eta\sigma_3}G^{-1}\Psi^\pm(x,t;\lambda)\,e^{\rmi(f(\lambda)y+g(\lambda)t)\sigma_3}.
\end{equation}
From \eqref{cj} we have
\begin{equation*}
M(y-\Delta,t;\lambda;\bphi)=M(y,t;\lambda;\bphi+\Delta\mathbf C^f),
\end{equation*}
with $\mathbf C^f=(C_1^f,\dots,C_n^f)$, and then
\begin{equation}\label{muasy}
\begin{aligned}
&\mu^-(x,t;\lambda)=e^{\rmi(f_0y+g_0t)\sigma_3}M(y,t;\lambda;\bphi)+o(1), && x\to-\infty,\\
&\mu^+(x,t;\lambda)=e^{(\eta_\Delta-\rmi f_0\Delta)\sigma_3}e^{\rmi(f_0y+g_0t)\sigma_3}M(y,t;\lambda;\bphi+\Delta\mathbf C^f)\,e^{\rmi f(\lambda)\Delta\sigma_3}+o(1), && x\to+\infty.
\end{aligned}
\end{equation}
The columns $\mu_1^-,\mu_2^+$ are holomorphic in $\mathbb C^+$ and $\mu_1^+,\mu_2^-$ in $\mathbb C^-$. All four extend continuously to $\mathbb R\setminus E$. The two Jost solutions are related by a scattering matrix $S(\lambda)$. Namely,
\begin{equation}\label{S}
\Psi^+(x,t;\lambda)=\Psi^-(x,t;\lambda)S(\lambda),\quad \lambda\in\mathbb R\setminus\bigcup_{j=0}^n[E_j,\hat E_j],\quad \det S(\lambda)=1,
\end{equation}
which lead to
\begin{equation}\label{Sentries}
S_{11}=|\Psi_1^+,\Psi_2^-|,\quad S_{22}=|\Psi_1^-,\Psi_2^+|,\quad S_{12}=|\Psi_2^+,\Psi_2^-|,\quad S_{21}=|\Psi_1^-,\Psi_1^+|.
\end{equation}
The diagonal entries $S_{11}$ and $S_{22}$ extend holomorphically to $\mathbb C^-$ and $\mathbb C^+$, respectively, and
\begin{equation}\label{Ssym}
S_{11}(\lambda)=\overline{S_{22}(\bar \lambda)},\quad S_{12}(\lambda)=\overline{S_{21}(\bar \lambda)}.
\end{equation}
We assume that $S_{11}$ has no transmission zeros on $\overline{\mathbb C^-}\setminus E$. The same nonvanishing holds at the nonendpoint boundary values of the bands, which are reached by continuation. Define the reflection coefficients by
\begin{equation}\label{r}
r_1(\lambda)=\frac{S_{12}(\lambda)}{S_{11}(\lambda)},\quad r_2(\lambda)=\frac{S_{21}(\lambda)}{S_{11}(\lambda)},\quad \lambda\in\mathbb R\setminus\bigcup_{j=0}^n[E_j,\hat E_j],
\end{equation}
they satisfy
\begin{equation*}
|r_{1}(\lambda)|<1, \quad |r_{2}(\lambda)|<1,
\end{equation*}
and
\begin{equation*}
r_{1}(\lambda)=\mathcal O(\lambda^2), \quad r_{2}(\lambda)=\mathcal O(\lambda^2), \quad \lambda\to0.
\end{equation*}
On each band $(E_j,\hat E_j)$, \eqref{JcaseA} yields
\begin{equation}\label{bandrel}
\Psi^-_{2,-}=\rmi e^{\rmi\phi_j}\Psi^-_{1,+},\quad \Psi^+_{1,-}=\rmi e^{-\rmi\phi_j}\Psi^+_{2,+},\quad S_{11,-}=S_{22,+}.
\end{equation}
If the perturbation is supported in $[a,b]$ at $t=0$, the scattering matrix has the closed form
\begin{equation}\label{S0}
S(\lambda)=\Psi_0(a,0;\lambda)^{-1}Y(a,b;\lambda)\Psi_0(b,0;\lambda),
\end{equation}
where $Y(a,b;\lambda)$ propagates the spatial equation of \eqref{lax} from $b$ to $a$. Entire in $\lambda$, the propagator satisfies
\begin{equation*}
S(\lambda)=I+\mathcal O(\lambda^2), \quad \lambda\to0
\end{equation*}
by virtue of $\int_a^b\big(Q-Q^{(AG)}\big)_x\rmd x=0$, in agreement with \eqref{r}. At a band endpoint $p\in E$, let $c_p$ denote the coefficient of $(\lambda-p)^{-1/2}$ in $S_{11}$ on the gap side of $p$. We call the endpoint generic if $c_p\neq0$. At $t=0$ the Jost functions are obtained from the analytic transfer matrices of \eqref{S0} applied to the background, so $c_p$ is the determinant of the leading singular columns.
\begin{proposition}\label{pro61}
Assume that every endpoint $p\in E$ is generic. Then, for $j=0,\dots,n$, the reflection coefficients take the endpoint values
\[
r_2(\hat E_j)=-e^{-\rmi\phi_j},\quad r_1(\hat E_j)=e^{\rmi\phi_j},\quad r_2(E_j)=e^{-\rmi\phi_j},\quad r_1(E_j)=-e^{\rmi\phi_j}.
\]
\end{proposition}

\begin{proof}
Under the nonresonance condition $c_p\neq0$, the leading singular coefficients of the Jost solutions at $p$ inherit the background relations $\Psi_2\sim\pm e^{\mathrm i\phi_j}\Psi_1$ through the analytic transfer matrices, so the endpoint values follow from the ratio of the leading $(\lambda-p)^{-1/2}$ coefficients in the Wronskian quotients $r_2=S_{21}/S_{11}$ and $r_1=S_{12}/S_{11}$.
\end{proof}

The nonresonance condition $c_p\neq0$, $p\in E$, is assumed from now on. According to $\det S=1$ and $|S_{11}|=|S_{22}|$ on a gap, this gives
\begin{equation}\label{edge}
1-|r_1(\lambda)|^2=1-|r_2(\lambda)|^2=|S_{11}(\lambda)|^{-2}\sim|c_p|^{-2}|\lambda-p|,
\end{equation}
which is the edge behaviour used in the scalar conjugation and in the local parametrices. For the exact background $S=I$ and $r_1=r_2=0$, so the endpoint values of Proposition~\ref{pro61} require this hypothesis. For the defocusing equation the derivatives of the phase functions factor as
\begin{equation}\label{fg}
f'(\lambda)=\frac{P_f(\lambda)}{w(\lambda)},\quad g'(\lambda)=\frac{c_g P_g(\lambda)}{\lambda^2w(\lambda)},\quad P_f(\lambda)=\prod_{k=0}^n(\lambda-\lambda_k^f),\quad P_g(\lambda)=\prod_{k=0}^n(\lambda-\lambda_k^g),
\end{equation}
the points $\lambda_k^f,\lambda_k^g\in(E_k,\hat E_k)$ being determined by
\begin{equation*}
\int_{E_j}^{\hat E_j}\rmd f=\int_{E_j}^{\hat E_j}\rmd g=0, \quad j=0,\dots,n,
\end{equation*}
and the constant by $c_gP_g(0)=-\frac14w(0)$. In particular $c_g<0$. The system $\int_{E_j}^{\hat E_j}\rmd f=\int_{E_j}^{\hat E_j}\rmd g=0$, $j=0,\dots,n$, comprising $n+1$ equations for $n+1$ unknowns, is square and uniquely solvable. Each vanishing moment forces a nonzero real polynomial of degree at most $n$ to change sign and hence to vanish in the interior of a different band. Such a polynomial can therefore have a zero moment on at most $n$ of the $n+1$ bands. Hence $P_f$ and $P_g$ are uniquely determined, and each has exactly one simple root in every band. The phase function is
\begin{equation}\label{thetadef}
\theta(\lambda)=\theta(\lambda;\xi)=-\big(f(\lambda)-f_0\big)\xi-\big(g(\lambda)-g_0\big),\quad \xi=\frac yt.
\end{equation}
and a direct computation shows
\begin{equation}\label{thetap}
\begin{aligned}
&\frac{\partial\theta(\lambda;\xi)}{\partial \lambda}=-\frac{T(\lambda;\xi)}{\lambda^2w(\lambda)},\quad \Xi(\lambda)=-\frac{g'(\lambda)}{f'(\lambda)},\\
&T(\lambda;\xi)=\xi \lambda^2P_f(\lambda)+c_gP_g(\lambda)=\lambda^2P_f(\lambda)\big(\xi-\Xi(\lambda)\big).
\end{aligned}
\end{equation}

\begin{lemma}\label{lem4}
On $\mathbb R\setminus\big(\bigcup_{j=0}^n[E_j,\hat E_j]\cup\{0\}\big)$, one has $f'>0$, $\Xi>0$ and
\begin{equation}\label{lem1sgn}
\operatorname{sgn}\Xi'(\lambda)=-\operatorname{sgn}\lambda,
\end{equation}
The same sign holds at $E_j,\hat E_j$. For every $\xi>0$, the polynomial $T(\cdot;\xi)$ has $n+3$ simple real zeros, one in the interior of each band and one further on each open half-line.
\end{lemma}

\begin{proof}
The position of the roots of $P_f$ and $P_g$ imply that $f'>0$, $\Xi>0$ and $c_g<0$. Since $\int_{E_j}^{\hat E_j}\rmd f=0$, the function $T(\cdot;\xi)$ changes sign at least once in the interior of every band. Let $N_+$ and $N_-$ count the bands on the positive and on the negative half-line. Notice that
\begin{equation*}
\operatorname{sgn}w(0)=(-1)^{N_+}, \quad T(0;\xi)=c_gP_g(0)=-\frac14w(0),
\end{equation*}
thus the number of sign changes on the positive half-line has parity $N_++1$, and that on the negative half-line has parity $N_-+1$, the latter following from the sign $(-1)^{n+3}$ of $T(\cdot;\xi)$ at $-\infty$. This yields at least $N_++1$ and $N_-+1$ distinct zeros of odd multiplicity, which account for the degree $n+3$, so all zeros are simple and the counts are exact.

Let $\lambda$ be a zero of $T(\cdot;\Xi(\lambda))$ in a gap or at a band endpoint. By the first part it is simple, hence
\begin{equation*}
T'(\lambda;\Xi(\lambda))=-\lambda^2P_f(\lambda)\Xi'(\lambda)\neq0 .
\end{equation*}
Every complete band on the same half-line contains one zero of each of $T(\cdot;\Xi(\lambda))$ and $P_f$, so $\operatorname{sgn}\big(T/P_f\big)$ does not change across that band. For increasing $\lambda$, this quotient changes sign at each of its zeros-from negative to positive at a positive zero and from positive to negative at a negative one, which gives \eqref{lem1sgn} on the gaps and at the endpoints.
\end{proof}

Two endpoints on the same half-line must have distinct critical values, for otherwise $T(\cdot;\xi)$ would have two further zeros on that half-line for the same value of $\xi$. Consequently
\begin{equation}\label{sgn}
\operatorname{sgn}\partial_\lambda\theta(\lambda;\xi)=\operatorname{sgn}\big(\Xi(\lambda)-\xi\big),\quad \lambda\in\mathbb R\setminus\Big(\bigcup_{j=0}^n[E_j,\hat E_j]\cup\{0\}\Big).
\end{equation}
The critical values are
\begin{equation}\label{xij}
\xi_j=\Xi(E_j)>0,\quad \hat\xi_j=\Xi(\hat E_j)>0,\quad T(E_j;\xi_j)=T(\hat E_j;\hat\xi_j)=0,\quad j=0,\dots,n.
\end{equation}
If $\lambda(\xi)$ is a simple real zero of $T(\cdot;\xi)$, then
\begin{equation}\label{dzdxi}
\frac{\rmd \lambda(\xi)}{\rmd\xi}=\frac{1}{\Xi'(\lambda(\xi))}.
\end{equation}

By \eqref{thetap}, the zeros of $T(\cdot;\xi)$  in a gap are precisely the solutions of $\Xi(\lambda)=\xi$, so $\mathcal Z(\xi)$ in \eqref{Zxi} is the set of zeros of $T(\cdot;\xi)$ in $\mathbb R\setminus\bigcup_j[E_j,\hat E_j]$. Since $\Xi>0$ on each gap, $\mathcal Z(\xi)=\emptyset$ for $\xi\le0$. Moreover, by Lemma~\ref{lem4}, $\Xi$ is strictly monotone on each gap, and hence each gap contains at most one point of $\mathcal Z(\xi)$. Finally, for $\xi\neq0$, the polynomial $T(\cdot;\xi)$ has degree $n+3$ and vanishes in every band, so $\mathcal Z(\xi)$ has at most two elements.

\section{Asymptotic analysis in transition region I}\label{sec5}
In this section, the transition window $|\xi-\hat\xi_{j_1}|t^{2/3}\le C$ is considered. We first treat in detail the half $-C\le\ve(\xi-\hat\xi_{j_1})t^{2/3}\le0$. The complementary half is treated in Section~\ref{sec5neg}. Set
\begin{equation*}
\ve=\operatorname{sgn}\hEj, \quad j_1\in\{0,\dots,n\}.
\end{equation*}
Let $\lambda_*$ be the zero of $T(\cdot;\xi)$ that merges with $\hEj$ as $\xi\to\hat\xi_{j_1}$. In the complementary half it lies in the band $(E_{j_1},\hEj)$ and $s<0$ in \eqref{zeta}. In the half treated first, when $t$ is large, the saddle point $\lambda_*\in\mathcal Z(\xi)$ obeys
\begin{equation*}
\lambda_*-\hEj=(\xi-\hat\xi_{j_1})/\Xi'(\hEj)+\mathcal O(t^{-4/3}), \quad \lambda_*\in[\hEj,b),
\end{equation*}
where
\begin{equation*}
b=\begin{cases}0, & 0\in(\hEj,E_{j_1+1}),\\E_{j_1+1}, & \text{otherwise},\end{cases}\quad E_{n+1}=+\infty.
\end{equation*}
Set
\begin{equation*}
y_\ve=\begin{cases}y, & \ve=-1,\\y-\Delta, & \ve=1,\end{cases}\quad
\end{equation*}
and introduce
\begin{equation}\label{M}
\Mc(\lambda)=\begin{cases}\Big(\mu_1^-,\ \dfrac{\mu_2^+}{S_{22}}\Big)e^{-\rmi(f_0y+g_0t)\sigma_3}, & \lambda\in\mathbb C^+,\\[10pt] \Big(\dfrac{\mu_1^+}{S_{11}},\ \mu_2^-\Big)e^{-\rmi(f_0y+g_0t)\sigma_3}, & \lambda\in\mathbb C^-,\end{cases}\quad \ve=-1,
\end{equation}
\begin{equation}\label{MM}
\Mc(\lambda)=\begin{cases}e^{-c_\ve\sigma_3}\Big(\dfrac{\mu_1^-}{S_{22}},\ \mu_2^+\Big)e^{-\rmi(f(\lambda)\Delta+f_0y+g_0t)\sigma_3}, & \lambda\in\mathbb C^+,\\[10pt] e^{-c_\ve\sigma_3}\Big(\mu_1^+,\ \dfrac{\mu_2^-}{S_{11}}\Big)e^{-\rmi(f(\lambda)\Delta+f_0y+g_0t)\sigma_3}, & \lambda\in\mathbb C^-,\end{cases}\quad \ve=1,
\end{equation}
with
\begin{equation*}
c_\ve=\begin{cases}0, & \ve=-1,\\\eta_\Delta-\rmi f_0\Delta, & \ve=1.\end{cases}\quad
\end{equation*}
The reflection coefficient is taken to be
\begin{equation*}
r=\begin{cases}r_2, & \ve=-1,\\r_1e^{2\rmi f(\lambda)\Delta}, & \ve=1.\end{cases}\quad
\end{equation*}

\begin{rhp}\label{rhp51}
The matrix $\Mc(\lambda)$ has the following properties.
\begin{conds}
\cond{1}{Normalization:} $\Mc(\lambda)=I+\mathcal O(\lambda^{-1})$ as $\lambda\to\infty$.
\cond{2}{Analyticity:} $\Mc(\lambda)$ is analytic in $\mathbb C\setminus\mathbb R$ and bounded near $\lambda=0$.
\cond{3}{Jump condition:} $\Mc_+(\lambda)=\Mc_-(\lambda)J_{\Mc}(\lambda)$, where
\begin{equation}\label{JM}
J_{\Mc}(\lambda)=e^{\rmi\left(f_0y+g_0t-\frac12(C_j^fy_\ve+C_j^gt-\phi_j)\right)\hat\sigma_3}\begin{pmatrix}0&-\rmi\\-\rmi&0\end{pmatrix},\quad \lambda\in(E_j,\hat E_j),\ j=0,\dots,n,
\end{equation}
\begin{equation}\label{JMgap}
J_{\Mc}(\lambda)=\begin{cases}\begin{pmatrix}1&\overline{r(\lambda)}e^{2\rmi t\theta(\lambda)}\\-r(\lambda)e^{-2\rmi t\theta(\lambda)}&1-|r(\lambda)|^2\end{pmatrix}, & \ve=-1,\\[12pt] \begin{pmatrix}1-|r(\lambda)|^2&r(\lambda)e^{2\rmi t\theta(\lambda)}\\-\overline{r(\lambda)}e^{-2\rmi t\theta(\lambda)}&1\end{pmatrix}, & \ve=1,\end{cases}\quad \lambda\in\mathbb R\setminus\bigcup_{j=0}^n[E_j,\hat E_j],
\end{equation}
with $C_0^f=C_0^g=\phi_0=0$.
\cond{4}{Local behavior:} For $p\in E$,
\begin{equation}\label{Mloc}
\Mc(\lambda)=\begin{cases}\big(\mathcal O((\lambda-p)^{\ve/4}),\ \mathcal O((\lambda-p)^{-\ve/4})\big), & \lambda\to p\ \text{from}\ \mathbb C^+,\\[2pt] \big(\mathcal O((\lambda-p)^{-\ve/4}),\ \mathcal O((\lambda-p)^{\ve/4})\big), & \lambda\to p\ \text{from}\ \mathbb C^-.\end{cases}
\end{equation}
\end{conds}
\end{rhp}

Let
\begin{equation*}
\begin{aligned}
&\hat g(\lambda)=g(\lambda)-\frac1{4\lambda}, \quad \tilde\Phi_\ve(\lambda)=(f(\lambda)-\lambda)y_\ve+\hat g(\lambda)t,\\
&\varphi_\ve=f(0)y_\ve-f_0y+(\hat g(0)-g_0)t
\end{aligned}
\end{equation*}
For $\lambda\to0$, one has
\begin{equation}\label{M0}
\Mc(\lambda)=e^{-c_\ve\sigma_3}e^{\eta\sigma_3}G^{-1}e^{\rmi\varphi_\ve\sigma_3}\Big\{I+\lambda\Big(e^{-\rmi\varphi_\ve\hat\sigma_3}Q+\rmi\big(y_\ve+\tilde\Phi_\ve'(0)-x\big)\sigma_3\Big)+\mathcal O(\lambda^2)\Big\}.
\end{equation}
Therefore, the solution is recovered through
\begin{equation}\label{recq5}
q(x,t)=e^{2\rmi\varphi_\ve}\lim_{\lambda\to0}\frac{\big(\Mc^{-1}(y,t;0)\Mc(y,t;\lambda)\big)_{12}}{\lambda},
\end{equation}
\begin{equation}\label{recx5}
x(y,t)=y_\ve+\tilde\Phi_\ve'(0)+\rmi\lim_{\lambda\to0}\frac{\big(\Mc^{-1}(y,t;0)\Mc(y,t;\lambda)\big)_{11}-1}{\lambda}.
\end{equation}

\begin{figure}[htbp]
\centering
\includegraphics[width=0.72\textwidth]{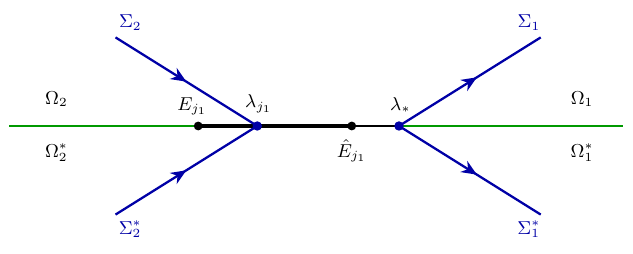}
\caption{The contour $\Sigma^{(1)}$ obtained after opening the lenses.}\label{fig9}
\end{figure}

To open lenses, introduce the sets
\[
\begin{aligned}
&S_1(\xi)=\Big\{\lambda\in\mathbb R\setminus\textstyle\bigcup_j[E_j,\hat E_j]:\ \ve\big(\Xi(\lambda)-\xi\big)<0\Big\},\\
& S_2(\xi)=\Big\{\lambda\in\mathbb R\setminus\textstyle\bigcup_j[E_j,\hat E_j]:\ \ve\big(\Xi(\lambda)-\xi\big)>0\Big\}\setminus(\hEj,\lambda_*).
\end{aligned}
\]
The set $S_2(\xi)$ is the integration set of the scalar function $\delta$ in \eqref{delta}. On the gaps, we write
\begin{equation*}
L(s)=\log\big(1-|r(s)|^2\big).
\end{equation*}
Since $|r_1|=|r_2|$ and $|e^{2\rmi f\Delta}|=1$ on the gaps, $L$ coincides with $\log(1-|r_2|^2)$ for both choices of $r$; by \eqref{edge}, $L$ has logarithmic singularities at the endpoints in $E$ and is integrable. Pick a point $\lambda_{j_1}\in(E_{j_1},\hEj)$, and set
\[
\begin{aligned}
&\Omega_1=\{\lambda:0\le\arg(\lambda-\lambda_*)\le\varphi_1\},&& \Sigma_1=\lambda_*+e^{\rmi\varphi_1}\mathbb R^+,\\
&\Omega_2=\{\lambda:\pi-\varphi_1\le\arg(\lambda-\lambda_{j_1})\le\pi\},&& \Sigma_2=\lambda_{j_1}+e^{\rmi(\pi-\varphi_1)}\mathbb R^+,
\end{aligned}
\]
with the angle $\varphi_1$ small enough that
\[
\ve\operatorname{Im}\theta(\lambda)<0,\ \lambda\in\Sigma_1\setminus\{\lambda_*\};\quad \ve\operatorname{Im}\theta(\lambda)>0,\ \lambda\in\Sigma_2\setminus\{\lambda_{j_1}\}.
\]
Outside a neighbourhood of $\hEj$, the domains $\Omega_1$ and $\Omega_2$ are extended to lenses over $S_1$ and $ S_2$, respectively, and carried through the bands with the help of (\ref{bandid}). Figure~\ref{fig9} shows the resulting contour. For a set $U$ we write $U^*$ for its reflection in the real axis. On the gaps, $J_{\Mc}$ factorizes as
\begin{equation}\label{fact}
\begin{aligned}
\ve=-1:\quad &J_{\Mc}=\psm{1&0\\-re^{-2\rmi t\theta}&1}\psm{1&\bar re^{2\rmi t\theta}\\0&1}, && \lambda\in S_1,\\
&J_{\Mc}=\psm{1&\frac{\bar re^{2\rmi t\theta}}{1-|r|^2}\\0&1}(1-|r|^2)^{-\sigma_3}\psm{1&0\\-\frac{re^{-2\rmi t\theta}}{1-|r|^2}&1}, && \lambda\in S_2\cup(\hEj,\lambda_*),\\
\ve=1:\quad &J_{\Mc}=\psm{1&re^{2\rmi t\theta}\\0&1}\psm{1&0\\-\bar re^{-2\rmi t\theta}&1}, && \lambda\in S_1,\\
&J_{\Mc}=\psm{1&0\\-\frac{\bar re^{-2\rmi t\theta}}{1-|r|^2}&1}(1-|r|^2)^{\sigma_3}\psm{1&\frac{re^{2\rmi t\theta}}{1-|r|^2}\\0&1}, && \lambda\in S_2\cup(\hEj,\lambda_*).
\end{aligned}
\end{equation}
On each band $(E_j,\hat E_j)$, one has
\begin{equation*}
J_{\Mc}=e^{\frac\rmi2(t(\theta_++\theta_-)+\tilde\phi_j)\hat\sigma_3}\psm{0&-\rmi\\-\rmi&0}, \quad \tilde\phi_j=\phi_j+(y-y_\ve)C_j^f.
\end{equation*}
Moreover, every function $\rho$ on $(E_j,\hat E_j)$ gives the identities
\begin{equation}\label{bandid}
J_{\Mc}=\begin{pmatrix}1&\rho e^{2\rmi t\theta_-}\\0&1\end{pmatrix}J_{\Mc}\begin{pmatrix}1&0\\-\rho e^{-2\rmi(t\theta_++\tilde\phi_j)}&1\end{pmatrix}=\begin{pmatrix}1&0\\-\rho e^{-2\rmi t\theta_-}&1\end{pmatrix}J_{\Mc}\begin{pmatrix}1&\rho e^{2\rmi(t\theta_++\tilde\phi_j)}\\0&1\end{pmatrix}.
\end{equation}

These factorizations suggest the transformation
\begin{equation}\label{M1}
\Mc^{(1)}(\lambda)=e^{\delta(\infty)\sigma_3}\Mc(\lambda) G(\lambda)e^{-\delta(\lambda)\sigma_3}.
\end{equation}
For $\ve=-1$,
\begin{equation}\label{G-1}
G=\left\{
\begin{aligned}
&\begin{pmatrix}
1&-\bar re^{2\rmi t\theta}\\0&1
\end{pmatrix},\quad &\lambda\in\Omega_1,\\
&\begin{pmatrix}
1&0\\-re^{-2\rmi t\theta}&1
\end{pmatrix},\quad\, \, \  &\lambda\in\Omega_1^*,\\
&\begin{pmatrix}
1&0\\\frac{re^{-2\rmi t\theta}}{1-|r|^2}&1
\end{pmatrix},\quad &\lambda\in\Omega_2,\\
&\begin{pmatrix}
1&\frac{\bar re^{2\rmi t\theta}}{1-|r|^2}\\0&1
\end{pmatrix}, \,\ \quad\ &\lambda\in\Omega_2^*,\\
\end{aligned}
\right.
\end{equation}
and for $\ve=1$,
\begin{equation}\label{G1}
G=\left\{
\begin{aligned}
&\begin{pmatrix}
1&0\\\bar re^{-2\rmi t\theta}&1
\end{pmatrix},\quad &\lambda\in\Omega_1,\\
&\begin{pmatrix}
1&re^{2\rmi t\theta}\\0&1
\end{pmatrix},\quad\, \, \  &\lambda\in\Omega_1^*,\\
&\begin{pmatrix}
1&-\frac{re^{2\rmi t\theta}}{1-|r|^2}\\0&1
\end{pmatrix},\quad &\lambda\in\Omega_2,\\
&\begin{pmatrix}
1&0\\-\frac{\bar re^{-2\rmi t\theta}}{1-|r|^2}&1
\end{pmatrix}, \,\ \quad\ &\lambda\in\Omega_2^*.\\
\end{aligned}
\right.
\end{equation}
Here the scalar function $\delta(\lambda)$ is given by
\begin{equation}\label{delta}
\delta(\lambda)=\frac{w(\lambda)}{2\pi\rmi}\left[\sum_{j=1}^n\delta_j\int_{E_j}^{\hat E_j}\frac{\rmi\,\rmd s}{w_+(s)(s-\lambda)}+\ve\int_{ S_2(\xi)}\frac{\log\big(1-|r(s)|^2\big)}{w(s)(s-\lambda)}\rmd s\right],
\end{equation}
and the constants $\delta_j\in\mathbb R$ are fixed by the linear system
\begin{equation}\label{deltasys}
\sum_{j=1}^n\delta_j\int_{E_j}^{\hat E_j}\frac{\rmi s^k}{w_+(s)}\rmd s=-\ve\int_{ S_2(\xi)}\frac{\log\big(1-|r(s)|^2\big)s^k}{w(s)}\rmd s,\quad k=0,1,\dots,n-1.
\end{equation}
For later use we abbreviate
\begin{equation*}
\mathcal I_m=\sum_{j=1}^n\delta_j\int_{E_j}^{\hat E_j}\frac{\rmi s^m}{w_+(s)}\rmd s+\ve\int_{ S_2}\frac{\log(1-|r(s)|^2)s^m}{w(s)}\rmd s, \quad m\ge0.
\end{equation*}

\begin{rhp}\label{rhp52}
The function $\delta(\lambda)$ solves the scalar RH problem below.
\begin{conds}
\cond{1}{Analyticity:} $\delta(\lambda)$ is holomorphic in $\mathbb C\setminus\big(\bigcup_{j=0}^n[E_j,\hat E_j]\cup\overline{ S_2}\big)$.
\cond{2}{Jump condition:} For $j=0,\dots,n$ and $\delta_0=0$,
\[
\delta_-(\lambda)=\begin{cases}\delta_+(\lambda)-\ve\log\big(1-|r(\lambda)|^2\big), & \lambda\in S_2,\\ -\delta_+(\lambda)+\rmi\delta_j, & \lambda\in(E_j,\hat E_j).\end{cases}
\]
\cond{3}{Normalization:} $\delta(\lambda)=\delta(\infty)+\delta^{(1)}\lambda^{-1}+\mathcal O(\lambda^{-2})$ as $\lambda\to\infty$, the coefficients being
\[
\delta(\infty)=-\frac{\mathcal I_n}{2\pi\rmi}\in\rmi\mathbb R,\quad \delta^{(1)}=-\frac12\delta(\infty)\sum_{j=0}^n(E_j+\hat E_j)-\frac{\mathcal I_{n+1}}{2\pi\rmi}.
\]
\cond{4}{Local behavior:} For $p\in E\cap\partial S_2$, we get
\begin{equation*}
e^{\delta(\lambda)}=\mathcal O((\lambda-p)^{\ve/2}), \quad \lambda\to p\ \text{from} \ \mathbb C^+.
\end{equation*}
In addition,
\begin{equation*}
\delta(\lambda)=\frac\rmi2\delta_{j_1}+\mathcal O((\lambda-\hEj)^{1/2}), \quad \lambda\to\hEj,
\end{equation*}
and
\begin{equation*}
\delta(\lambda)=\mp\rmi\ve\nu_k\log(\lambda_k-\lambda)+\mathcal O(1), \quad \lambda\to \lambda_k\in\partial S_2\cap\mathcal Z(\xi),
\end{equation*}
where
\begin{equation*}
\nu_k=-\frac1{2\pi}\log(1-|r(\lambda_k)|^2).
\end{equation*}
\cond{5}{Symmetry:} $\delta(\lambda)=-\overline{\delta(\bar \lambda)}$.
\end{conds}
\end{rhp}

\begin{rhp}\label{rhp53}
In terms of $\Mc^{(1)}(\lambda)$, the corresponding RH problem takes the following form.
\begin{conds}
\cond{1}{Analyticity:} $\Mc^{(1)}(\lambda)$ is holomorphic in $\mathbb C\setminus\Sigma^{(1)}$, where
\begin{equation*}
\Sigma^{(1)}=\bigcup_{j=0}^n[E_j,\hat E_j]\cup[\hEj,\lambda_*]\cup\Sigma_1\cup\Sigma_1^*\cup\Sigma_2\cup\Sigma_2^*.
\end{equation*}
The rays $\Sigma_1,\Sigma_1^*$ point away from $\lambda_*$, while $\Sigma_2,\Sigma_2^*$ point towards $\lambda_{j_1}$.
\cond{2}{Jump condition:} For $j=0,\dots,n$, $\Mc^{(1)}_+(\lambda)=\Mc^{(1)}_-(\lambda)J^{(1)}(\lambda)$, where
\[
J^{(1)}(\lambda)=e^{\frac\rmi2\left(t(\theta_+(\lambda)+\theta_-(\lambda))+\tilde\phi_j+\delta_j\right)\hat\sigma_3}\begin{pmatrix}0&-\rmi\\-\rmi&0\end{pmatrix},\quad \lambda\in(E_j,\hat E_j).
\]
In addition, for $\ve=-1$,
\[
J^{(1)}=\begin{cases}\psm{1&\bar re^{2\rmi t\theta+2\delta}\\0&1}, & \lambda\in\Sigma_1,\\[4pt] \psm{1&0\\-re^{-2\rmi t\theta-2\delta}&1}, & \lambda\in\Sigma_1^*,\\[4pt] \psm{1&0\\-\frac{re^{-2\rmi t\theta-2\delta}}{1-|r|^2}&1}, & \lambda\in\Sigma_2,\\[4pt] \psm{1&\frac{\bar re^{2\rmi t\theta+2\delta}}{1-|r|^2}\\0&1}, & \lambda\in\Sigma_2^*,\\[4pt] \psm{1&\bar re^{2\rmi t\theta+2\delta}\\-re^{-2\rmi t\theta-2\delta}&1-|r|^2}, & \lambda\in(\hEj,\lambda_*),\end{cases}
\]
and for $\ve=1$,
\[
J^{(1)}=\begin{cases}\psm{1&0\\-\bar re^{-2\rmi t\theta-2\delta}&1}, & \lambda\in\Sigma_1,\\[4pt] \psm{1&re^{2\rmi t\theta+2\delta}\\0&1}, & \lambda\in\Sigma_1^*,\\[4pt] \psm{1&\frac{re^{2\rmi t\theta+2\delta}}{1-|r|^2}\\0&1}, & \lambda\in\Sigma_2,\\[4pt] \psm{1&0\\-\frac{\bar re^{-2\rmi t\theta-2\delta}}{1-|r|^2}&1}, & \lambda\in\Sigma_2^*,\\[4pt] \psm{1-|r|^2&re^{2\rmi t\theta+2\delta}\\-\bar re^{-2\rmi t\theta-2\delta}&1}, & \lambda\in(\hEj,\lambda_*).\end{cases}
\]
\cond{3}{Normalization:} $\Mc^{(1)}(\lambda)=I+\mathcal O(\lambda^{-1})$ as $\lambda\to\infty$.
\cond{4}{Local behavior:} For $p\in E\setminus\{\hEj\}$, $\Mc^{(1)}(\lambda)$ exhibits asymptotic behavior as follows
\begin{equation*}
\Mc^{(1)}(\lambda)=\mathcal O((\lambda-p)^{-1/4}), \quad \lambda\to p.
\end{equation*}
Near $\lambda=0$, $\Mc^{(1)}$ stays bounded and is non-tangentially bounded near each $\lambda_k\in\mathcal Z(\xi)\setminus\{\lambda_*\}$.
\end{conds}
\end{rhp}

\subsection{Global parametrix}
Dropping all jumps except those on the bands leads to the following RH problem.

\begin{rhp}\label{rhp54}
Find $\Mg(\lambda)$ such that:
\begin{conds}
\cond{1}{Analyticity:} $\Mg(\lambda)$ is holomorphic in $\mathbb C\setminus\bigcup_{j=0}^n[E_j,\hat E_j]$.
\cond{2}{Jump condition:} For $j=0,\dots,n$, $\Mg_+(\lambda)=\Mg_-(\lambda)J^{(glo)}(\lambda)$, where
\begin{equation}\label{Jglo}
J^{(glo)}(\lambda)=e^{\rmi\left[f_0y+g_0t-\frac12\left(C_j^fy_\ve+C_j^gt-\phi_j-\delta_j\right)\right]\hat\sigma_3}\begin{pmatrix}0&-\rmi\\-\rmi&0\end{pmatrix},\quad \lambda\in(E_j,\hat E_j).
\end{equation}
\cond{3}{Normalization:} $\Mg(\lambda)=I+\mathcal O(\lambda^{-1})$ as $\lambda\to\infty$.
\cond{4}{Local behavior:} $\Mg(\lambda)=\mathcal O((\lambda-p)^{-1/4})$ as $\lambda\to p\in E$.
\end{conds}
\end{rhp}

A solution is provided by the planar Baker-Akhiezer matrix ,
\begin{equation}\label{Mglo}
\Mg(\lambda)=e^{\rmi(f_0y+g_0t)\sigma_3}D^{(glo)}(\infty)^{-1}D^{(glo)}(\lambda)e^{-\rmi(f_0y+g_0t)\sigma_3},
\end{equation}
in which $D^{(glo)}=(D^{(glo)}_{sl})_{s,l=1,2}$ has the entries
\[
\begin{aligned}
&D^{(glo)}_{s1}(\lambda)=\frac12\big(\tau(\lambda)+(-1)^{s-1}\tau(\lambda)^{-1}\big)\frac{\Theta(\mathbf A(\lambda)+\mathbf c^*+\mathbf d_s)}{\Theta(\mathbf A(\lambda)+\mathbf d_s)},\\
&D^{(glo)}_{s2}(\lambda)=\frac12\big(\tau(\lambda)+(-1)^{s}\tau(\lambda)^{-1}\big)\frac{\Theta(-\mathbf A(\lambda)+\mathbf c^*+\mathbf d_s)}{\Theta(-\mathbf A(\lambda)+\mathbf d_s)},
\end{aligned}
\]
where the phase vector is $\mathbf c^*=(c_1^*,\dots,c_n^*)^T$ with
\begin{equation*}
c_j^*=-\dfrac{y_\ve C_j^f+tC_j^g-\phi_j-\delta_j}{2\pi}.
\end{equation*}
The corresponding finite-gap solution reads
\begin{equation}\label{qglo}
q^{(AG)}(x,t;\mathbf E,\hat{\mathbf E},\bphi+\boldsymbol\delta)=e^{2\rmi\tilde\Phi_\ve(0)}\lim_{\lambda\to0}\frac{\big(D^{(glo)}(0)^{-1}D^{(glo)}(\lambda)\big)_{12}}{\lambda},
\end{equation}
together with
\begin{equation}\label{xglo}
x(y,t)=y_\ve+\tilde\Phi_\ve'(0)+\rmi\lim_{\lambda\to0}\frac{\big(D^{(glo)}(0)^{-1}D^{(glo)}(\lambda)\big)_{11}-1}{\lambda},
\end{equation}
where $\boldsymbol\delta=(\delta_1,\dots,\delta_n)$.

\subsection{Local parametrix}

Take the disk $U_1=\{\lambda\in\mathbb C:|\lambda-\hEj|<\rho_1\}$ with a $t$-independent radius $\rho_1>0$ satisfying
\begin{equation}\label{U1}
\rho_1<\frac12\min\Big\{|\hEj|,\ \hEj-\lambda_{j_1},\ b-\hEj,\ \operatorname{dist}\big(\hEj,\mathcal Z(\hat\xi_{j_1})\setminus\{\hEj\}\big)\Big\}.
\end{equation}
For $t$ large enough, $\lambda_*\in U_1$. In particular $0\notin\overline{U_1}$, so that the expansion of $E(\lambda)$ at $\lambda=0$ below involves only the residue at $\hEj$.

\begin{rhp}\label{rhp55}
The local parametrix $\Ml(\lambda)$ is required to satisfy:
\begin{conds}
\cond{1}{Analyticity:} $\Ml(\lambda)$ is holomorphic in $U_1\setminus\Sigma^{(1)}$.
\cond{2}{Jump condition:} $\Ml_+(\lambda)=\Ml_-(\lambda)J^{(1)}(\lambda)$, $\lambda\in\Sigma^{(1)}\cap U_1$.
\cond{3}{Matching condition:} $\Ml(\lambda)\Mg(\lambda)^{-1}=I+\mathcal O(t^{-1/3})$ uniformly for $\lambda\in\partial U_1$.
\cond{4}{Local behavior:} $\Ml(\lambda)=\mathcal O((\lambda-\hEj)^{-1/4})$ as $\lambda\to\hEj$.
\end{conds}
\end{rhp}

Nptice that
\begin{equation*}
w(\lambda)=(\lambda-\hEj)^{1/2}\hat w_{j_1}(\lambda), \quad \hat w_{j_1}(\lambda)=\big[\prod_{k=0}^n(\lambda-E_k)\prod_{k\neq j_1}(\lambda-\hat E_k)\big]^{1/2}.
\end{equation*}
For fixed $\xi$ and as $\lambda\to\hEj$, the phase has the expansion,
\begin{equation}\label{thetaexp}
\theta(\lambda;\xi)=\theta^{(0,\hat j_1)}+(\xi-\hat\xi_{j_1})\theta^{(1,\hat j_1)}(\lambda-\hEj)^{1/2}+\frac23\theta^{(3,\hat j_1)}(\lambda-\hEj)^{3/2}+\mathcal O\big((\lambda-\hEj)^{5/2}\big),
\end{equation}
where
\[
\theta^{(0,\hat j_1)}=\Big(f_0-\frac12C^f_{j_1}\Big)\xi+g_0-\frac12C^g_{j_1},\quad \theta^{(1,\hat j_1)}=-\frac{2P_f(\hEj)}{\hat w_{j_1}(\hEj)}<0,
\]
\[
\begin{aligned}
&\theta^{(3,\hat j_1)}=\frac{T(\hEj;\xi)\big(2\hat w_{j_1}(\hEj)+\hEj\hat w_{j_1}'(\hEj)\big)-\hEj\,\partial_\lambda T(\hEj;\xi)\,\hat w_{j_1}(\hEj)}{\hEj^3\hat w_{j_1}(\hEj)^2},\\
&\theta^{(3,\hat j_1)}(\hat\xi_{j_1})=-\frac12\theta^{(1,\hat j_1)}\Xi'(\hEj)\neq0.
\end{aligned}
\]
One can find a conformal map $\zeta=\zeta(\lambda)$ on $U_1$ and a real parameter $s=s(\xi,t)\in\mathbb R$ with
\begin{equation}\label{zeta}
\rmi t\big(\theta(\lambda)-\theta(\hEj)\big)=-\frac23\zeta(\lambda)^{3/2}-s\,\zeta(\lambda)^{1/2},\quad \lambda\in U_1.
\end{equation}
$\zeta^{1/2}$ of above equation denotes the branch analytic in $\mathbb C\setminus[0,+\infty)$ with $\zeta^{1/2}=-\rmi\ve|\zeta|^{1/2}$ for $\zeta<0$, and $\zeta^{3/2}=\zeta\cdot\zeta^{1/2}$. This branch equals the principal one in $\mathbb C^{-\ve}$ and minus the principal one in $\mathbb C^{\ve}$. In addition,
\begin{equation}\label{zetaprop}
\begin{aligned}
&\zeta(\hEj)=0,\quad \zeta(\lambda_*)=-\frac s2,\quad \zeta'(\hEj)=-\big|\theta^{(3,\hat j_1)}\big|^{2/3}t^{2/3}\big(1+\mathcal O(\xi-\hat\xi_{j_1})\big),\\
&s=2\Big(\frac34t\big|\theta(\lambda_*)-\theta(\hEj)\big|\Big)^{2/3}\ge0.
\end{aligned}
\end{equation}
The parameter $s$ behaves like
\begin{equation*}
s=\ve(\hat\xi_{j_1}-\xi)t^{2/3}|\theta^{(1,\hat j_1)}||\theta^{(3,\hat j_1)}|^{-1/3}\big(1+\mathcal O(\xi-\hat\xi_{j_1})\big).
\end{equation*}
Writing $\hat\theta(\zeta)=\frac23\zeta^{3/2}+s\zeta^{1/2}$ with principal branches, and using that $\zeta$ maps $U_1\cap\mathbb C^\pm$ into $\mathbb C^\mp$ by \eqref{zetaprop}, we obtain
\begin{equation}\label{thetahat}
\rmi t\theta(\lambda)=\rmi t\theta(\hEj)\mp\ve\hat\theta(\zeta(\lambda)),\quad \lambda\in U_1\cap\mathbb C^\pm.
\end{equation}

\subsection{The model RH problem}
In what follows, we will use the Painlev\'e XXXIV parametrix $M^{(P34)}(\zeta)=M^{(P34)}(\zeta;s,\alpha,\omega)$ defined by RH Problem \ref{rhpA1} to construct the function $\Ml(\lambda)$ associated with the model RH problem. We first introduce
\[
\tau_0=t\theta(\hEj),\quad \vartheta=\tilde\phi_{j_1}+\delta_{j_1},\quad \kappa=\tau_0+\frac\vartheta2,\quad \zeta_*=-\frac s2,\quad R=\begin{cases}\overline{r(\hEj)}e^{\rmi\delta_{j_1}}, & \ve=-1,\\ r(\hEj)e^{\rmi\delta_{j_1}}, & \ve=1.\end{cases}
\]

\begin{rhp}\label{rhp56}
We seek a model function $\hat\Psi(\zeta)$ with the following properties.
\begin{conds}
\cond{1}{Analyticity:} $\hat\Psi(\zeta)$ is holomorphic in $\mathbb C\setminus\Sigma^{(loc,\zeta)}$, where
\begin{equation*}
\Sigma^{(loc,\zeta)}=\mathbb R^+\cup[\zeta_*,0]\cup\Gamma_+\cup\Gamma_-, \quad \Gamma_\pm=\zeta_*+e^{\pm\frac{2\pi\rmi}3}\mathbb R^+.
\end{equation*}
Since $\zeta'(\hEj)<0$, $\Sigma^{(loc,\zeta)}$ carries the orientation induced by $\lambda\mapsto\zeta$: $\mathbb R^+$ and $[\zeta_*,0]$ run from right to left, and $\Gamma_\pm$ run outward from $\zeta_*$. The jumps below refer to this orientation.
Below $\hat\theta_\pm$ denote the boundary values of $\hat\theta$ from $\mathbb C^\pm$.
\cond{2}{Jump condition:} $\hat\Psi_+(\zeta)=\hat\Psi_-(\zeta)J^{(loc)}(\zeta)$, with $J^{(loc)}=e^{\rmi\kappa\hat\sigma_3}\psm{0&-\rmi\\-\rmi&0}$ on $\mathbb R^+$. For $\ve=-1$
\begin{equation}
J^{(loc)}=e^{\rmi\tau_0\hat\sigma_3}\left\{
\begin{aligned}
&\begin{pmatrix}
1&0\\-\bar Re^{2\hat\theta}&1
\end{pmatrix},\quad &\zeta\in\Gamma_+,\\
&\begin{pmatrix}
1&Re^{2\hat\theta}\\0&1
\end{pmatrix},\quad\, \, \  &\zeta\in\Gamma_-,\\
&\begin{pmatrix}
1&0\\-\bar Re^{2\hat\theta_+}&1
\end{pmatrix}\begin{pmatrix}
1&Re^{2\hat\theta_-}\\0&1
\end{pmatrix},\quad\, \, \  &\zeta\in(\zeta_*,0),\\
\end{aligned}
\right.
\end{equation}
and for $\ve=-1$,
\begin{equation}
J^{(loc)}=e^{\rmi\tau_0\hat\sigma_3}\left\{
\begin{aligned}
&\begin{pmatrix}
1&Re^{2\hat\theta}\\0&
\end{pmatrix},\quad &\zeta\in\Gamma_+,\\
&\begin{pmatrix}
1&0\\-\bar Re^{2\hat\theta}&1
\end{pmatrix},\quad\, \, \  &\zeta\in\Gamma_-,\\
&\begin{pmatrix}
1&Re^{2\hat\theta_+}\\0&1
\end{pmatrix}\begin{pmatrix}
1&0\\-\bar Re^{2\hat\theta_-}&1
\end{pmatrix},\quad\, \, \  &\zeta\in(\zeta_*,0).
\end{aligned}
\right.
\end{equation}

\[
\begin{aligned}
\ve=-1:\quad J^{(loc)}&=e^{\rmi\tau_0\hat\sigma_3}\psm{1&0\\-\bar Re^{2\hat\theta}&1},\ \ \zeta\in\Gamma_+;\quad e^{\rmi\tau_0\hat\sigma_3}\psm{1&Re^{2\hat\theta}\\0&1},\ \ \zeta\in\Gamma_-;\\
J^{(loc)}&=e^{\rmi\tau_0\hat\sigma_3}\psm{1&0\\-\bar Re^{2\hat\theta_+}&1}\psm{1&Re^{2\hat\theta_-}\\0&1},\ \ \zeta\in(\zeta_*,0);\\
\ve=1:\quad J^{(loc)}&=e^{\rmi\tau_0\hat\sigma_3}\psm{1&Re^{2\hat\theta}\\0&1},\ \ \zeta\in\Gamma_+;\quad e^{\rmi\tau_0\hat\sigma_3}\psm{1&0\\-\bar Re^{2\hat\theta}&1},\ \ \zeta\in\Gamma_-;\\
J^{(loc)}&=e^{\rmi\tau_0\hat\sigma_3}\psm{1&Re^{2\hat\theta_+}\\0&1}\psm{1&0\\-\bar Re^{2\hat\theta_-}&1},\ \ \zeta\in(\zeta_*,0).
\end{aligned}
\]
\cond{3}{Asymptotics at infinity:} As $\zeta\to\infty$,
\begin{equation*}
\hat\Psi(\zeta)=\big(I+\mathcal O(\zeta^{-1})\big)\dfrac{\zeta^{-\frac14\sigma_3}}{\sqrt2}\begin{pmatrix}1&\rmi\\\rmi&1\end{pmatrix}G_1(\zeta),
\end{equation*}
where, for $\ve=1$,
\begin{equation}
G_1(\zeta)=\begin{cases}-\rmi\sigma_2e^{\frac{\pi\rmi}4\sigma_3}e^{-\rmi\kappa\sigma_3}, & \zeta\in\mathbb C^+,\\ e^{\frac{\pi\rmi}4\sigma_3}e^{-\rmi\kappa\sigma_3}, & \zeta\in\mathbb C^-,\end{cases}
\end{equation}
and for $\ve=-1$,
\begin{equation}
G_1(\zeta)=\begin{cases}e^{\frac{\pi\rmi}4\sigma_3}e^{-\rmi\kappa\sigma_3}, & \zeta\in\mathbb C^+,\\ \rmi\sigma_2e^{\frac{\pi\rmi}4\sigma_3}e^{-\rmi\kappa\sigma_3}, & \zeta\in\mathbb C^-.\end{cases}
\end{equation}
\cond{4}{Local behavior:} $\hat\Psi(\zeta)=\mathcal O(\zeta^{-1/4})$ as $\zeta\to0$.
\end{conds}
\end{rhp}

The explicit solution of RH problem \eqref{rhp56} can be expressed as
\begin{equation}\label{Psihat}
\hat\Psi(\zeta)=\begin{pmatrix}1&0\\\rmi a(s)&1\end{pmatrix}M^{(P34)}\Big(\zeta;s,\frac14,0\Big)\,e^{\hat\theta(\zeta)\sigma_3}G_1(\zeta)G_2(\zeta),
\end{equation}
and for $\ve=1$,
\begin{equation}
G_2=e^{\rmi\tau_0\hat\sigma_3}\left\{
\begin{aligned}
&\begin{pmatrix}
1&-Re^{2\hat\theta}\\0&1
\end{pmatrix},\quad &\zeta\in D_+,\\
&\begin{pmatrix}
1&0\\-\bar Re^{2\hat\theta}&1
\end{pmatrix},\quad\, \, \  &\zeta\in D_-,\\
\end{aligned}
\right.
\end{equation}
for $\ve=-1$,
\begin{equation}
G_2=e^{\rmi\tau_0\hat\sigma_3}\left\{
\begin{aligned}
&\begin{pmatrix}
1&0\\\bar Re^{2\hat\theta}&1
\end{pmatrix},\quad &\zeta\in D_+,\\
&\begin{pmatrix}
1&Re^{2\hat\theta}\\0&1
\end{pmatrix},\quad\, \, \  &\zeta\in D_-.
\end{aligned}
\right.
\end{equation}

With $\omega=0$, the factor $M^{(P34)}e^{\hat\theta\sigma_3}$ has no jump on $\mathbb R^+$, and $G_1(\zeta)$ produces the band jump $J^{(loc)}$ there. On $\mathbb R^-$, the jump $\rmi\sigma_2$ of $M^{(P34)}$ on $\Sigma_3^P$ is compensated by the jump of $G_1$, because $G_1|_{\mathbb C^-}G_1|_{\mathbb C^+}^{-1}=\rmi\sigma_2$ in both cases. Furthermore, $\hat\Psi$ is continuous across $\Sigma_2^P$ and $\Sigma_4^P$ if and only if
\begin{equation*}
\begin{aligned}
&G_2=G_1^{-1}\begin{pmatrix}1&0\\e^{2\alpha\pi\rmi}e^{2\hat\theta}&1\end{pmatrix}G_1,\quad \zeta\in D_+,\\
&G_2=G_1^{-1}\begin{pmatrix}1&0\\-e^{-2\alpha\pi\rmi}e^{2\hat\theta}&1\end{pmatrix}G_1,\quad \zeta\in D_-.
\end{aligned}
\end{equation*}
A direct computation shows that $G_2|_{D_+}^{-1}$ and $G_2|_{D_-}$ have the triangular form of $J^{(loc)}$ on $\Gamma_+$ and $\Gamma_-$, respectively, and agree with $J^{(loc)}$ there if and only if
\begin{equation*}
e^{2\alpha\pi\rmi}=\begin{cases}-\rmi\,\bar R\,e^{\rmi\vartheta}, & \ve=-1,\\ \rmi\,R\,e^{-\rmi\vartheta}, & \ve=1.\end{cases}
\end{equation*}
For $\ve=-1$, we have
\begin{equation*}
R=\overline{r_2(\hEj)}e^{\rmi\delta_{j_1}}, \quad \vartheta=\phi_{j_1}+\delta_{j_1}.
\end{equation*}
For $\ve=1$, we also obtain
\begin{equation*}
R=r_1(\hEj)e^{2\rmi f(\hEj)\Delta}e^{\rmi\delta_{j_1}}, \quad f(\hEj)=\frac12C^f_{j_1}, \quad \vartheta=\phi_{j_1}+\Delta C^f_{j_1}+\delta_{j_1}.
\end{equation*}
Proposition~\ref{pro61} then gives
\begin{equation*}
\begin{aligned}
&\bar Re^{\rmi\vartheta}=-1, \quad \ve=-1,\\
&Re^{-\rmi\vartheta}=1, \quad \ve=1,
\end{aligned}
\end{equation*}
hence $e^{2\alpha\pi\rmi}=\rmi$ in both cases. The local behavior (4) of RH Problem~\ref{rhp56} together with \eqref{MP340b} selects $\alpha=\frac14$. The jump on $(\zeta_*,0)$ is then $G_2|_{D_+}^{-1}G_2|_{D_-}$, which is the stated one. Therefore, the parameters of the Painlev\'e XXXIV parametrix are
\begin{equation}\label{gamma}
(\alpha,\omega)=\Big(\frac{1}{4},0\Big),
\end{equation}
which implies that
\begin{equation}\label{as}
a(s)=a\Big(s;\frac14,0\Big)=\int_{-\infty}^s\Big(v(\tau)+\frac\tau2\Big)\rmd\tau,
\end{equation}
where $D_+$ denotes the sector bounded by $\Sigma_2^P$ and $\Gamma_+$, $D_-$ the one bounded by $\Sigma_4^P$ and $\Gamma_-$, and $G_2=I$ outside $D_+\cup D_-$. According to \eqref{P34eq}-\eqref{aint}, $v$ is the solution of the Painlev\'e XXXIV equation
\begin{equation}\label{P34}
v''(s)=4v^2(s)+2sv(s)+\frac{v'(s)^2-\frac14}{2v(s)}
\end{equation}
singled out by the following asymptotics
\begin{equation*}
\begin{aligned}
&v(s)=\frac14s^{-1/2}+\mathcal O(s^{-2}), \quad s\to+\infty,\\
&v(s)=-\frac s2+\mathcal O(s^{-2}), \quad s\to-\infty.
\end{aligned}
\end{equation*}
As $\alpha=\frac14\ge0$, the local form of $M^{(P34)}$ at $\zeta=0$ is \eqref{MP340b}, from which $\hat\Psi(\zeta)=\mathcal O(\zeta^{-1/4})$ follows.

Let
\begin{equation*}
U_1'=\{|\lambda-\hEj|<\rho_1/2\},\quad U_1''=\{|\lambda-\hEj|<3\rho_1/4\},
\end{equation*}
with a cutoff $\chi\in C^\infty(\mathbb C)$, $0\le\chi\le1$, $\chi|_{U_1'}=1$, $\chi|_{\mathbb C\setminus U_1''}=0$. For $t$ large, $\lambda_*\in U_1'$, and by \eqref{zetaprop}
\begin{equation*}
\zeta(U_1'')\subset\{|\zeta|<c_1t^{2/3}\},\quad \zeta(\partial U_1)\subset\{|\zeta|>c_2t^{2/3}\},\quad 0<c_1<c_2,
\end{equation*}
uniformly in $s\in[-C',C']$. On $\Gamma_\pm$, $J^{(1)}$ and $J^{(loc)}$ are triangular with amplitudes $\varrho(\lambda)$ and $R=\varrho(\hEj)$, where
\begin{equation*}
\varrho(\lambda)-R=\mathcal O(|\lambda-\hEj|^{1/2}).
\end{equation*}
Let $J^{(\chi)}$ be $J^{(loc)}$ with $R$ replaced by $R+\chi(\varrho-R)$ on $\Gamma_\pm$, with $J^{(\chi)}=J^{(1)}$ on $[\zeta_*,0]\cup\mathbb R^+$. Then $J^{(\chi)}=J^{(1)}$ on $\zeta(U_1')$, $J^{(\chi)}=J^{(loc)}$ off $\zeta(U_1'')$, and $J^{(\chi)}J^{(loc)-1}-I\to0$ continuously at $\partial\zeta(U_1'')$.

\begin{rhp}\label{rhp57}
The local problem in the new variable is formulated for a matrix $\Mc^{(loc,0)}(\zeta)$ subject to:
\begin{conds}
\cond{1}{Analyticity:} $\Mc^{(loc,0)}(\zeta)$ is holomorphic in $\mathbb C\setminus\Sigma^{(loc,\zeta)}$.
\cond{2}{Jump condition:} $\Mc^{(loc,0)}_+(\zeta)=\Mc^{(loc,0)}_-(\zeta)J^{(\chi)}(\zeta)$ for $\zeta\in\Sigma^{(loc,\zeta)}$.
\cond{3}{Normalization:} As $\zeta\to\infty$,
\begin{equation*}
\Mc^{(loc,0)}(\zeta)=\big(I+\mathcal O(\zeta^{-1})\big)\frac{\zeta^{-\frac14\sigma_3}}{\sqrt2}\begin{pmatrix}1&\rmi\\\rmi&1\end{pmatrix}G_1(\zeta).
\end{equation*}
\cond{4}{Local behavior:} $\Mc^{(loc,0)}\hat\Psi^{-1}$ and its inverse belong to $L^b_{\rm loc}(\mathbb C)$ for every $b<6$.
\end{conds}
\end{rhp}

We seek $\Mc^{(loc,0)}=\mathcal R\hat\Psi$. The jumps $J^{(\chi)}$ and $J^{(loc)}$ coincide on $\mathbb R^+$, so $\mathcal R$ jumps only on $\Sigma_{\mathcal R}=(\Gamma_+\cup\Gamma_-\cup[\zeta_*,0])\cap\zeta(U_1'')$, with
\begin{equation*}
\mathcal R_+=\mathcal R_-V,\quad V=\hat\Psi_-J^{(\chi)}J^{(loc)-1}\hat\Psi_-^{-1}.
\end{equation*}
The off-diagonal entry of $J^{(\chi)}J^{(loc)-1}-I$ is $\eta e^{\pm2\hat\theta}$ with $\eta=\mathcal O(t^{-1/3}|\zeta|^{1/2})$. Conjugation by $\hat\Psi_-=\mathcal O(|\zeta|^{-1/4})$ removes $e^{\pm2\hat\theta}$, giving
\begin{equation*}
\|V-I\|_{L^p(\Sigma_{\mathcal R})}=\mathcal O(t^{-1/3}),\quad p\in[1,\infty],
\end{equation*}
pointwise up to $\zeta=0$, with $V-I=0$ on $\partial\zeta(U_1'')$. The contour $\Sigma_{\mathcal R}$ is uniformly Lipschitz in $s\in[-C',C']$. Hence $\mu=I+\mathcal C_-(\mu(V-I))$ has a unique solution with $\|\mu-I\|_{L^6}=\mathcal O(t^{-1/3})$. Put
\begin{equation*}
\mathcal R(\zeta)=I+\frac1{2\pi\rmi}\int_{\Sigma_{\mathcal R}}\frac{\mu(s)(V(s)-I)}{s-\zeta}\rmd s,\quad \mathcal R^{\pm1}\in L^b_{\rm loc},\ b<6.
\end{equation*}
Then $\Mc^{(loc,0)}=\mathcal R\hat\Psi$ solves RH Problem~\ref{rhp57}, uniformly in $s\in[-C',C']$ \cite{IKO08,IKO09}. No bound at $\zeta_*$ or $\zeta=0$ is assumed (see after \eqref{E}). On $\zeta(\overline{\mathbb C\setminus U_1})$,
\begin{equation*}
|\mathcal R(\zeta)-I|\le\frac{\|\mu-I\|_{L^6}\|V-I\|_{L^{6/5}}+\|V-I\|_{L^1}}{2\pi(c_2-c_1)t^{2/3}}=\mathcal O(t^{-1}).
\end{equation*}
Thus
\begin{equation}\label{Xi}
\mathcal R(\zeta)=I+\mathcal O(t^{-1/3}),\quad \mathcal R(\zeta)-I=\mathcal O\big(t^{-1/3}|\zeta|^{-1}\big),\ |\zeta|\to\infty,
\end{equation}
so that
\begin{equation}\label{Mloc0}
\begin{aligned}
&\Mc^{(loc,0)}(\zeta)=\Big(I+\frac{\Mc^{(loc,0)}_1}{\zeta}+\mathcal O(\zeta^{-2})\Big)\frac{\zeta^{-\frac14\sigma_3}}{\sqrt2}\begin{pmatrix}1&\rmi\\\rmi&1\end{pmatrix}G_1(\zeta),\\
&\big(\Mc^{(loc,0)}_1\big)_{12}=\rmi a(s)+\mathcal O(t^{-1/3}).
\end{aligned}
\end{equation}
The local parametrix is now taken as
\begin{equation}\label{Mlocdef}
\Ml(\lambda)=H_1(\lambda)t^{\sigma_3/6}\Mc^{(loc,0)}(\zeta(\lambda)),
\end{equation}
where $H_1$ is holomorphic in $U_1$, and its expression is
\begin{equation*}
H_1(\lambda)=\Mg(\lambda)G_1(\zeta(\lambda))^{-1}\frac1{\sqrt2}\begin{pmatrix}1&-\rmi\\-\rmi&1\end{pmatrix}\big(t^{-2/3}\zeta(\lambda)\big)^{\sigma_3/4}.
\end{equation*}
Then, as $t\to\infty$
\begin{equation}\label{match}
\Ml(\lambda)\Mg(\lambda)^{-1}=I+\frac{t^{1/3}}{\zeta(\lambda)}H_1(\lambda)\begin{pmatrix}0&\rmi a(s)\\0&0\end{pmatrix}H_1(\lambda)^{-1}+\mathcal O(t^{-2/3}).
\end{equation}

\subsection{The small-norm RH problem}

For every $\lambda_k\in\mathcal Z(\xi)\setminus\{\lambda_*\}$, let $U_k$ be a disk of fixed radius about $\lambda_k$, chosen as in \eqref{rho3} so that $0\notin\overline{U_k}$ and the disks are pairwise disjoint and disjoint from $\overline{U_1}$, and let $\Mc^{(pc,k)}$ denote the parabolic-cylinder parametrix in $U_k$ constructed in Section~\ref{sec7}, with $r$, $\delta$ of this section and $\beta_k$ frozen at $\lambda_k$; since $\delta$ jumps in $U_k$ only on one side of $\lambda_k$, \eqref{betaHolder} and \eqref{lensmismatch} hold in $U_k$ with the same proof. The error matrix is
\begin{equation}\label{E}
E(\lambda)=\begin{cases}\Mc^{(1)}(\lambda)\Mg(\lambda)^{-1}, & \lambda\in\mathbb C\setminus\big(U_1\cup\bigcup_kU_k\big),\\ \Mc^{(1)}(\lambda)\Ml(\lambda)^{-1}, & \lambda\in U_1,\\ \Mc^{(1)}(\lambda)\Mc^{(pc,k)}(\lambda)^{-1}, & \lambda\in U_k.\end{cases}
\end{equation}
In $U_1'$ the jumps of $\Mc^{(1)}$ and $\Ml$ coincide, so $E$ is holomorphic in $U_1'\setminus\{\hEj,\lambda_*\}$. From $\Mc^{(1)}=\mathcal O(|\lambda-\hEj|^{-1/4})$, $\hat\Psi^{-1}=\mathcal O(|\zeta|^{-1/4})$ and $\mathcal R^{-1}\in L^b_{\rm loc}$, H\"older's inequality gives $E\in L^q_{\rm loc}$, $q<12/5$. These isolated singularities are therefore removable, and
\begin{equation*}
\Ml=E^{-1}\Mc^{(1)}=\mathcal O\big(|\lambda-\hEj|^{-1/4}\big),\quad \lambda\to\hEj.
\end{equation*}

\begin{figure}[htbp]
\centering
\includegraphics[width=0.72\textwidth]{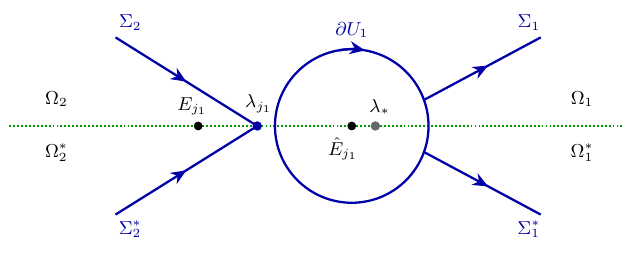}
\caption{The jump contour $\Sigma^{(E)}$ of the error matrix.}\label{fig10}
\end{figure}

\begin{rhp}\label{rhp58}
The error matrix $E(\lambda)$ solves the RH problem below.
\begin{conds}
\cond{1}{Analyticity:} $E(\lambda)$ is holomorphic in $\mathbb C\setminus\Sigma^{(E)}$, where
\begin{equation*}
\Sigma^{(E)}=\partial U_1\cup\bigcup_k\partial U_k\cup\big((\Sigma_1\cup\Sigma_1^*\cup\Sigma_2\cup\Sigma_2^*)\setminus U_1'\big),
\end{equation*}
as shown in Figure~\ref{fig10}.
\cond{2}{Jump condition:} $E_+(\lambda)=E_-(\lambda)J_E(\lambda)$ with
\begin{equation}\label{JE}
J_E(\lambda)=\begin{cases}\Mg(\lambda)J^{(1)}(\lambda)\Mg(\lambda)^{-1}, & \lambda\in\Sigma^{(E)}\setminus\big(\overline{U_1}\cup\bigcup_k\overline{U_k}\big),\\ \Ml_-(\lambda)J^{(1)}(\lambda)J^{(\chi)}(\zeta(\lambda))^{-1}\Ml_-(\lambda)^{-1}, & \lambda\in\Sigma^{(E)}\cap(U_1\setminus\overline{U_1'}),\\ \Mc^{(pc,k)}_-(\lambda)J^{(1)}(\lambda)J^{(pc,k)}(\lambda)^{-1}\Mc^{(pc,k)}_-(\lambda)^{-1}, & \lambda\in\Sigma^{(E)}\cap U_k,\\ \Ml(\lambda)\Mg(\lambda)^{-1}, & \lambda\in\partial U_1,\\ \Mc^{(pc,k)}(\lambda)\Mg(\lambda)^{-1}, & \lambda\in\partial U_k,\end{cases}
\end{equation}
where $J^{(pc,k)}$ denotes the jump of $\Mc^{(pc,k)}$, i.e.\ $J^{(1)}$ with $r$ and $\beta_k$ frozen at $\lambda_k$.
\cond{3}{Asymptotics at infinity:} $E(\lambda)=I+\mathcal O(\lambda^{-1})$ as $\lambda\to\infty$.
\end{conds}
\end{rhp}

A direct computation gives, for some $c>0$ and all $p\in[1,\infty]$,
\begin{equation}\label{JEest}
\|J_E-I\|_{L^p}=\begin{cases}\mathcal O(e^{-ct}), & \text{on }\Sigma^{(E)}\setminus\big(\bigcup_k\overline{U_k}\cup\partial U_1\big),\\ \mathcal O(t^{-\frac12-\frac1{2p}}\log t), & \text{on }\Sigma^{(E)}\cap U_k,\\ \mathcal O(t^{-1/3}), & \text{on }\partial U_1,\\ \mathcal O(t^{-1/2}), & \text{on }\partial U_k.\end{cases}
\end{equation}
On $\Sigma^{(E)}\cap(U_1\setminus\overline{U_1'})$, the jump equals $I+(1-\chi)(\cdots)$, which is $\mathcal O(e^{-ct})$ by using $|\zeta|\ge c_1t^{2/3}$. The bound on $\Sigma^{(E)}\cap U_k$ is \eqref{lensmismatch}.
Hence the small-norm theory of \cite{DZ93} applies and, for $t$ sufficiently large, the RH problem \ref{rhp58} admits a unique solution, which is given by
\begin{equation}\label{Erep}
E(\lambda)=I+\frac1{2\pi\rmi}\int_{\Sigma^{(E)}}\frac{\varpi(\varrho)\big(J_E(\varrho)-I\big)}{\varrho-\lambda}\rmd\varrho,
\end{equation}
where $\varpi\in I+L^2(\Sigma^{(E)})$ denotes the unique solution of the singular integral equation
\begin{equation}\label{varpi}
\begin{aligned}
&\varpi=I+\mathcal C_E\varpi,\quad \mathcal C_E(f)=\mathcal C_-\big(f(J_E-I)\big),\\
&\mathcal C_\pm f(\lambda)=\lim_{\lambda'\to \lambda\in\Sigma^{(E)}}\frac1{2\pi\rmi}\int_{\Sigma^{(E)}}\frac{f(\varrho)}{\varrho-\lambda'}\rmd\varrho,
\end{aligned}
\end{equation}
Expanding as $\lambda\to0$, we get
\begin{equation}\label{E0}
E(\lambda)=E(0)+E_1\lambda+\mathcal O(\lambda^2),\\
\end{equation}
where
\begin{equation}
\begin{aligned}
&E(0)=I+\frac1{2\pi\rmi}\int_{\Sigma^{(E)}}\frac{\varpi(\varrho)(J_E(\varrho)-I)}{\varrho}\rmd\varrho,\\
&E_1=\frac1{2\pi\rmi}\int_{\Sigma^{(E)}}\frac{\varpi(\varrho)(J_E(\varrho)-I)}{\varrho^2}\rmd\varrho.
\end{aligned}
\end{equation}
Combining (\ref{match}) with the residue theorem (inside $U_1$ the leading term of $J_E-I$ has a single simple pole, at $\hEj$, where $\zeta$ vanishes with $\zeta'(\hEj)$ given in \eqref{zetaprop}. $\lambda=0\notin\overline{U_1}$), we find for $t\to\infty$
\begin{equation}\label{E0asy}
\begin{aligned}
E(0)&=I+\frac{t^{-1/3}}{|\theta^{(3,\hat j_1)}(\hat\xi_{j_1})|^{2/3}\hEj}H_1(\hEj)\begin{pmatrix}0&\rmi a(s)\\0&0\end{pmatrix}H_1(\hEj)^{-1}+\sum_k E_k(0)+\mathcal O(t^{-2/3}),\\
E_1&=\frac{t^{-1/3}}{|\theta^{(3,\hat j_1)}(\hat\xi_{j_1})|^{2/3}\hEj^2}H_1(\hEj)\begin{pmatrix}0&\rmi a(s)\\0&0\end{pmatrix}H_1(\hEj)^{-1}+\sum_k E_{k,1}+\mathcal O(t^{-2/3}),
\end{aligned}
\end{equation}
in which the terms $ E_k(0), E_{k,1}=\mathcal O(t^{-1/2})$ come from the circles $\partial U_k$ and from $\Sigma^{(E)}\cap U_k$; the latter contribute $\mathcal O(t^{-1}\log t)$, which is absorbed in the remainder $\mathcal O(t^{-2/3})$. In particular,
\begin{equation}\label{E0inv}
E(0)^{-1}=I+\mathcal O(t^{-1/3}).
\end{equation}

\subsection{The complementary half}\label{sec5neg}
Let $0<\ve(\xi-\hat\xi_{j_1})t^{2/3}\le C$. As $\Xi$ is analytic at $\hEj$ with $\Xi'(\hEj)\neq0$, the zero $\lambda_*$ of $T(\cdot;\xi)$ near $\hEj$ satisfies
\begin{equation*}
\lambda_*-\hEj=\frac{\xi-\hat\xi_{j_1}}{\Xi'(\hEj)}+\mathcal O(t^{-4/3}),\quad \lambda_*\in(E_{j_1},\hEj),
\end{equation*}
by Lemma~\ref{lem4}. Because $\Xi$ is monotone on $(\hEj,b)$ with $\Xi(\hEj)=\hat\xi_{j_1}$, thus
\begin{equation*}
\ve\big(\Xi(\lambda)-\xi\big)<0,\quad \lambda\in(\hEj,b).
\end{equation*}
Hence $(\hEj,b)\subset S_1(\xi)$, and no stationary point lies in this gap. We keep all definitions of this section, with $\lambda_*$ replaced by $\hEj$ in $S_2$, $\Omega_1$, $\Sigma_1$ and $\Sigma^{(1)}$. Thus
\begin{equation*}
S_2(\xi)=\Big\{\lambda\in\mathbb R\setminus\textstyle\bigcup_j[E_j,\hat E_j]:\ \ve\big(\Xi(\lambda)-\xi\big)>0\Big\},\quad \Sigma_1=\hEj+e^{\rmi\varphi_1}\mathbb R^+,
\end{equation*}
and the segment $[\hEj,\lambda_*]$ is absent from $\Sigma^{(1)}$. The lens is admissible at its vertex. For $\lambda-\hEj=re^{\rmi\varphi}$, $0<\varphi\le\varphi_1$, \eqref{thetaexp} gives
\begin{equation*}
\operatorname{Im}\theta(\lambda)=(\xi-\hat\xi_{j_1})\theta^{(1,\hat j_1)}r^{1/2}\sin\frac\varphi2+\frac23\theta^{(3,\hat j_1)}r^{3/2}\sin\frac{3\varphi}2+\mathcal O(r^{5/2}),
\end{equation*}
where
\begin{equation*}
\operatorname{sgn}\big((\xi-\hat\xi_{j_1})\theta^{(1,\hat j_1)}\big)=\operatorname{sgn}\theta^{(3,\hat j_1)}=-\ve.
\end{equation*}
Thus $\ve\operatorname{Im}\theta<0$ on $\Sigma_1\setminus\{\hEj\}$, and likewise $\ve\operatorname{Im}\theta>0$ on $\Sigma_1^*\setminus\{\hEj\}$. In the half treated above, the two terms have opposite signs, which is why the vertex was placed at $\lambda_*$. Since $\hEj\notin\partial S_2$, RH Problems~\ref{rhp52} and \ref{rhp53} hold verbatim, with
\begin{equation*}
\delta(\lambda)=\frac\rmi2\delta_{j_1}+\mathcal O((\lambda-\hEj)^{1/2}).
\end{equation*}

The conformal map $\zeta$ of \eqref{zeta} exists uniformly for $s$ in compact subsets of $\mathbb R$. It maps $\lambda_*$ into the image of the band:
\begin{equation*}
\zeta(\lambda_*)=-\frac s2>0,\quad s=-2\Big(\frac34t\big|\theta_+(\lambda_*)-\theta(\hEj)\big|\Big)^{2/3}<0.
\end{equation*}
The asymptotic formula for $s$ following \eqref{zetaprop} is unchanged. The vertex $\hEj$ of the lens is mapped to $\zeta=0$, so in RH Problem~\ref{rhp56} we replace $\zeta_*$ by $0$, drop the segment $[\zeta_*,0]$, and take
\begin{equation*}
\Gamma_\pm=e^{\pm\frac{2\pi\rmi}3}\mathbb R^+=\Sigma_2^P,\ \Sigma_4^P.
\end{equation*}
On these rays
\begin{equation*}
\operatorname{Re}\hat\theta(\zeta)=-\frac23|\zeta|^{3/2}+\frac s2|\zeta|^{1/2}<0,\quad \zeta\neq0.
\end{equation*}
The sectors $D_\pm$ are empty, $G_2=I$, and the model solution is
\begin{equation*}
\hat\Psi(\zeta)=\begin{pmatrix}1&0\\\rmi a(s)&1\end{pmatrix}M^{(P34)}\Big(\zeta;s,\frac14,0\Big)\,e^{\hat\theta(\zeta)\sigma_3}G_1(\zeta).
\end{equation*}
Its jumps on $\Sigma_2^P,\Sigma_4^P$ are $G_1^{-1}\psm{1&0\\ \pm e^{\pm2\alpha\pi\rmi}e^{2\hat\theta}&1}G_1$. By the computation following \eqref{Psihat}, these coincide with $J^{(loc)}$ on $\Gamma_\pm$ exactly when $\alpha=\frac14$. RH Problem~\ref{rhp57}, the estimate \eqref{Xi}, the definitions \eqref{Mlocdef}, the matching \eqref{match} and RH Problem~\ref{rhp58} carry over without change. The jump of $\mathcal R$ is now supported on $\Gamma_\pm\cap\zeta(U_1'')$. Consequently \eqref{E0asy} and \eqref{E0inv} hold uniformly for $|\xi-\hat\xi_{j_1}|t^{2/3}\le C$. Here $a(s)$ is defined by \eqref{as} for every $s\in\mathbb R$.

\section{Asymptotic analysis in transition region II}\label{sec6}
We now turn to the basic RH problem in transition region II, i.e.\ $|\xi-\xi_{j_1}|t^{2/3}\le C$. The half $0\le\ve(\xi-\xi_{j_1})t^{2/3}\le C$ is treated in detail, and the complementary half in Section~\ref{sec6neg}, where
\begin{equation*}
\ve=\operatorname{sgn}\Ej, \quad j_1\in\{0,\dots,n\}.
\end{equation*}
For $t$ large the saddle point $\lambda_*\in\mathcal Z(\xi)$ that merges with $\Ej$ lies in $(a,\Ej]$, where
\begin{equation*}
a=\begin{cases}0, & 0\in(\hat E_{j_1-1},\Ej),\\ \hat E_{j_1-1}, & \text{otherwise},\end{cases}\quad \hat E_{-1}=-\infty.
\end{equation*}
 Let
\[
\hat y_\ve=\begin{cases}y, & \ve=1,\\ y-\Delta, & \ve=-1,\end{cases}\quad \hat c_\ve=\begin{cases}0, & \ve=1,\\ \eta_\Delta-\rmi f_0\Delta, & \ve=-1,\end{cases}\quad r=\begin{cases}r_2, & \ve=1,\\ r_1e^{2\rmi f(\lambda)\Delta}, & \ve=-1.\end{cases}
\]
In terms of these quantities we introduce
\begin{equation}\label{N1}
\Nc(\lambda)=\begin{cases}\Big(\mu_1^-,\ \dfrac{\mu_2^+}{S_{22}}\Big)e^{-\rmi(f_0y+g_0t)\sigma_3}, & \lambda\in\mathbb C^+,\\[10pt] \Big(\dfrac{\mu_1^+}{S_{11}},\ \mu_2^-\Big)e^{-\rmi(f_0y+g_0t)\sigma_3}, & \lambda\in\mathbb C^-,\end{cases}\quad \ve=1,
\end{equation}
\begin{equation}\label{N2}
\Nc(\lambda)=\begin{cases}e^{-\hat c_\ve\sigma_3}\Big(\dfrac{\mu_1^-}{S_{22}},\ \mu_2^+\Big)e^{-\rmi(f(\lambda)\Delta+f_0y+g_0t)\sigma_3}, & \lambda\in\mathbb C^+,\\[10pt] e^{-\hat c_\ve\sigma_3}\Big(\mu_1^+,\ \dfrac{\mu_2^-}{S_{11}}\Big)e^{-\rmi(f(\lambda)\Delta+f_0y+g_0t)\sigma_3}, & \lambda\in\mathbb C^-,\end{cases}\quad \ve=-1.
\end{equation}

\begin{rhp}\label{rhp61}
The matrix $\Nc(\lambda)$ solves the RH problem below.
\begin{conds}
\cond{1}{Normalization:} $\Nc(\lambda)=I+\mathcal O(\lambda^{-1})$ as $\lambda\to\infty$.
\cond{2}{Analyticity:} $\Nc(\lambda)$ is analytic in $\mathbb C\setminus\mathbb R$ and bounded near $\lambda=0$.
\cond{3}{Jump condition:} $\Nc_+(\lambda)=\Nc_-(\lambda)J_{\Nc}(\lambda)$, where
\begin{equation}\label{JNband}
J_{\Nc}(\lambda)=e^{\rmi\left(f_0y+g_0t-\frac12(C_j^f\hat y_\ve+C_j^gt-\phi_j)\right)\hat\sigma_3}\begin{pmatrix}0&-\rmi\\-\rmi&0\end{pmatrix},\quad \lambda\in(E_j,\hat E_j),\ j=0,\dots,n,
\end{equation}
\begin{equation}\label{JNgap}
J_{\Nc}(\lambda)=\begin{cases}\begin{pmatrix}1&\overline{r(\lambda)}e^{2\rmi t\theta(\lambda)}\\-r(\lambda)e^{-2\rmi t\theta(\lambda)}&1-|r(\lambda)|^2\end{pmatrix}, & \ve=1,\\[12pt] \begin{pmatrix}1-|r(\lambda)|^2&r(\lambda)e^{2\rmi t\theta(\lambda)}\\-\overline{r(\lambda)}e^{-2\rmi t\theta(\lambda)}&1\end{pmatrix}, & \ve=-1,\end{cases}\quad \lambda\in\mathbb R\setminus\bigcup_{j=0}^n[E_j,\hat E_j],
\end{equation}
with $C_0^f=C_0^g=\phi_0=0$.
\cond{4}{Local behavior:} For $p\in E$,
\begin{equation}\label{Nloc}
\Nc(\lambda)=\begin{cases}\big(\mathcal O((\lambda-p)^{-\ve/4}),\ \mathcal O((\lambda-p)^{\ve/4})\big), & \lambda\to p\ \text{from}\ \mathbb C^+,\\[2pt] \big(\mathcal O((\lambda-p)^{\ve/4}),\ \mathcal O((\lambda-p)^{-\ve/4})\big), & \lambda\to p\ \text{from}\ \mathbb C^-.\end{cases}
\end{equation}
\end{conds}
\end{rhp}

With $\hat\varphi_\ve=f(0)\hat y_\ve-f_0y+(\hat g(0)-g_0)t$ and $\hat\Phi_\ve(\lambda)=(f(\lambda)-\lambda)\hat y_\ve+\hat g(\lambda)t$, one has, as $\lambda\to0$,
\begin{equation}\label{N0}
\Nc(\lambda)=e^{-\hat c_\ve\sigma_3}e^{\eta\sigma_3}G^{-1}e^{\rmi\hat\varphi_\ve\sigma_3}\Big\{I+\lambda\Big(e^{-\rmi\hat\varphi_\ve\hat\sigma_3}Q+\rmi\big(\hat y_\ve+\hat\Phi_\ve'(0)-x\big)\sigma_3\Big)+\mathcal O(\lambda^2)\Big\},
\end{equation}
and consequently
\begin{equation}\label{recq6}
q(x,t)=e^{2\rmi\hat\varphi_\ve}\lim_{\lambda\to0}\frac{\big(\Nc^{-1}(y,t;0)\Nc(y,t;\lambda)\big)_{12}}{\lambda},
\end{equation}
\begin{equation}\label{recx6}
x(y,t)=\hat y_\ve+\hat\Phi_\ve'(0)+\rmi\lim_{\lambda\to0}\frac{\big(\Nc^{-1}(y,t;0)\Nc(y,t;\lambda)\big)_{11}-1}{\lambda}.
\end{equation}

\begin{figure}[htbp]
\centering
\includegraphics[width=0.72\textwidth]{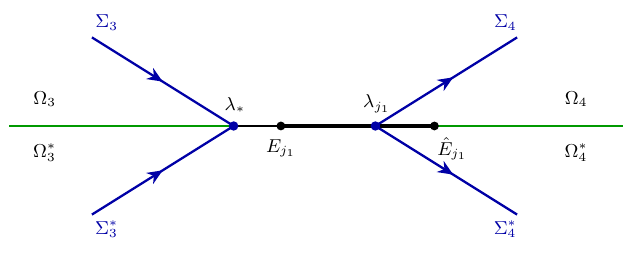}
\caption{Contour $\Sigma^{(1)}$ after the lens opening in transition region II.}\label{fig11}
\end{figure}

For the lens opening we introduce the following subsets of the gaps:
\[
\begin{aligned}
&S_1(\xi)=\Big\{\lambda\in\mathbb R\setminus\textstyle\bigcup_j[E_j,\hat E_j]:\ \ve\big(\Xi(\lambda)-\xi\big)>0\Big\},\\
& S_2(\xi)=\Big\{\lambda\in\mathbb R\setminus\textstyle\bigcup_j[E_j,\hat E_j]:\ \ve\big(\Xi(\lambda)-\xi\big)<0\Big\}\setminus(\lambda_*,\Ej).
\end{aligned}
\]
As in Section~\ref{sec5}, $S_2(\xi)$ is the integration set of $\delta$ in \eqref{delta6}, and $L(s)=\log(1-|r(s)|^2)$.
Next, pick $\lambda_{j_1}\in(\Ej,\hat E_{j_1})$ and put
\[
\begin{aligned}
&\Omega_3=\{\lambda:\pi-\varphi_2\le\arg(\lambda-\lambda_*)\le\pi\},&& \Sigma_3=\lambda_*+e^{\rmi(\pi-\varphi_2)}\mathbb R^+,\\
&\Omega_4=\{\lambda:0\le\arg(\lambda-\lambda_{j_1})\le\varphi_2\},&& \Sigma_4=\lambda_{j_1}+e^{\rmi\varphi_2}\mathbb R^+,
\end{aligned}
\]
with the angle $\varphi_2$ fixed so that
\begin{equation}\label{Imtheta6}
\begin{aligned}
&\ve\operatorname{Im}\theta(\lambda)>0,\ \lambda\in\Sigma_3\setminus\{\lambda_*\},\\
&\ve\operatorname{Im}\theta(\lambda)<0,\ \lambda\in\Sigma_4\setminus\{\lambda_{j_1}\}.
\end{aligned}
\end{equation}
Outside a neighborhood of $\Ej$, the sectors $\Omega_3$ and $\Omega_4$ are extended to lenses along $S_1$ and $ S_2$, respectively, and carried over the bands with the help of \eqref{bandid}. Figure~\ref{fig11} shows the resulting contour. On the gaps, (\ref{JNgap}) splits as follows
\begin{equation}\label{fact61}
J_{\Nc}(\lambda)=\begin{cases}\begin{pmatrix}1&0\\-re^{-2\rmi t\theta}&1\end{pmatrix}\begin{pmatrix}1&\bar re^{2\rmi t\theta}\\0&1\end{pmatrix}, & \lambda\in S_1,\\[12pt] \begin{pmatrix}1&\frac{\bar re^{2\rmi t\theta}}{1-|r|^2}\\0&1\end{pmatrix}(1-|r|^2)^{-\sigma_3}\begin{pmatrix}1&0\\-\frac{re^{-2\rmi t\theta}}{1-|r|^2}&1\end{pmatrix}, & \lambda\in S_2\cup(\lambda_*,\Ej),\end{cases}\quad \ve=1,
\end{equation}
\begin{equation}\label{fact62}
J_{\Nc}(\lambda)=\begin{cases}\begin{pmatrix}1&re^{2\rmi t\theta}\\0&1\end{pmatrix}\begin{pmatrix}1&0\\-\bar re^{-2\rmi t\theta}&1\end{pmatrix}, & \lambda\in S_1,\\[12pt] \begin{pmatrix}1&0\\-\frac{\bar re^{-2\rmi t\theta}}{1-|r|^2}&1\end{pmatrix}(1-|r|^2)^{\sigma_3}\begin{pmatrix}1&\frac{re^{2\rmi t\theta}}{1-|r|^2}\\0&1\end{pmatrix}, & \lambda\in S_2\cup(\lambda_*,\Ej),\end{cases}\quad \ve=-1.
\end{equation}

These factorizations motivate the transformation
\begin{equation}\label{N1def}
\Nc^{(1)}(\lambda)=e^{\delta(\infty)\sigma_3}\Nc(\lambda) G(\lambda)e^{-\delta(\lambda)\sigma_3},
\end{equation}
in which $ G=I$ off $\Omega_3\cup\Omega_3^*\cup\Omega_4\cup\Omega_4^*$ and for $\ve=-1$,
\begin{equation}\label{G-6}
G=\left\{
\begin{aligned}
&\begin{pmatrix}
1&0\\\bar re^{-2\rmi t\theta}&1
\end{pmatrix},\quad &\lambda\in\Omega_3,\\
&\begin{pmatrix}
1&re^{2\rmi t\theta}\\0&1
\end{pmatrix},\quad\, \, \  &\lambda\in\Omega_3^*,\\
&\begin{pmatrix}
1&-\frac{re^{2\rmi t\theta}}{1-|r|^2}\\0&1
\end{pmatrix},\quad &\lambda\in\Omega_4,\\
&\begin{pmatrix}
1&0\\-\frac{\bar re^{-2\rmi t\theta}}{1-|r|^2}&1
\end{pmatrix}, \,\ \quad\ &\lambda\in\Omega_4^*,\\
\end{aligned}
\right.
\end{equation}
for $\ve=1$,
\begin{equation}\label{G6}
G=\left\{
\begin{aligned}
&\begin{pmatrix}
1&-\bar re^{2\rmi t\theta}\\0&1
\end{pmatrix},\quad &\lambda\in\Omega_3,\\
&\begin{pmatrix}
1&0\\-re^{-2\rmi t\theta}&1
\end{pmatrix},\quad\, \, \  &\lambda\in\Omega_3^*,\\
&\begin{pmatrix}
1&0\\\frac{re^{-2\rmi t\theta}}{1-|r|^2}&1
\end{pmatrix},\quad &\lambda\in\Omega_4,\\
&\begin{pmatrix}
1&\frac{\bar re^{2\rmi t\theta}}{1-|r|^2}\\0&1
\end{pmatrix}, \,\ \quad\ &\lambda\in\Omega_4^*.\\
\end{aligned}
\right.
\end{equation}
Here the scalar function $\delta(\lambda)$ is given by
\begin{equation}\label{delta6}
\delta(\lambda)=\frac{w(\lambda)}{2\pi\rmi}\left[\sum_{j=1}^n\delta_j\int_{E_j}^{\hat E_j}\frac{\rmi\,\rmd s}{w_+(s)(s-\lambda)}-\ve\int_{ S_2(\xi)}\frac{\log\big(1-|r(s)|^2\big)}{w(s)(s-\lambda)}\rmd s\right],
\end{equation}
and the constants $\delta_j\in\mathbb R$ solve the linear system
\begin{equation}\label{deltasys6}
\sum_{j=1}^n\delta_j\int_{E_j}^{\hat E_j}\frac{\rmi s^k}{w_+(s)}\rmd s=\ve\int_{ S_2(\xi)}\frac{\log\big(1-|r(s)|^2\big)s^k}{w(s)}\rmd s,\quad k=0,1,\dots,n-1.
\end{equation}

\begin{rhp}\label{rhp62}
The function $\delta(\lambda)$ has the properties listed below.
\begin{conds}
\cond{1}{Analyticity:} $\delta(\lambda)$ is analytic in $\mathbb C\setminus\big(\bigcup_{j=0}^n[E_j,\hat E_j]\cup\overline{ S_2}\big)$.
\cond{2}{Jump condition:}
\[
\delta_-(\lambda)=\begin{cases}\delta_+(\lambda)+\ve\log\big(1-|r(\lambda)|^2\big), & \lambda\in S_2,\\ -\delta_+(\lambda)+\rmi\delta_j, & \lambda\in(E_j,\hat E_j),\ j=0,\dots,n,\ \delta_0=0.\end{cases}
\]
\cond{3}{Normalization:} $\delta(\lambda)=\delta(\infty)+\delta^{(1)}\lambda^{-1}+\mathcal O(\lambda^{-2})$ as $\lambda\to\infty$, where
\[
\delta(\infty)=-\frac{\mathcal I_n}{2\pi\rmi}\in\rmi\mathbb R,\quad \delta^{(1)}=-\frac12\delta(\infty)\sum_{j=0}^n(E_j+\hat E_j)-\frac{\mathcal I_{n+1}}{2\pi\rmi}.
\]
\cond{4}{Local behavior:} For $p\in E\cap\partial S_2$, we have
\begin{equation*}
e^{\delta(\lambda)}=\mathcal O((\lambda-p)^{-\ve/2}), \quad \lambda\to p.
\end{equation*}
Furthermore,
\begin{equation*}
\delta(\lambda)=\frac\rmi2\delta_{j_1}+\mathcal O((\lambda-\Ej)^{1/2}), \quad \lambda\to\Ej,
\end{equation*}
and
\begin{equation*}
\delta(\lambda)=\pm\rmi\ve\nu_k\log(\lambda_k-\lambda)+\mathcal O(1), \quad \lambda\to \lambda_k\in\partial S_2\cap\mathcal Z(\xi).
\end{equation*}
\cond{5}{Symmetry:} $\delta(\lambda)=-\overline{\delta(\bar \lambda)}$.
\end{conds}
\end{rhp}

\begin{rhp}\label{rhp63}
The function $\Nc^{(1)}(\lambda)$ satisfies the RH problem below.
\begin{conds}
\cond{1}{Analyticity:} $\Nc^{(1)}(\lambda)$ is analytic in $\mathbb C\setminus\Sigma^{(1)}$, where
\begin{equation*}
\Sigma^{(1)}=\bigcup_{j=0}^n[E_j,\hat E_j]\cup[\lambda_*,\Ej]\cup\Sigma_3\cup\Sigma_3^*\cup\Sigma_4\cup\Sigma_4^*.
\end{equation*}
\cond{2}{Jump condition:} $\Nc^{(1)}_+(\lambda)=\Nc^{(1)}_-(\lambda)J^{(1)}_{\Nc}(\lambda)$ with
\[
J^{(1)}_{\Nc}(\lambda)=e^{\frac\rmi2\left(t(\theta_+(\lambda)+\theta_-(\lambda))+\tilde\phi_j+\delta_j\right)\hat\sigma_3}\begin{pmatrix}0&-\rmi\\-\rmi&0\end{pmatrix},\quad \lambda\in(E_j,\hat E_j),\ j=0,\dots,n,
\]
and for $\ve=1$,
\[
J^{(1)}_{\Nc}=\begin{cases}\psm{1&\bar re^{2\rmi t\theta+2\delta}\\0&1}, & \lambda\in\Sigma_3,\\[4pt] \psm{1&0\\-re^{-2\rmi t\theta-2\delta}&1}, & \lambda\in\Sigma_3^*,\\[4pt] \psm{1&0\\-\frac{re^{-2\rmi t\theta-2\delta}}{1-|r|^2}&1}, & \lambda\in\Sigma_4,\\[4pt] \psm{1&\frac{\bar re^{2\rmi t\theta+2\delta}}{1-|r|^2}\\0&1}, & \lambda\in\Sigma_4^*,\\[4pt] \psm{1&\bar re^{2\rmi t\theta+2\delta}\\-re^{-2\rmi t\theta-2\delta}&1-|r|^2}, & \lambda\in(\lambda_*,\Ej);\end{cases}
\]
for $\ve=-1$,
\[
J^{(1)}_{\Nc}=\begin{cases}\psm{1&0\\-\bar re^{-2\rmi t\theta-2\delta}&1}, & \lambda\in\Sigma_3,\\[4pt] \psm{1&re^{2\rmi t\theta+2\delta}\\0&1}, & \lambda\in\Sigma_3^*,\\[4pt] \psm{1&\frac{re^{2\rmi t\theta+2\delta}}{1-|r|^2}\\0&1}, & \lambda\in\Sigma_4,\\[4pt] \psm{1&0\\-\frac{\bar re^{-2\rmi t\theta-2\delta}}{1-|r|^2}&1}, & \lambda\in\Sigma_4^*,\\[4pt] \psm{1-|r|^2&re^{2\rmi t\theta+2\delta}\\-\bar re^{-2\rmi t\theta-2\delta}&1}, & \lambda\in(\lambda_*,\Ej).\end{cases}
\]
\cond{3}{Normalization:} For $\lambda\to\infty$, $\Nc^{(1)}(\lambda)=I+\mathcal O(\lambda^{-1})$.
\cond{4}{Local behavior:} For $p\in E\setminus\{\Ej\}$, we arrive at
\begin{equation*}
\Nc^{(1)}(\lambda)=\mathcal O((\lambda-p)^{-1/4}), \quad \lambda\to p.
\end{equation*}
Moreover, $\Nc^{(1)}$ stays bounded near $\lambda=0$ and is non-tangentially bounded near $\lambda_k\in\mathcal Z(\xi)\setminus\{\lambda_*\}$.
\end{conds}
\end{rhp}

\subsection{Global parametrix}

For every $\lambda_k\in\mathcal Z(\xi)\setminus\{\lambda_*\}$ we place a disk $U_k$ of fixed radius centered at $\lambda_k$, in which $\Nc^{(pc,k)}$ denotes the parabolic-cylinder parametrix of Section~\ref{sec7}, built with $r$, $\delta$ of this section and $\beta_k$ frozen at $\lambda_k$, so that \eqref{betaHolder}-\eqref{lensmismatch} hold in $U_k$. The solution is then written as
\begin{equation}\label{decomp6}
\Nc^{(1)}(\lambda)=\begin{cases}E(\lambda)\Ng(\lambda), & \lambda\in\mathbb C\setminus\big(U_2\cup\bigcup_kU_k\big),\\ E(\lambda)\Nl(\lambda), & \lambda\in U_2,\\ E(\lambda)\Nc^{(pc,k)}(\lambda), & \lambda\in U_k,\end{cases}
\end{equation}
with $U_2=\{\lambda\in\mathbb C:|\lambda-\Ej|<\rho_2\}$, the radius $\rho_2>0$ being independent of $t$ and subject to
\begin{equation}\label{U2}
\rho_2<\frac12\min\Big\{|\Ej|,\ \lambda_{j_1}-\Ej,\ \Ej-a,\ \operatorname{dist}\big(\Ej,\mathcal Z(\xi_{j_1})\setminus\{\Ej\}\big)\Big\}.
\end{equation}
For $t$ large, this guarantees $\lambda_*\in U_2$ and $0\notin\overline{U_2}$. In the same way, one can infer that
\begin{equation}\label{Nglo6}
\Ng(\lambda)=e^{\rmi(f_0y+g_0t)\sigma_3}D^{(glo)}(\infty)^{-1}D^{(glo)}(\lambda)e^{-\rmi(f_0y+g_0t)\sigma_3},\quad c_j^*=-\frac{\hat y_\ve C_j^f+tC_j^g-\phi_j-\delta_j}{2\pi},
\end{equation}
with $D^{(glo)}$ as introduced after \eqref{Mglo}.

\begin{rhp}\label{rhp64}
The local parametrix $\Nl(\lambda)$ is required to solve the RH problem below.
\begin{conds}
\cond{1}{Analyticity:} $\Nl(\lambda)$ is analytic in $U_2\setminus\Sigma^{(1)}$.
\cond{2}{Jump condition:} $\Nl_+(\lambda)=\Nl_-(\lambda)J^{(1)}_{\Nc}(\lambda)$, $\lambda\in\Sigma^{(1)}\cap U_2$.
\cond{3}{Matching condition:} $\Nl(\lambda)\Ng(\lambda)^{-1}=I+\mathcal O(t^{-1/3})$ uniformly for $\lambda\in\partial U_2$.
\cond{4}{Local behavior:} $\Nl(\lambda)=\mathcal O((\lambda-\Ej)^{-1/4})$ as $\lambda\to\Ej$.
\end{conds}
\end{rhp}

\subsection{Local parametrix}

To isolate the square-root singularity at the endpoint, we write
\begin{equation*}
\hat w^{(j_1)}(\lambda)=\big[-\prod_{k=0}^n(\lambda-\hat E_k)\prod_{k\neq j_1}(\lambda-E_k)\big]^{1/2},
\end{equation*}
which gives
\begin{equation*}
w(\lambda)=(\Ej-\lambda)^{1/2}\hat w^{(j_1)}(\lambda),
\end{equation*}
with $(\Ej-\lambda)^{1/2}>0$ for $\lambda<\Ej$. For fixed $\xi$, the phase has the expansion, as $\lambda\to\Ej$,
\begin{equation}\label{thetaexp6}
\theta(\lambda;\xi)=\theta^{(0,j_1)}+(\xi-\xi_{j_1})\theta^{(1,j_1)}(\Ej-\lambda)^{1/2}+\frac23\theta^{(3,j_1)}(\Ej-\lambda)^{3/2}+\mathcal O\big((\Ej-\lambda)^{5/2}\big),
\end{equation}
where
\[
\theta^{(0,j_1)}=\Big(f_0-\frac12C^f_{j_1}\Big)\xi+g_0-\frac12C^g_{j_1},\quad \theta^{(1,j_1)}=\frac{2P_f(\Ej)}{\hat w^{(j_1)}(\Ej)}>0,
\]
\[
\begin{aligned}
&\theta^{(3,j_1)}=\frac{T(\Ej;\xi)\big(2\hat w^{(j_1)}(\Ej)+\Ej(\hat w^{(j_1)})'(\Ej)\big)-\Ej\,\partial_\lambda T(\Ej;\xi)\,\hat w^{(j_1)}(\Ej)}{\Ej^3\hat w^{(j_1)}(\Ej)^2},\\
&\theta^{(3,j_1)}(\xi_{j_1})=\frac12\theta^{(1,j_1)}\Xi'(\Ej),\quad \operatorname{sgn}\theta^{(3,j_1)}=-\ve.
\end{aligned}
\]
One can find a conformal map $\zeta=\zeta(\lambda)$ on $U_2$ together with $s=s(\xi,t)\in\mathbb R$ for which
\begin{equation}\label{zeta6}
\rmi t\big(\theta(\lambda)-\theta(\Ej)\big)=-\frac23\zeta(\lambda)^{3/2}-s\,\zeta(\lambda)^{1/2},\quad \lambda\in U_2,
\end{equation}
where, as in \eqref{zeta}, $\zeta^{1/2}$ is analytic in $\mathbb C\setminus[0,+\infty)$ with $\zeta^{1/2}=-\rmi\ve|\zeta|^{1/2}$ for $\zeta<0$. In addition,
\begin{equation}\label{zetaprop6}
\begin{aligned}
&\zeta(\Ej)=0,\quad \zeta(\lambda_*)=-\frac s2,\quad \zeta'(\Ej)=\big|\theta^{(3,j_1)}\big|^{2/3}t^{2/3}\big(1+\mathcal O(\xi-\xi_{j_1})\big)>0,\\
&s=2\Big(\frac34t\big|\theta(\lambda_*)-\theta(\Ej)\big|\Big)^{2/3}=\ve(\xi-\xi_{j_1})\,t^{2/3}\,\theta^{(1,j_1)}|\theta^{(3,j_1)}|^{-1/3}\big(1+\mathcal O(\xi-\xi_{j_1})\big)\ge0,
\end{aligned}
\end{equation}
and, writing $\hat\theta(\zeta)=\frac23\zeta^{3/2}+s\zeta^{1/2}$ with principal branches and noting that $\zeta$ maps $U_2\cap\mathbb C^\pm$ into $\mathbb C^\pm$,
\begin{equation}\label{thetahat6}
\rmi t\theta(\lambda)=\rmi t\theta(\Ej)\pm\ve\hat\theta(\zeta(\lambda)),\quad \lambda\in U_2\cap\mathbb C^\pm.
\end{equation}

To build the Painlev\'e XXXIV parametrix, we introduce
\[
\begin{gathered}
\tau_0=t\theta(\Ej),\quad \vartheta=\tilde\phi_{j_1}+\delta_{j_1},\quad \kappa=\tau_0+\frac\vartheta2,\quad \zeta_*=-\frac s2,\quad \Gamma_\pm=\zeta_*+e^{\pm\frac{2\pi\rmi}3}\mathbb R^+,\\ R=\begin{cases}\overline{r(\Ej)}e^{\rmi\delta_{j_1}}, & \ve=1,\\ r(\Ej)e^{\rmi\delta_{j_1}}, & \ve=-1,\end{cases},
\end{gathered}
\]
where both rays $\Gamma_\pm$ are directed towards $\zeta_*$ and $\tilde\phi_{j}=\phi_j+(y-\hat y_\ve)C_j^f$. Since $\zeta'(\Ej)>0$, the map $\lambda\mapsto\zeta(\lambda)$ preserves orientation: $\mathbb R^+$ and $[\zeta_*,0]$ are oriented from left to right. Freezing $r(\lambda)e^{2\delta(\lambda)}$ at $\Ej$ in the jumps of RH Problem~\ref{rhp63}, one obtains the model jumps for $\ve=1$
\begin{equation*}
J^{(loc)}=\begin{cases}e^{\rmi\kappa\hat\sigma_3}\psm{0&-\rmi\\-\rmi&0}, & \zeta\in\mathbb R^+,\\[3pt]
e^{\rmi\tau_0\hat\sigma_3}\psm{1&Re^{2\hat\theta}\\0&1}, & \zeta\in\Gamma_+,\\[3pt]
e^{\rmi\tau_0\hat\sigma_3}\psm{1&0\\-\bar Re^{2\hat\theta}&1}, \quad & \zeta\in\Gamma_-,\\
e^{\rmi\tau_0\hat\sigma_3}\psm{1&0\\-\bar Re^{2\hat\theta_-}&1}\psm{1&Re^{2\hat\theta_+}\\0&1}, & \zeta\in(\zeta_*,0).
\end{cases}
\end{equation*}
In addition, for $\ve=-1$,
\begin{equation*}
J^{(loc)}=\begin{cases}e^{\rmi\kappa\hat\sigma_3}\psm{0&-\rmi\\-\rmi&0}, & \zeta\in\mathbb R^+,\\[3pt]
e^{\rmi\tau_0\hat\sigma_3}\psm{1&0\\-\bar Re^{2\hat\theta}&1}, \quad &\zeta\in\Gamma_+,\\
e^{\rmi\tau_0\hat\sigma_3}\psm{1&Re^{2\hat\theta}\\0&1}, & \zeta\in\Gamma_-\ ,\\[3pt]
e^{\rmi\tau_0\hat\sigma_3}\psm{1&Re^{2\hat\theta_-}\\0&1}\psm{1&0\\-\bar Re^{2\hat\theta_+}&1}, & \zeta\in(\zeta_*,0)\ ,
\end{cases}
\end{equation*}
where $\hat\theta_\pm$ are the boundary values from $\mathbb C^\pm$. As in Section~\ref{sec5}, the parameters are
\begin{equation}\label{gamma6}
(\alpha,\omega)=\Big(\frac14,0\Big).
\end{equation}
Denote by $D_+$ the region enclosed by $\Sigma_2^P$ and $\Gamma_+$, and by $D_-$ the one enclosed by $\Sigma_4^P$ and $\Gamma_-$. The model function is
\begin{equation}\label{Psihat6}
\hat\Psi(\zeta)=\begin{pmatrix}1&0\\\rmi a(s)&1\end{pmatrix}M^{(P34)}\Big(\zeta;s,\frac14,0\Big)\,e^{\hat\theta(\zeta)\sigma_3}G_1(\zeta)G_2(\zeta),
\end{equation}
with $a(s)$ from \eqref{as}. Here, for $\ve=1$,
\begin{equation}
G_1(\zeta)=\begin{cases}-\rmi\sigma_2e^{-\frac{\pi\rmi}4\sigma_3}e^{-\rmi\kappa\sigma_3}, & \zeta\in\mathbb C^+,\\ e^{-\frac{\pi\rmi}4\sigma_3}e^{-\rmi\kappa\sigma_3}, & \zeta\in\mathbb C^-,\end{cases}
\end{equation}
\begin{equation}
G_2=e^{\rmi\tau_{0}\hat{\sigma}_{3}}\left\{
\begin{aligned}
&\begin{pmatrix}
1&Re^{2\hat\theta}\\0&1
\end{pmatrix},\quad &\zeta\in D_+,\\
&\begin{pmatrix}
1&0\\\bar Re^{2\hat\theta}&1
\end{pmatrix},\quad\, \, \  &\zeta\in D_-,
\end{aligned}
\right.
\end{equation}
and for $\ve=-1$
\begin{equation}
G_1(\zeta)=\begin{cases}e^{-\frac{\pi\rmi}4\sigma_3}e^{-\rmi\kappa\sigma_3}, & \zeta\in\mathbb C^+,\\ \rmi\sigma_2e^{-\frac{\pi\rmi}4\sigma_3}e^{-\rmi\kappa\sigma_3}, & \zeta\in\mathbb C^-,\end{cases}
\end{equation}
\begin{equation}
G_2=e^{\rmi\tau_{0}\hat{\sigma}_{3}}\left\{
\begin{aligned}
&\begin{pmatrix}
1&0\\-\bar Re^{2\hat\theta}&1
\end{pmatrix},\quad &\zeta\in D_+,\\
&\begin{pmatrix}
1&-Re^{2\hat\theta}\\0&1
\end{pmatrix},\quad\, \, \  &\zeta\in D_-.
\end{aligned}
\right.
\end{equation}
With $U_2'$, $U_2''$ of radii $\rho_2/2$, $3\rho_2/4$ about $\Ej$, a cutoff $\chi$ and $J^{(\chi)}$ as in Section~\ref{sec5}, put $\Nc^{(loc,0)}=\mathcal R\hat\Psi$, where $\mathcal R$ is the $L^6$ small-norm solution with jump
\begin{equation*}
\hat\Psi_-J^{(\chi)}J^{(loc)-1}\hat\Psi_-^{-1}.
\end{equation*}
Then
\begin{equation*}
\Nc^{(loc,0)}(\zeta)=(I+\mathcal O(\zeta^{-1}))\zeta^{-\sigma_3/4}\frac1{\sqrt2}\psm{1&\rmi\\\rmi&1}G_1(\zeta), \quad \zeta\to\infty,
\end{equation*}
uniformly in $s\in[-C',C']$. The fourth-root bound at $\Ej$ follows as after \eqref{E}, and
\begin{equation*}
\Nc^{(loc,0)}=(I+\mathcal O(t^{-1/3}))\hat\Psi.
\end{equation*}
The local parametrix is
\begin{equation}\label{Nlocdef}
\begin{aligned}
&\Nl(\lambda)=\tilde H_1(\lambda)t^{\sigma_3/6}\Nc^{(loc,0)}(\zeta(\lambda)),\\
&\tilde H_1(\lambda)=\Ng(\lambda)G_1(\zeta(\lambda))^{-1}\frac1{\sqrt2}\begin{pmatrix}1&-\rmi\\-\rmi&1\end{pmatrix}\big(t^{-2/3}\zeta(\lambda)\big)^{\sigma_3/4}.
\end{aligned}
\end{equation}
The prefactor $\tilde H_1$ is analytic in $U_2$, and for $t\to\infty$,
\begin{equation}\label{match6}
\Nl(\lambda)\Ng(\lambda)^{-1}=I+\frac{t^{1/3}}{\zeta(\lambda)}\tilde H_1(\lambda)\begin{pmatrix}0&\rmi a(s)\\0&0\end{pmatrix}\tilde H_1(\lambda)^{-1}+\mathcal O(t^{-2/3}).
\end{equation}

\subsection{The small-norm RH problem}

\begin{figure}[htbp]
\centering
\includegraphics[width=0.72\textwidth]{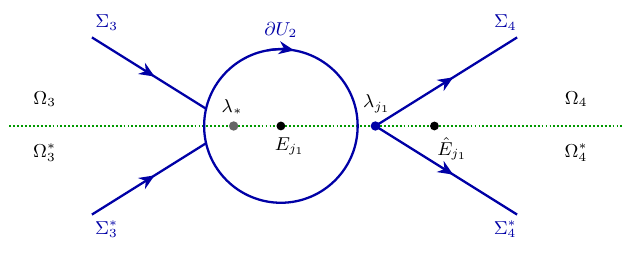}
\caption{Jump contour $\Sigma^{(E)}$ of the error matrix.}\label{fig12}
\end{figure}

\begin{rhp}\label{rhp65}
The error matrix $E(\lambda)$ is then determined by the conditions below.
\begin{conds}
\cond{1}{Analyticity:} $E(\lambda)$ is analytic in $\mathbb C\setminus\Sigma^{(E)}$ (See Figure~\ref{fig12}), where
\begin{equation*}
\Sigma^{(E)}=\partial U_2\cup\bigcup_k\partial U_k\cup\big((\Sigma_3\cup\Sigma_3^*\cup\Sigma_4\cup\Sigma_4^*)\setminus U_2'\big).
\end{equation*}
\cond{2}{Jump condition:} $E_+(\lambda)=E_-(\lambda)J_E(\lambda)$ with
\begin{equation}\label{JE6}
J_E(\lambda)=\begin{cases}\Ng(\lambda)J^{(1)}_{\Nc}(\lambda)\Ng(\lambda)^{-1}, & \lambda\in\Sigma^{(E)}\setminus\big(\overline{U_2}\cup\bigcup_k\overline{U_k}\big),\\ \Nl_-J^{(1)}_{\Nc}J^{(\chi)-1}(\Nl_-)^{-1}, & \lambda\in\Sigma^{(E)}\cap(U_2\setminus\overline{U_2'}),\\ \Nc^{(pc,k)}_-J^{(1)}_{\Nc}J^{(pc,k)-1}(\Nc^{(pc,k)}_-)^{-1}, & \lambda\in\Sigma^{(E)}\cap U_k,\\ \Nl(\lambda)\Ng(\lambda)^{-1}, & \lambda\in\partial U_2,\\ \Nc^{(pc,k)}(\lambda)\Ng(\lambda)^{-1}, & \lambda\in\partial U_k.\end{cases}
\end{equation}
\cond{3}{Asymptotics at infinity:} For $\lambda\to\infty$, $E(\lambda)=I+\mathcal O(\lambda^{-1})$.
\end{conds}
\end{rhp}

For every $p\in[1,\infty]$, (\ref{match6}) implies that
\begin{equation*}
\begin{aligned}
&\|J_E-I\|_{L^p(\Sigma^{(E)}\setminus(\partial U_2\cup\bigcup_k\overline{U_k})})\lesssim e^{-ct},\\
&\|J_E-I\|_{L^p(\Sigma^{(E)}\cap U_k)}\lesssim t^{-\frac12-\frac1{2p}}\log t,\\
&\|J_E-I\|_{L^p(\partial U_2)}\lesssim t^{-1/3},\\
&\|J_E-I\|_{L^p(\partial U_k)}\lesssim t^{-1/2}.
\end{aligned}
\end{equation*}
For large $t$, small-norm theory then yields a unique solution of RH problem \ref{rhp65} as $\lambda\to0$,
\begin{equation}\label{E06}
E(\lambda)=E(0)+E_1\lambda+\mathcal O(\lambda^2),
\end{equation}
where
\begin{equation*}
\begin{aligned}
&E(0)=I-\frac{t^{-1/3}}{|\theta^{(3,j_1)}(\xi_{j_1})|^{2/3}\Ej}\tilde H_1(\Ej)\begin{pmatrix}0&\rmi a(s)\\0&0\end{pmatrix}\tilde H_1(\Ej)^{-1}+\sum_k E_k(0)+\mathcal O(t^{-2/3}),\\
&E_1=-\frac{t^{-1/3}}{|\theta^{(3,j_1)}(\xi_{j_1})|^{2/3}\Ej^2}\tilde H_1(\Ej)\begin{pmatrix}0&\rmi a(s)\\0&0\end{pmatrix}\tilde H_1(\Ej)^{-1}+\sum_k E_{k,1}+\mathcal O(t^{-2/3}).
\end{aligned}
\end{equation*}
The terms $ E_k(0), E_{k,1}=\mathcal O(t^{-1/2})$ coming from the disks $U_k$ are given in (\ref{E07}) below. The lens pieces $\Sigma^{(E)}\cap U_k$ add $\mathcal O(t^{-1}\log t)$, and the pieces in $U_2\setminus\overline{U_2'}$ add $\mathcal O(e^{-ct})$. The leading terms are the residues at $\Ej$ of the clockwise integrals of \eqref{match6}; the sign opposite to \eqref{E0asy} comes from $\zeta'(\Ej)>0$ in \eqref{zetaprop6}, and $\lambda=0\notin\overline{U_2}$. Finally,
\begin{equation*}
E(0)^{-1}=I+\mathcal O(t^{-1/3}).
\end{equation*}

\subsection{The complementary half}\label{sec6neg}
Let $-C\le\ve(\xi-\xi_{j_1})t^{2/3}<0$. The zero $\lambda_*$ of $T(\cdot;\xi)$ near $\Ej$ now lies in the band $(\Ej,\hat E_{j_1})$, with
\begin{equation*}
\lambda_*-\Ej=\frac{\xi-\xi_{j_1}}{\Xi'(\Ej)}+\mathcal O(t^{-4/3}).
\end{equation*}
By monotonicity of $\Xi$ on $(a,\Ej)$,
\begin{equation*}
\ve\big(\Xi(\lambda)-\xi\big)>0,\quad \lambda\in(a,\Ej),
\end{equation*}
so $(a,\Ej)\subset S_1(\xi)$. As in Section~\ref{sec5neg}, $\lambda_*$ is replaced by $\Ej$ in $S_2$, $\Omega_3$, $\Sigma_3$ and $\Sigma^{(1)}$. Thus $\Sigma_3=\Ej+e^{\rmi(\pi-\varphi_2)}\mathbb R^+$, and the segment $[\lambda_*,\Ej]$ is absent. For $\Ej-\lambda=re^{-\rmi\varphi}$, $0<\varphi\le\varphi_2$, \eqref{thetaexp6} gives
\begin{equation*}
\operatorname{Im}\theta(\lambda)=-(\xi-\xi_{j_1})\theta^{(1,j_1)}r^{1/2}\sin\frac\varphi2-\frac23\theta^{(3,j_1)}r^{3/2}\sin\frac{3\varphi}2+\mathcal O(r^{5/2}),
\end{equation*}
where both coefficients have the sign of $\ve$. Hence \eqref{Imtheta6} holds on $\Sigma_3\setminus\{\Ej\}$. The map $\zeta$ of \eqref{zeta6} sends $\lambda_*$ to
\begin{equation*}
\zeta(\lambda_*)=-\frac s2>0,\quad s=-2\Big(\frac34t\big|\theta_+(\lambda_*)-\theta(\Ej)\big|\Big)^{2/3}<0,
\end{equation*}
and the asymptotic formula for $s$ in \eqref{zetaprop6} is unchanged. The rays $\Gamma_\pm$ are replaced by $\Sigma_2^P$, $\Sigma_4^P$, directed towards $0$, and the segment $[\zeta_*,0]$ is dropped. Then $\operatorname{Re}\hat\theta<0$ on these rays, $G_2=I$, and
\begin{equation*}
\hat\Psi(\zeta)=\begin{pmatrix}1&0\\\rmi a(s)&1\end{pmatrix}M^{(P34)}\Big(\zeta;s,\frac14,0\Big)\,e^{\hat\theta(\zeta)\sigma_3}G_1(\zeta),
\end{equation*}
with $G_1$ as in \eqref{Psihat6}. The remaining steps are those of the half treated above. Hence the expansions of $E(0)$, $E_1$ and $E(0)^{-1}$ obtained there hold uniformly for $|\xi-\xi_{j_1}|t^{2/3}\le C$.
 
\section[Asymptotic analysis in regions III and IV]{Asymptotic analysis in the Zakharov-Manakov region III\\ and fast-decay region IV}\label{sec7}

We now treat regions III and IV. In region III it is assumed in addition that $\operatorname{dist}(\mathcal Z(\xi), E)\ge c>0$. Put
\[
S_\pm(\xi)=\Big\{\lambda\in\mathbb R\setminus\textstyle\bigcup_j[E_j,\hat E_j]:\ \pm\big(\Xi(\lambda)-\xi\big)>0\Big\}.
\]

\subsection{The sign of $\operatorname{Im}\theta$ on the bands}\label{sec7sign}
On each band, $\theta_-=\overline{\theta_+}$ and $\theta_++\theta_-$ is constant. In particular, $\operatorname{Im}\theta_+$ vanishes at $E_j,\hat E_j$, and by \eqref{thetap},
\begin{equation}\label{dImtheta}
\frac{\rmd}{\rmd\lambda}\operatorname{Im}\theta_+(\lambda)=-\frac{T(\lambda;\xi)}{\lambda^2}\operatorname{Im}\frac1{w_+(\lambda)},\quad \lambda\in(E_j,\hat E_j),
\end{equation}
with $\operatorname{Im}(1/w_+)\neq0$. Put $\sigma_\xi=\operatorname{sgn}(\Xi-\xi)$ on the gaps and $\sigma_\xi=\operatorname{sgn}\operatorname{Im}\theta_+$ on the bands.

\begin{lemma}\label{lem71}
Let $\xi>0$, $\xi\notin\{\xi_p\}_{p\in E}$.
\begin{enumerate}[label=\textup{(\roman*)},leftmargin=2em]
\item On $(E_j,\hat E_j)$, $\operatorname{sgn}\operatorname{Im}\theta_+=\operatorname{sgn}(\xi_j-\xi)$ near $E_j$ and $\operatorname{sgn}(\hat\xi_j-\xi)$ near $\hat E_j$. Hence $\sigma_\xi$ does not change sign at $E$.
\item Each band contains one or two zeros of $T(\cdot;\xi)$. In the former case, $\operatorname{Im}\theta_+\neq0$ throughout the band. In the latter, $\operatorname{Im}\theta_+$ has exactly one zero $\lambda_j^{sw}$ in the band, and this zero is simple \textup{(}switching band\textup{)}.
\item Each half-line contains exactly one point of $\mathcal Z(\xi)\cup\{\lambda_j^{sw}\}$, denoted $c_-<0<c_+$, and
\begin{equation}\label{sigmaxi}
\sigma_\xi=+1\ \text{on}\ (c_-,c_+),\quad \sigma_\xi=-1\ \text{on}\ \mathbb R\setminus[c_-,c_+].
\end{equation}
\end{enumerate}
\end{lemma}

\begin{proof}
By \eqref{thetaexp} and \eqref{thetaexp6}, together with $(E_j-\lambda)^{1/2}_+=-\rmi(\lambda-E_j)^{1/2}$ and $(\lambda-\hat E_j)^{1/2}_+=\rmi(\hat E_j-\lambda)^{1/2}$ for $\lambda$ in the band,
\begin{equation*}
\begin{aligned}
\operatorname{Im}\theta_+\sim(\xi_j-\xi)\theta^{(1,j)}(\lambda-E_j)^{1/2}, \quad\lambda\to E_j,\\
\operatorname{Im}\theta_+\sim(\xi-\hat\xi_j)\theta^{(1,\hat j)}(\hat E_j-\lambda)^{1/2}, \quad\lambda\to\hat E_j,
\end{aligned}
\end{equation*}
with $\theta^{(1,j)}>0>\theta^{(1,\hat j)}$. On the other hand, the critical points of $\operatorname{Im}\theta_+$ in a band are the one or two simple zeros of $T$ by Lemma~\ref{lem4} and \eqref{dImtheta}. Note that $\operatorname{Im}\theta_+$ vanishes at both endpoints, this gives no interior zero or exactly one simple zero, respectively. Finally, The extra zero of $T$ on each half-line lies in a gap, where it belongs to $\mathcal Z(\xi)$, or in a band, which is then a switching band. The sign $\sigma_\xi$ changes only at these points, and $\sigma_\xi=-1$ near $\pm\infty$, $\sigma_\xi=+1$ near $0$.
\end{proof}

The function $\operatorname{Im}\theta+\xi\operatorname{Im}\lambda-\operatorname{Im}\lambda/(4|\lambda|^2)$ is bounded and harmonic in $\mathbb C^+$, vanishes on the gaps and equals $\operatorname{Im}\theta_+$ on the bands. Hence
\begin{equation}\label{Poisson7}
\operatorname{Im}\theta(\lambda)=-\xi\operatorname{Im}\lambda+\frac{\operatorname{Im}\lambda}{4|\lambda|^2}+\frac1\pi\sum_{j=0}^n\int_{E_j}^{\hat E_j}\operatorname{Im}\theta_+(s)\,\frac{\operatorname{Im}\lambda}{|s-\lambda|^2}\,\rmd s,\quad \lambda\in\mathbb C^+.
\end{equation}

\begin{lemma}\label{lem72}
If $\xi\le0$, then $\operatorname{Im}\theta_+>0$ on every band, and
\begin{equation*}
\operatorname{Im}\theta(\lambda)\ge|\xi|\operatorname{Im}\lambda+\frac{\operatorname{Im}\lambda}{4|\lambda|^2},\quad \lambda\in\mathbb C^+.
\end{equation*}
\end{lemma}

\begin{proof}
We have
\begin{equation*}
\operatorname{Im}\theta_+=|\xi|\operatorname{Im}f_++\operatorname{Im}(-g)_+.
\end{equation*}
Because $f'>0$ and $(-g)'=\Xi f'>0$ on the gaps, and $P_f$, $P_g$ have one zero per band, the argument of Lemma~\ref{lem71}(i)-(ii) gives $\operatorname{Im}f_+>0$ and $\operatorname{Im}(-g)_+>0$. Then apply \eqref{Poisson7}.
\end{proof}

\subsection{Opening of the lenses}\label{sec7lens}
By \eqref{S0}, $S$ extends analytically off the bands. Then, the continuations
\begin{equation*}
\bar r_2\to r_2^*=\frac{S_{12}}{S_{22}},\quad r_2\to\frac{S_{21}}{S_{11}},\quad \frac{r_2}{1-|r_2|^2}\to S_{21}S_{22},\quad \frac{\bar r_2}{1-|r_2|^2}\to S_{12}S_{11}
\end{equation*}
are analytic in $\mathbb C^+$ (respectively $\mathbb C^-$) off the bands. The first two are bounded near $E$ by the nonresonance condition, while \eqref{edge} gives
\begin{equation*}
|S_{21}S_{22}|,\ |S_{12}S_{11}|\sim|c_p|^2|\lambda-p|^{-1},\quad \lambda\to p\in E,
\end{equation*}
compensated by $e^{\pm2\delta}$ (see after RH Problem~\ref{rhp71}). From $S_{22}(\lambda)e^{\rmi\lambda\Delta}\to e^{-\eta_\Delta}$ in $\overline{\mathbb C^+}$, $S_{22}\neq0$ in $\overline{\mathbb C^+}\setminus E$ and $|Y(a,b;\lambda)|\le e^{C|\lambda|}$ in \eqref{S0}, all four continuations are $\mathcal O(e^{C|\lambda|})$ away from $E$. It follows from \eqref{bandrel} that $(r_2^*)_+=e^{2\rmi\phi_j}(r_2)_-$ on the bands. The same holds for the last two. This yields precisely the compatibility required in \eqref{bandid}. Let $\xi>0$. If $c_\pm\in\mathcal Z(\xi)$, we use the sectors
\begin{equation}\label{Omega7}
\begin{aligned}
&\Omega_1^{(k)}=\{\lambda:0\le\arg(\lambda-\lambda_k)\le\varphi_1\},\quad \Omega_2^{(k)}=\{\lambda:\pi-\varphi_1\le\arg(\lambda-\lambda_k)\le\pi\}, && \lambda_k=c_-<0,\\
&\Omega_1^{(k)}=\{\lambda:\pi-\varphi_1\le\arg(\lambda-\lambda_k)\le\pi\},\quad \Omega_2^{(k)}=\{\lambda:0\le\arg(\lambda-\lambda_k)\le\varphi_1\}, && \lambda_k=c_+>0,
\end{aligned}
\end{equation}
and set $a_\pm=b_\pm=c_\pm$. If $c_\pm=\lambda_j^{sw}$, the lenses must avoid $\lambda_j^{sw}$, where $\operatorname{Im}\theta_+=0$. We take
\begin{equation*}
a_\pm=\lambda_j^{sw}\mp d,\quad b_\pm=\lambda_j^{sw}\pm d,\quad 0<d<\tfrac12\operatorname{dist}(\lambda_j^{sw},E),
\end{equation*}
so that $\operatorname{Im}\theta_+(a_\pm)>0>\operatorname{Im}\theta_+(b_\pm)$. Let $\Omega_1$ be a lens over $[a_-,a_+]$, $\Omega_2$ two lenses over $(-\infty,b_-]$ and $[b_+,\infty)$, asymptotic to $\arg\lambda=\pi-\varphi_1,\varphi_1$. Their upper boundaries $\Sigma_1,\Sigma_2$, oriented from left to right, meet the bands vertically at $a_\pm,b_\pm$. The segments $(a_\pm,b_\pm)$ carry only the band jump. For $\xi\le-c_0$, $\Omega_2=\emptyset$ and $\Omega_1=\{0<\operatorname{Im}\lambda<h(1+|\operatorname{Re}\lambda|)\}$, so that $\Sigma_1=\{x+\rmi h(1+|x|):x\in\mathbb R\}$.

\begin{lemma}\label{lem73}
Let $K$ be a compact subset of $\{\xi>0:\mathcal Z(\xi)\neq\emptyset\}\setminus\{\xi_p\}_{p\in E}$ or of $\{\xi>0:\mathcal Z(\xi)=\emptyset\}\setminus\{\xi_p\}_{p\in E}$. The lenses can be chosen so that
\begin{equation}\label{Imtheta7}
\begin{aligned}
&\operatorname{Im}\theta(\lambda)\ge c,\ \lambda\in\Sigma_1\setminus D_\rho, &&\operatorname{Im}\theta(\lambda)\le-c(1+|\lambda|),\ \lambda\in\Sigma_2\setminus D_\rho,\\
&\operatorname{Im}\theta(\lambda)\ge c|\lambda-\lambda_k|^2,\ \lambda\in\Sigma_1\cap D_\rho, &&\operatorname{Im}\theta(\lambda)\le-c|\lambda-\lambda_k|^2,\ \lambda\in\Sigma_2\cap D_\rho,
\end{aligned}
\end{equation}
where $c,\rho>0$ and
\begin{equation*}
D_\rho=\bigcup_{\lambda_k\in\mathcal Z(\xi)}\{|\lambda-\lambda_k|<\rho\}.
\end{equation*}
If $\xi\le-c_0$, then $\operatorname{Im}\theta\ge c_0h(1+|\operatorname{Re}\lambda|)$ on $\Sigma_1$.
\end{lemma}

\begin{proof}
The case $\xi\le-c_0$ is Lemma~\ref{lem72}. For $\xi\in K$, $\sigma_\xi=+1$ on $[a_-,a_+]$ and $-1$ on $(-\infty,b_-]\cup[b_+,\infty)$ by \eqref{sigmaxi}. Each real point of these intervals has a neighbourhood in which $\sigma_\xi\operatorname{Im}\theta>0$ in $\mathbb C^+$, it is seen that
\begin{equation*}
\operatorname{Im}\theta=\begin{cases}\eta\,\theta'(x)+\mathcal O(\eta^2), & \lambda=x+\rmi\eta,\ x\ \text{in a gap},\\ \operatorname{Im}\theta_+(x)+\mathcal O(|\lambda-x|), & x\ \text{in a band},\\ \sigma_\xi(p)|\xi-\xi_p||\theta^{(1)}(p)|\,r^{1/2}\chi_p(\varphi)\big(1+\mathcal O(r)\big), & \lambda=p+re^{\rmi\varphi},\ p\in E,\\ \frac{\operatorname{Im}\lambda}{4|\lambda|^2}\big(1+\mathcal O(\lambda)\big), & \lambda\to0,\\ -\xi\operatorname{Im}\lambda+\mathcal O(|\lambda|^{-1}), & \lambda\to\infty,\end{cases}
\end{equation*}
where $\chi_p(\varphi)=\cos\frac\varphi2$ at $E_j$ and $\sin\frac\varphi2$ at $\hat E_j$. By compactness these neighbourhoods contain a strip of height $h$, in which we place $\Sigma_1,\Sigma_2$. Near $\lambda_k$, \eqref{thetaexp7} gives the quadratic bounds for $\varphi_1<\pi/4$. All quantities depend continuously on $\xi\in K$, so the constants are uniform.
\end{proof}
On $\Sigma_1$ for $\xi\le-c_0$ we have $|\lambda|\le(1+h)(1+|\operatorname{Re}\lambda|)$. Lemma~\ref{lem73} with the bound $\mathcal O(e^{C|\lambda|})$ gives
\begin{equation*}
\begin{aligned}
&|r_2^*e^{2\rmi t\theta}|\le Ce^{-(2c_0ht-C(1+h))(1+|\operatorname{Re}\lambda|)}, && \lambda\in\Sigma_1,\ \xi\le-c_0,\\
&|S_{21}S_{22}\,e^{-2\rmi t\theta}|\le Ce^{-(2ct-C)(1+|\lambda|)}, && \lambda\in\Sigma_2\setminus D_\rho,
\end{aligned}
\end{equation*}
with the conjugate bounds on $\Sigma_1^*,\Sigma_2^*$. The lens jumps off $D_\rho$ are thus $\mathcal O(e^{-ct})$ in $L^1\cap L^\infty$, uniformly in $\xi$.
Next, we introduce the transformation
\begin{equation}\label{N1def7}
\Nc^{(1)}(\lambda)=e^{\delta(\infty)\sigma_3}\Nc(\lambda) G(\lambda)e^{-\delta(\lambda)\sigma_3},
\end{equation}
where $ G=I$ outside $\Omega_1\cup\Omega_1^*\cup\Omega_2\cup\Omega_2^*$ and
\begin{equation}\label{G7}
G=\left\{
\begin{aligned}
&\begin{pmatrix}
1&-\bar r_2e^{2\rmi t\theta}\\0&1
\end{pmatrix},\quad &\lambda\in \Omega_1,\\
&\begin{pmatrix}
1&0\\-r_2e^{-2\rmi t\theta}&1
\end{pmatrix},\quad\, \, \  &\lambda\in \Omega_1^{\ast},\\
&\begin{pmatrix}
1&0\\\frac{r_2e^{-2\rmi t\theta}}{1-|r_2|^2}&1
\end{pmatrix},\quad\, \, \  &\lambda\in \Omega_2,\\
&\begin{pmatrix}
1&\frac{\bar r_2e^{2\rmi t\theta}}{1-|r_2|^2}\\0&1
\end{pmatrix},\quad\, \, \  &\lambda\in \Omega_2^{\ast},\\
\end{aligned}
\right.
\end{equation}
Here $\delta(\lambda)$ is the scalar function
\begin{equation}\label{delta7}
\delta(\lambda)=\frac{w(\lambda)}{2\pi\rmi}\left[\sum_{j=1}^n\delta_j\int_{E_j}^{\hat E_j}\frac{\rmi\,\rmd s}{w_+(s)(s-\lambda)}-\int_{S_-(\xi)}\frac{\log\big(1-|r_2(s)|^2\big)}{w(s)(s-\lambda)}\rmd s\right],
\end{equation}
with the constants $\delta_j\in\mathbb R$ fixed by the linear system
\begin{equation}\label{deltasys7}
\sum_{j=1}^n\delta_j\int_{E_j}^{\hat E_j}\frac{\rmi s^k}{w_+(s)}\rmd s=\int_{S_-(\xi)}\frac{\log\big(1-|r_2(s)|^2\big)s^k}{w(s)}\rmd s,\quad k=0,1,\dots,n-1.
\end{equation}
If $\xi<0$, then $\delta\equiv0$. We also write
\begin{equation*}
\begin{aligned}
&L(s)=\log(1-|r_2(s)|^2), \quad \nu_k=-\frac1{2\pi}L(\lambda_k),\\
&\mathcal I_m=\sum_{j=1}^n\delta_j\int_{E_j}^{\hat E_j}\frac{\rmi s^m}{w_+(s)}\rmd s-\int_{S_-}\frac{L(s)s^m}{w(s)}\rmd s, \quad m\ge0.
\end{aligned}
\end{equation*}

\begin{rhp}\label{rhp71}
$\delta(\lambda)$ solves the following scalar RH problem.
\begin{conds}
\cond{1}{Analyticity:} $\delta(\lambda)$ is analytic in $\mathbb C\setminus\big(\bigcup_{j=0}^n[E_j,\hat E_j]\cup\overline{S_-}\big)$.
\cond{2}{Jump condition:}
\begin{equation}\label{deltajump7}
\delta_-(\lambda)=\begin{cases}\delta_+(\lambda)+L(\lambda), & \lambda\in S_-(\xi),\\ -\delta_+(\lambda)+\rmi\delta_j, & \lambda\in(E_j,\hat E_j),\ j=0,\dots,n,\ \delta_0=0.\end{cases}
\end{equation}
\cond{3}{Normalization:} $\delta(\lambda)=\delta(\infty)+\delta^{(1)}\lambda^{-1}+\mathcal O(\lambda^{-2})$ as $\lambda\to\infty$, where
\begin{equation}\label{deltainf7}
\delta(\infty)=-\frac{\mathcal I_n}{2\pi\rmi}\in\rmi\mathbb R,\quad \delta^{(1)}=-\frac12\delta(\infty)\sum_{j=0}^n(E_j+\hat E_j)-\frac{\mathcal I_{n+1}}{2\pi\rmi}.
\end{equation}
\cond{4}{Local behavior:} For $p\in E\cap\partial S_-$, then
\begin{equation*}
|e^{\delta(\lambda)}|\asymp|\lambda-p|^{-1/2}, \quad \lambda\to p \ \text{from} \ \mathbb C^+,
\end{equation*}
and
\begin{equation*}
\delta(\lambda)=\operatorname{sgn}(\lambda_k)\,\rmi\nu_k\log(\lambda_k-\lambda)+\mathcal O(1), \quad \lambda\to \lambda_k\in\mathcal Z(\xi).
\end{equation*}
\cond{5}{Symmetry:} $\delta(\lambda)=-\overline{\delta(\bar \lambda)}$.
\end{conds}
\end{rhp}

By \eqref{edge}, $L(s)=\log|s-p|+\mathcal O(1)$ near $p$. The Cauchy integral in \eqref{delta7} then gives
\begin{equation*}
\delta(\lambda)=-\tfrac12\log(\lambda-p)+\mathcal O(1),\quad \lambda\to p\ \text{from}\ \mathbb C^+,
\end{equation*}
which is (4). By (5), $|e^{\delta}|\asymp|\lambda-p|^{1/2}$ from $\mathbb C^-$. In $\Omega_2\cup\Omega_2^*$ near $p\in E\cap\partial S_-$,
\begin{equation*}
\Nc^{(1)}=e^{\delta(\infty)\sigma_3}\big(\Nc e^{-\delta\sigma_3}\big)\big(e^{\delta\sigma_3}Ge^{-\delta\sigma_3}\big),\quad e^{\delta\sigma_3}Ge^{-\delta\sigma_3}=\begin{cases}\psm{1&0\\S_{21}S_{22}e^{-2\rmi t\theta-2\delta}&1}, & \Omega_2,\\[3pt] \psm{1&S_{12}S_{11}e^{2\rmi t\theta+2\delta}\\0&1}, & \Omega_2^*.\end{cases}
\end{equation*}
The entries $S_{21}S_{22}e^{-2\delta}$, $S_{12}S_{11}e^{2\delta}$ are $\mathcal O(|\lambda-p|^{-1}\cdot|\lambda-p|)=\mathcal O(1)$. By \eqref{Nloc} with $\ve=1$,
\begin{equation*}
\Nc e^{-\delta\sigma_3}=\big(\mathcal O(|\lambda-p|^{\mp1/4\pm1/2}),\ \mathcal O(|\lambda-p|^{\pm1/4\mp1/2})\big)=\mathcal O(|\lambda-p|^{-1/4}),\quad \lambda\to p\ \text{from}\ \mathbb C^\pm,
\end{equation*}
hence 
\begin{equation*}
\Nc^{(1)}=\mathcal O(|\lambda-p|^{-1/4}).
\end{equation*}
At the remaining endpoints $e^{\pm\delta}$, $r_2^*$, $S_{21}/S_{11}$ are bounded. The same holds for \eqref{M1}, \eqref{N1def}, with $|e^{\delta}|\asymp|\lambda-p|^{\pm\ve/2}$ from $\mathbb C^+$.

\begin{figure}[htbp]
\centering
\includegraphics[width=0.8\textwidth]{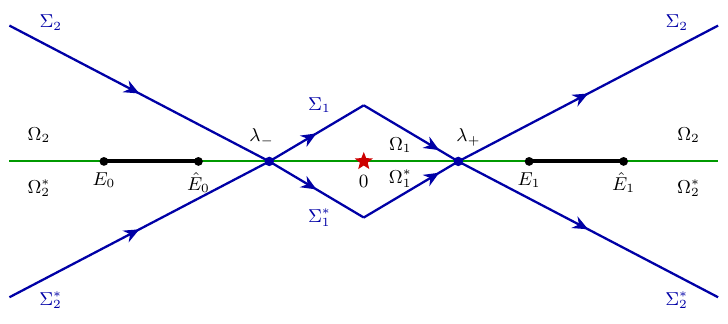}
\caption{Jump contour $\Sigma^{(1)}$ (Zakharov-Manakov region).}\label{fig13}
\end{figure}

\begin{figure}[htbp]
\centering
\includegraphics[width=0.8\textwidth]{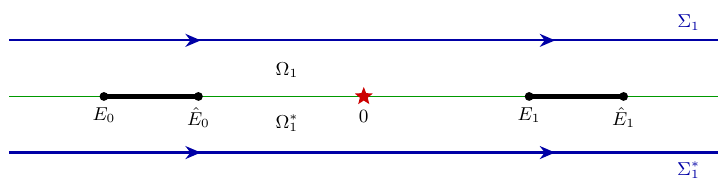}
\caption{Jump contour $\Sigma^{(1)}$ in the fast-decay region for $\xi<0$.}\label{fig14}
\end{figure}

\begin{figure}[htbp]
\centering
\includegraphics[width=0.8\textwidth]{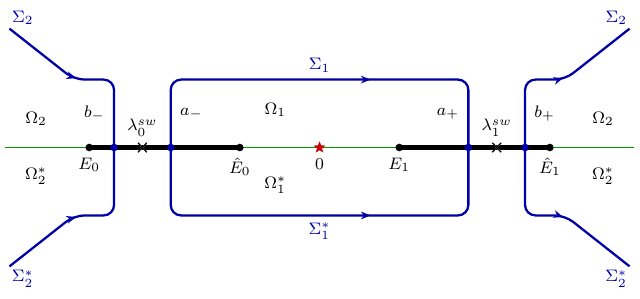}
\caption{Jump contour $\Sigma^{(1)}$ in the fast-decay region for $\xi>0$.}\label{fig16}
\end{figure}

\begin{rhp}\label{rhp72}
The matrix $\Nc^{(1)}(\lambda)$ then solves the following RH problem.
\begin{conds}
\cond{1}{Analyticity:} $\Nc^{(1)}(\lambda)$ is analytic in $\mathbb C\setminus\Sigma^{(1)}$ (See Figures~\ref{fig13}, \ref{fig14} and \ref{fig16}), where
\begin{equation*}
\Sigma^{(1)}=\bigcup_{j=0}^n[E_j,\hat E_j]\cup\Sigma_1\cup\Sigma_1^*\cup\Sigma_2\cup\Sigma_2^*.
\end{equation*}
\cond{2}{Jump condition:} $\Nc^{(1)}_+(\lambda)=\Nc^{(1)}_-(\lambda)J^{(1)}_{\Nc}(\lambda)$, where
\begin{equation}\label{J17}
J^{(1)}_{\Nc}(\lambda)=\left\{
\begin{aligned}
&\begin{pmatrix}
1&\bar r_2e^{2\rmi t\theta+2\delta}\\0&1
\end{pmatrix},&\quad &\lambda\in \Sigma_1,\\
&\begin{pmatrix}
1&0\\-r_2e^{-2\rmi t\theta-2\delta}&1
\end{pmatrix},&\quad\, \, \  &\lambda\in\Sigma_1^*,\\
&\begin{pmatrix}
1&0\\-\frac{r_2e^{-2\rmi t\theta-2\delta}}{1-|r_2|^2}&1
\end{pmatrix},&\quad\, \, \  &\lambda\in\Sigma_2,\\
&\begin{pmatrix}
1&\frac{\bar r_2e^{2\rmi t\theta+2\delta}}{1-|r_2|^2}\\0&1
\end{pmatrix},&\quad\, \, \  &\lambda\in\Sigma_2^*,\\
&e^{\frac\rmi2\left(t(\theta_+(\lambda)+\theta_-(\lambda))+\phi_j+\delta_j\right)\hat\sigma_3}\begin{pmatrix}
0&-\rmi\\-\rmi&0
\end{pmatrix},&\quad\, \, \, &\lambda\in(E_j,\hat E_j).\\
\end{aligned}
\right.
\end{equation}
\cond{3}{Normalization:} $\Nc^{(1)}(\lambda)=I+\mathcal O(\lambda^{-1})$ as $\lambda\to\infty$.
\cond{4}{Local behavior:} For $p\in E$, we obtain
\begin{equation*}
\Nc^{(1)}(\lambda)=\mathcal O((\lambda-p)^{-1/4}), \quad \lambda\to p.
\end{equation*}
Moreover, $\Nc^{(1)}$ remains bounded in a neighborhood of $\lambda=0$, and it is non-tangentially bounded at each $\lambda_k\in\mathcal Z(\xi)$.
\end{conds}
\end{rhp}


The solution is factorized as
\begin{equation}\label{decomp7}
\Nc^{(1)}(\lambda)=\begin{cases}E(\lambda)\Ng(\lambda), & \lambda\in\mathbb C\setminus U_3,\\ E(\lambda)\Nl(\lambda), & \lambda\in U_3,\end{cases}
\end{equation}
where
\begin{equation}\label{U3}
U_3=\begin{cases}\bigcup_{\lambda_k\in\mathcal Z(\xi)}U_3^{(k)},\quad U_3^{(k)}=\{\lambda:|\lambda-\lambda_k|<\rho_3\}, & \xi\in III,\\ \emptyset, & \xi\in IV,\end{cases}
\end{equation}
with
\begin{equation}\label{rho3}
\rho_3=\frac12\min_{\lambda_k\in\mathcal Z(\xi)}\Big\{\operatorname{dist}(\lambda_k, E),\ |\lambda_k|,\ \min_{\lambda_l\neq \lambda_k}|\lambda_k-\lambda_l|\Big\}.
\end{equation}
The outer parametrix is obtained as in transition region I: $\Ng$ solves RH Problem~\ref{rhp54} with $y_\ve=y$ and with $\delta_j$ taken from (\ref{deltasys7}), so that \eqref{Mglo} applies with $c_j^*=-(yC_j^f+tC_j^g-\phi_j-\delta_j)/(2\pi)$. If $\xi<0$, then
\[
\Ng=e^{\rmi(f_0y+g_0t)\sigma_3}M(y,t;\lambda;\bphi)e^{-\rmi(f_0y+g_0t)\sigma_3}.
\]
Local parametrices are needed only in the Zakharov-Manakov region.

\begin{rhp}\label{rhp73}
We seek $\Nl(\lambda)$ with the following properties.
\begin{conds}
\cond{1}{Analyticity:} $\Nl(\lambda)$ is analytic in $U_3\setminus\Sigma^{(1)}$.
\cond{2}{Jump condition:} $\Nl_+(\lambda)=\Nl_-(\lambda)J^{(1)}_{\Nc}(\lambda)$, $\lambda\in\Sigma^{(1)}\cap U_3$.
\cond{3}{Matching condition:} $\Nl(\lambda)\Ng(\lambda)^{-1}=I+\mathcal O(t^{-1/2})$ uniformly for $\lambda\in\partial U_3$.
\cond{4}{Local behavior:} $\Nl(\lambda)$ is non-tangentially bounded near each $\lambda_k\in\mathcal Z(\xi)$.
\end{conds}
\end{rhp}

Set $\ve_k=\operatorname{sgn}\lambda_k$. Then, as $\lambda\to \lambda_k$,
\begin{equation}\label{thetaexp7}
\theta(\lambda)=\theta(\lambda_k)-\ve_k\kappa_k(\lambda-\lambda_k)^2+\mathcal O\big((\lambda-\lambda_k)^3\big),
\end{equation}
where
\begin{equation}\label{kappa7}
\kappa_k=\frac{|\partial_\lambda T(\lambda_k;\xi)|}{2\lambda_k^2|w(\lambda_k)|}=\frac12f'(\lambda_k)|\Xi'(\lambda_k)|>0.
\end{equation}
On the disk $U_3^{(k)}$ we introduce the conformal map $\zeta_k$ given by
\begin{equation}\label{zeta7}
\frac{\zeta_k^2}{4}=-\ve_kt\big(\theta(\lambda)-\theta(\lambda_k)\big),\quad \zeta_k(\lambda)=2\sqrt{\kappa_kt}\,(\lambda-\lambda_k)\big(1+\mathcal O(\lambda-\lambda_k)\big).
\end{equation}
Splitting off the logarithmic singularity, we take
\begin{equation*}
\delta(\lambda)=\rmi\ve_k\nu_k\log(\ve_k(\lambda_k-\lambda))+\beta_k(\lambda),
\end{equation*}
with
\[
\beta_k(\lambda_k)=-\rmi\ve_k\nu_k\log\rho_3+\frac{w(\lambda_k)}{2\pi\rmi}\left[\sum_{j=1}^n\delta_j\int_{E_j}^{\hat E_j}\frac{\rmi\,\rmd s}{w_+(s)(s-\lambda_k)}-\int_{\mathbb R}\frac{\chi_{S_-}(s)\frac{L(s)}{w(s)}-\chi_k(s)\frac{L(\lambda_k)}{w(\lambda_k)}}{s-\lambda_k}\rmd s\right],
\]
where $\chi_k$ denotes the indicator function of $(\lambda_k,\lambda_k+\rho_3)$ if $\lambda_k>0$ and of $(\lambda_k-\rho_3,\lambda_k)$ if $\lambda_k<0$. The function $\beta_k$ is not analytic in $U_3^{(k)}$. By \eqref{deltajump7}, it is analytic in $U_3^{(k)}\setminus S_-$, continuous up to $S_-$, and on $S_-\cap U_3^{(k)}$ it has the jump
\begin{equation}\label{betajump}
(\beta_k)_-(s)-(\beta_k)_+(s)=L(s)-L(\lambda_k).
\end{equation}
Since $\overline{U_3^{(k)}}$ is disjoint from $E$, $r_2$ extends analytically to a neighborhood of the real segment $\overline{U_3^{(k)}}\cap\mathbb R$ with $|r_2|<1$ there, so $L$ is real-analytic on that segment. Hence the jump \eqref{betajump} vanishes linearly at $\lambda_k$, and the Plemelj–Privalov estimate for the Cauchy integral gives
\begin{equation}\label{betaHolder}
|\beta_k(\lambda)-\beta_k(\lambda_k)|\le C|\lambda-\lambda_k|\big(1+\big|\log|\lambda-\lambda_k|\big|\big),\quad \lambda\in\overline{U_3^{(k)}},
\end{equation}
uniformly for $\xi$ in compact subsets of region III. Moreover $\beta_k(\lambda_k)\in\rmi\mathbb R$ by the symmetry of $\delta$. Define further
\[
\rho_k=\begin{cases}r_2(\lambda_k)e^{-2\rmi t\theta(\lambda_k)-2\beta_k(\lambda_k)}\big(2\sqrt{\kappa_kt}\big)^{2\rmi\nu_k}, & \lambda_k>0,\\[2pt] u_-^{(k)}=\,\overline{r_2(\lambda_k)}e^{2\rmi t\theta(\lambda_k)+2\beta_k(\lambda_k)}\big(2\sqrt{\kappa_kt}\big)^{2\rmi\nu_k}, & \lambda_k<0,\end{cases}
\]
so that $|\rho_k|=|r_2(\lambda_k)|$, and
\begin{equation}\label{beta7}
\beta_{12}^{(k)}=\frac{\sqrt{2\pi}e^{\frac{\pi\rmi}4}e^{-\frac{\pi\nu_k}2}}{\rho_k\Gamma(-\rmi\nu_k)},\quad \beta_{21}^{(k)}=\frac{\nu_k}{\beta_{12}^{(k)}}=\overline{\beta_{12}^{(k)}},\quad A_k=\begin{cases}\psm{0&-\rmi\beta_{12}^{(k)}\\\rmi\overline{\beta_{12}^{(k)}}&0}, & \lambda_k>0,\\[4pt] \psm{0&\rmi\overline{\beta_{12}^{(k)}}\\-\rmi\beta_{12}^{(k)}&0}, & \lambda_k<0.\end{cases}
\end{equation}
Here $\beta_{21}^{(k)}=\overline{\beta_{12}^{(k)}}$ by $|\Gamma(\rmi\nu)|^2=\pi/(\nu\sinh\pi\nu)$. For $\lambda_k<0$, $\beta_{12}^{(k)}$ is evaluated at $u_-^{(k)}$, not at its conjugate. If $r_2(\lambda_k)=0$, then $A_k=0$, the limit of \eqref{beta7} as $\rho_k\to0$.

Let
\begin{equation*}
P(\zeta;\rho)=I+P_1(\rho)\zeta^{-1}+\mathcal O(\zeta^{-2}), \quad \nu=-\frac1{2\pi}\log(1-|\rho|^2), \quad |\rho|<1
\end{equation*}
solve the RH problem with jumps
\begin{equation*}
\begin{aligned}
&\psm{1&\bar\rho(-\zeta)^{2\rmi\nu}e^{-\rmi\zeta^2/2}\\0&1},\ \arg\zeta=\pi-\varphi_1,\quad &&\psm{1&0\\-\rho(-\zeta)^{-2\rmi\nu}e^{\rmi\zeta^2/2}&1},\ \arg\zeta=-\pi+\varphi_1,\\
&\psm{1&0\\-\frac{\rho(-\zeta)^{-2\rmi\nu}e^{\rmi\zeta^2/2}}{1-|\rho|^2}&1},\ \arg\zeta=\varphi_1,\quad &&\psm{1&\frac{\bar\rho(-\zeta)^{2\rmi\nu}e^{-\rmi\zeta^2/2}}{1-|\rho|^2}\\0&1},\ \arg\zeta=-\varphi_1.
\end{aligned}
\end{equation*}
By \cite{DZ93},
\begin{equation*}
P_1(\rho)=\begin{pmatrix}0&\rmi\beta_{12}(\rho)\\-\rmi\overline{\beta_{12}(\rho)}&0\end{pmatrix},\quad \beta_{12}(\rho)=\frac{\sqrt{2\pi}e^{\frac{\pi\rmi}4}e^{-\frac{\pi\nu}2}}{\rho\,\Gamma(-\rmi\nu)}.
\end{equation*}
Under \eqref{zeta7}, the jumps of RH Problem~\ref{rhp72} in $U_3^{(k)}$ with $r_2$ frozen at $\lambda_k$ are those of $P(\zeta;\rho_k)$ for $\lambda_k>0$ and of $\sigma_1P(-\zeta;u_-^{(k)})\sigma_1$ for $\lambda_k<0$; in both cases the $1/\zeta$ coefficient is $-\ve_kA_k$.

Since the disks have fixed radius, we keep the exact conformal factor, and we freeze $\beta_k$ at $\lambda_k$. Let
\begin{equation*}
h_k(\lambda)=\frac{\zeta_k(\lambda)}{2\sqrt{\kappa_kt}\,(\lambda-\lambda_k)},\quad \omega_k(\lambda)=-\rmi\ve_k\nu_k\log h_k(\lambda),
\end{equation*}
where $h_k$ is independent of $t$, analytic and zero-free in $U_3^{(k)}$, with $h_k(\lambda_k)=1$, and $\log h_k(\lambda_k)=0$. Thus $\omega_k$ is analytic in $U_3^{(k)}$ and $\omega_k(\lambda_k)=0$. In $U_3^{(k)}$ we have exactly
\begin{equation*}
e^{2\delta}=e^{2(\beta_k(\lambda)-\beta_k(\lambda_k))}\,e^{2\omega_k}\,e^{2\beta_k(\lambda_k)}(2\sqrt{\kappa_kt})^{-2\rmi\ve_k\nu_k}(-\ve_k\zeta_k)^{2\rmi\ve_k\nu_k}.
\end{equation*}
The local parametrix is obtained by dropping the first factor
\begin{equation*}
\Nl(\lambda)=\Ng(\lambda)e^{\omega_k(\lambda)\sigma_3}P_k(\zeta_k(\lambda))e^{-\omega_k(\lambda)\sigma_3},\quad P_k(\zeta)=\begin{cases}P(\zeta;\rho_k), & \lambda_k>0,\\ \sigma_1P(-\zeta;u_-^{(k)})\sigma_1, & \lambda_k<0.\end{cases}
\end{equation*}
It is analytic in $U_3^{(k)}$ off the four lens rays; in particular it has no jump on $S_-\cap U_3^{(k)}$. On the other hand, $\Nc^{(1)}$ has no jump on $\mathbb R\cap U_3^{(k)}$ by RH Problem~\ref{rhp72}, since the jump of $\delta$ on $S_-$ cancels the diagonal factor $(1-|r_2|^2)^{\pm\sigma_3}$ of the factorization. On the lens rays inside $U_3^{(k)}$, the jumps of $\Nc^{(1)}$ and $\Nl$ differ by freezing $r_2$ at $\lambda_k$ and by the factor $e^{\pm2(\beta_k(\lambda)-\beta_k(\lambda_k))}$. By \eqref{betaHolder} and $|e^{\pm2\rmi t\theta}|\le e^{-ct|\lambda-\lambda_k|^2}$ on these rays, the difference is
\begin{equation}\label{lensmismatch}
\mathcal O\big(|\lambda-\lambda_k|\big(1+\big|\log|\lambda-\lambda_k|\big|\big)e^{-ct|\lambda-\lambda_k|^2}\big),
\end{equation}
whose $L^1$, $L^2$ and $L^\infty$ norms are $\mathcal O(t^{-1}\log t)$, $\mathcal O(t^{-3/4}\log t)$ and $\mathcal O(t^{-1/2}\log t)$, respectively. As $t\to\infty$, uniformly on the fixed circle $\partial U_3^{(k)}$,
\begin{equation}\label{match7}
\Nl(\lambda)\Ng(\lambda)^{-1}=I-\frac{\ve_k}{\zeta_k(\lambda)}\Ng(\lambda)e^{\omega_k(\lambda)\sigma_3}A_ke^{-\omega_k(\lambda)\sigma_3}\Ng(\lambda)^{-1}+\mathcal O(t^{-1}).
\end{equation}
On fixed circles, $1/\zeta_k$ and $e^{\omega_k\sigma_3}$ are not linearized, since that would cost $\mathcal O(t^{-1/2})$. Only $\zeta_k'(\lambda_k)=2\sqrt{\kappa_kt}$ and $\omega_k(\lambda_k)=0$ enter the residues.

\begin{figure}[htbp]
\centering
\includegraphics[width=0.8\textwidth]{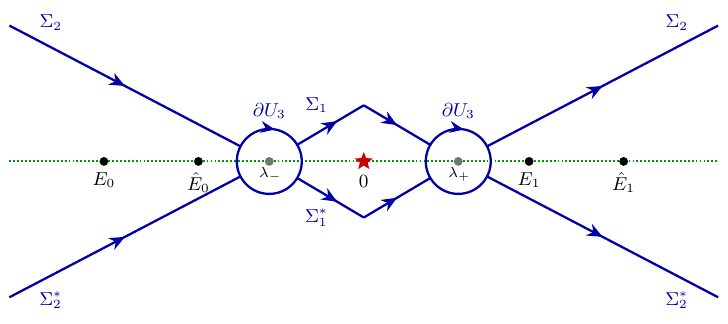}
\caption{Jump contour $\Sigma^{(E)}$ of the error problem (Zakharov-Manakov region).}\label{fig15}
\end{figure}

\begin{rhp}\label{rhp74}
The error function $E(\lambda)$ is characterized by the following RH problem.
\begin{conds}
\cond{1}{Analyticity:} $E(\lambda)$ is analytic in $\mathbb C\setminus\Sigma^{(E)}$, where
\begin{equation*}
\Sigma^{(E)}=\partial U_3\cup\big(\Sigma_1\cup\Sigma_1^*\cup\Sigma_2\cup\Sigma_2^*\big),
\end{equation*}
as shown in Figure~\ref{fig15}.
\cond{2}{Jump condition:} $E_+(\lambda)=E_-(\lambda)J_E(\lambda)$, where
\begin{equation}\label{JE7}
J_E(\lambda)=\begin{cases}\Ng(\lambda)J^{(1)}_{\Nc}(\lambda)\Ng(\lambda)^{-1}, & \lambda\in\Sigma^{(E)}\setminus\overline{U_3},\\ \Nl_-(\lambda)J^{(1)}_{\Nc}(\lambda)J^{(pc)}(\lambda)^{-1}\Nl_-(\lambda)^{-1}, & \lambda\in\Sigma^{(E)}\cap U_3,\\ \Nl(\lambda)\Ng(\lambda)^{-1}, & \lambda\in\partial U_3,\end{cases}
\end{equation}
where $J^{(pc)}$ is the jump of $\Nl$.
\cond{3}{Asymptotics at infinity:} $E(\lambda)=I+\mathcal O(\lambda^{-1})$ as $\lambda\to\infty$.
\cond{4}{Local behavior:} $E(\lambda)$ remains bounded in neighborhoods of $ E$ and of $\lambda=0$.
\end{conds}
\end{rhp}

By Lemma~\ref{lem73}, we have, for $p\in[1,\infty]$,
\begin{equation}\label{JEest7}
\|J_E-I\|_{L^p(\Sigma^{(E)})}=\begin{cases}\mathcal O(e^{-ct}), & \text{on }\Sigma^{(E)}\setminus\overline{U_3},\\ \mathcal O(t^{-\frac12-\frac1{2p}}\log t), & \text{on }\Sigma^{(E)}\cap U_3,\\ \mathcal O(t^{-1/2}), & \text{on }\partial U_3.\end{cases}
\end{equation}
Consequently RH problem \ref{rhp74} has a unique solution for $t$ large. If $\xi$ lies in the fast-decay region, then $U_3=\emptyset$, all jumps of $E$ are $\mathcal O(e^{-ct})$ by Lemma~\ref{lem73}, and
\begin{equation}\label{Efd}
E(\lambda)=I+\mathcal O(e^{-ct}), \quad \lambda\to0.
\end{equation}
For $\xi$ in the Zakharov-Manakov region one finds instead, by \eqref{Erep}, \eqref{match7} and residues at $\lambda_k$ on the clockwise circles ($0\notin\overline{U_3}$). The lens pieces in $U_3$ contribute $\mathcal O(t^{-1}\log t)$ by \eqref{lensmismatch},
\begin{equation}\label{E07}
E(\lambda)=E(0)+E_1\lambda+\mathcal O(\lambda^2),
\end{equation}
where
\begin{equation}
\begin{aligned}
&E(0)=I+\sum_{\lambda_k\in\mathcal Z(\xi)}\frac{t^{-1/2}}{2\sqrt{\kappa_k}\,|\lambda_k|}\Ng(\lambda_k)A_k\Ng(\lambda_k)^{-1}+\mathcal O(t^{-1}\log t),\\
&E_1=\sum_{\lambda_k\in\mathcal Z(\xi)}\frac{\ve_k\,t^{-1/2}}{2\sqrt{\kappa_k}\,\lambda_k^2}\Ng(\lambda_k)A_k\Ng(\lambda_k)^{-1}+\mathcal O(t^{-1}\log t), \quad \lambda\to0,
\end{aligned}
\end{equation}
whence
\begin{equation*}
E(0)^{-1}=I+\mathcal O(t^{-1/2}).
\end{equation*}

\section{The asymptotic behaviors of the defocusing CSP equation}\label{sec8}

For regions III and IV we have $y_\ve=y$ and $\varphi_\ve=(f(0)-f_0)y+(\hat g(0)-g_0)t$. Letting $\lambda\to0$ along $\rmi\mathbb R^+$, we see from \eqref{G-1}, \eqref{G1}, \eqref{G6}, \eqref{G-6} and \eqref{G7}, in view of
\begin{equation*}
\operatorname{Im}\theta(\lambda)=\frac{\operatorname{Im}\lambda}{4|\lambda|^2}(1+\mathcal O(\lambda)), \quad \lambda\to0.
\end{equation*}
that
\begin{equation*}
G(\lambda)=I+\mathcal O(e^{-ct/|\lambda|}).
\end{equation*}
Therefore
\begin{equation}\label{M80}
M(\lambda)=e^{-\delta(\infty)\sigma_3}E(\lambda)\Mg(\lambda)e^{\delta(\lambda)\sigma_3}\big(I+\mathcal O(e^{-ct/|\lambda|})\big).
\end{equation}
Introducing
\begin{equation*}
K=\Mg(0)^{-1}\partial_\lambda\Mg(0), \quad F=\Mg(0)^{-1}E(0)^{-1}E_1\Mg(0),
\end{equation*}
we get
\begin{equation}\label{M81}
M(0)^{-1}M(\lambda)=e^{-\delta(0)\sigma_3}\Big[I+\lambda\big(K+F+\delta'(0)\sigma_3\big)+\mathcal O(\lambda^2)\Big]e^{\delta(0)\sigma_3},
\end{equation}
where $\delta(0)\in\rmi\mathbb R$ and $\rmi\delta'(0)\in\mathbb R$ are given below, with $(c_*, S_*)=(\ve, S_2)$, $(-\ve, S_2)$ and $(-1,S_-)$ for regions I, II and III-IV, respectively
\begin{equation}\label{delta0}
\begin{aligned}
&\delta(0)=\frac{w(0)}{2\pi\rmi}\Big[\sum_{j=1}^n\delta_j\int_{E_j}^{\hat E_j}\frac{\rmi\,\rmd s}{w_+(s)s}+c_*\int_{ S_*}\frac{L(s)}{w(s)s}\rmd s\Big],\\
&\delta'(0)=\frac{w'(0)}{w(0)}\delta(0)+\frac{w(0)}{2\pi\rmi}\Big[\sum_{j=1}^n\delta_j\int_{E_j}^{\hat E_j}\frac{\rmi\,\rmd s}{w_+(s)s^2}+c_*\int_{ S_*}\frac{L(s)}{w(s)s^2}\rmd s\Big].
\end{aligned}
\end{equation}
Combining these expansions with \eqref{recq5}, \eqref{recx5}, \eqref{recq6}, \eqref{recx6}, \eqref{qglo} and \eqref{xglo}, we arrive at
\begin{equation}\label{q8}
q(x,t)=e^{-2\delta(0)}\Big[q^{(AG)}\big(x^{(glo)},t;\mathbf E,\hat{\mathbf E},\bphi+\bdelta\big)+e^{2\rmi\varphi_\ve}F_{12}\Big],
\end{equation}
\begin{equation}\label{x8}
x(y,t)=x^{(glo)}(y,t)+\rmi\delta'(0)+\rmi F_{11},\quad x^{(glo)}(y,t)=y_\ve+\tilde\Phi_\ve'(0)+\rmi K_{11}.
\end{equation}

\begin{remark}[Physical rays]\label{rem81}
Since $\tilde\Phi_\ve(\lambda)=(f(\lambda)-\lambda)y_\ve+\hat g(\lambda)t$ and $y-y_\ve\in\{0,\Delta\}$, \eqref{x8} gives
\begin{equation*}
x(y,t)=f'(0)\,y+\hat g'(0)\,t+\mathcal O(1),\quad t\to\infty,
\end{equation*}
uniformly on each of the sets of Theorem~\ref{thm14}. The term $\rmi K_{11}$ is a continuous function of the phase vector $\mathbf c^*$ on the torus $\mathbb R^n/\mathbb Z^n$, because the denominators in $D^{(glo)}$ do not depend on $\mathbf c^*$; hence it is bounded. The term $\rmi\delta'(0)$ is bounded by \eqref{delta0}, and $F_{11}\to0$. Thus the ray $y=\xi t$ in the reciprocal plane corresponds to the physical ray $x=v(\xi)t+\mathcal O(1)$ with speed
\begin{equation*}
v(\xi)=f'(0)\,\xi+\hat g'(0)=\frac{P_f(0)}{w(0)}\,\xi+\hat g'(0).
\end{equation*}
\end{remark}

i) Transition region I. Here \eqref{E0asy} gives
\begin{equation}\label{F1}
F=\frac{t^{-1/3}}{|\theta^{(3,\hat j_1)}(\hat\xi_{j_1})|^{2/3}\hat E_{j_1}^2}\Mg(0)^{-1}H_1(\hat E_{j_1})\begin{pmatrix}0&\rmi a(s)\\0&0\end{pmatrix}H_1(\hat E_{j_1})^{-1}\Mg(0)+\mathcal F^{pc}+\mathcal O(t^{-2/3}).
\end{equation}

ii) Transition region II. From (\ref{E06}) we get
\begin{equation}\label{F2}
F=-\frac{t^{-1/3}}{|\theta^{(3,j_1)}(\xi_{j_1})|^{2/3}\Ej^2}\Ng(0)^{-1}\tilde H_1(\Ej)\begin{pmatrix}0&\rmi a(s)\\0&0\end{pmatrix}\tilde H_1(\Ej)^{-1}\Ng(0)+\mathcal F^{pc}+\mathcal O(t^{-2/3}).
\end{equation}

iii) Zakharov-Manakov region III. By (\ref{E07}),
\begin{equation}\label{F3}
F=\mathcal J+\mathcal O(t^{-1}\log t),\quad \mathcal J=\sum_{\lambda_k\in\mathcal Z(\xi)}\frac{\ve_kt^{-1/2}}{2\sqrt{\kappa_k}\lambda_k^2}\Ng(0)^{-1}\Ng(\lambda_k)A_k\Ng(\lambda_k)^{-1}\Ng(0).
\end{equation}
In i) and ii), $\mathcal F^{pc}=\mathcal O(t^{-1/2})$ denotes the contribution of the parabolic-cylinder parametrices in $U_k$, $\lambda_k\in\mathcal Z(\xi)\setminus\{\lambda_*\}$, built with $r,\delta$ of Sections~\ref{sec5} and \ref{sec6}. Only its order is used.

iv) Fast-decay region IV. Here (\ref{Efd}) implies $F=\mathcal O(e^{-ct})$. In particular, for $\xi<0$,
\begin{equation}\label{F4}
q(x,t)=q^{(AG)}(x,t;\mathbf E,\hat{\mathbf E},\bphi)+\mathcal O(e^{-ct}).
\end{equation}

Inserting (\ref{F1})-(\ref{F4}) into (\ref{q8})-(\ref{x8}), we obtain Theorem~\ref{thm14}. \hfill$\square$

\appendix
\renewcommand{\thesection}{A}
\section{The Painlev\'e XXXIV parametrix}\label{appA}

Sections \ref{sec5} and \ref{sec6} use local parametrices modeled on the Painlev\'e XXXIV parametrix $M^{(P34)}(\zeta)=M^{(P34)}(\zeta;s,\alpha,\omega)$.

\begin{figure}[htbp]
\centering
\includegraphics{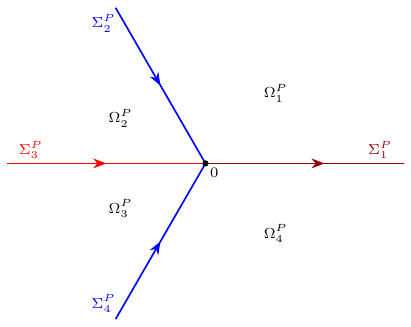}
\caption{RH Problem \ref{rhpA1}: the rays $\Sigma_k^P$ and the sectors $\Omega_k^P$, $k=1,2,3,4$.}\label{figP34}
\end{figure}

\begin{rhp}\label{rhpA1}
We seek $M^{(P34)}(\zeta)$ with the following properties.
\begin{conds}
\cond{a}{Analyticity:} $M^{(P34)}(\zeta)$ is analytic for $\zeta\in\mathbb C\setminus\big(\bigcup_{k=1}^4\Sigma_k^P\cup\{0\}\big)$, with the rays (See Figure \ref{figP34})
\begin{equation}\label{SigP}
\Sigma_1^P=\mathbb R^+,\quad \Sigma_2^P=e^{\frac{2\pi\rmi}3}\mathbb R^+,\quad \Sigma_3^P=e^{\pi\rmi}\mathbb R^+,\quad \Sigma_4^P=e^{-\frac{2\pi\rmi}3}\mathbb R^+.
\end{equation}
\cond{b}{Jump condition:} $M^{(P34)}_+(\zeta)=M^{(P34)}_-(\zeta)J^{(P34)}(\zeta)$, where
\begin{equation}\label{JP34}
J^{(P34)}(\zeta)=\begin{cases}
\begin{pmatrix}1&\omega\\0&1\end{pmatrix},&\zeta\in\Sigma_1^P,\\[8pt]
\begin{pmatrix}1&0\\e^{2\alpha\pi\rmi}&1\end{pmatrix},&\zeta\in\Sigma_2^P,\\[8pt]
\begin{pmatrix}0&1\\-1&0\end{pmatrix},&\zeta\in\Sigma_3^P,\\[8pt]
\begin{pmatrix}1&0\\e^{-2\alpha\pi\rmi}&1\end{pmatrix},&\zeta\in\Sigma_4^P.
\end{cases}
\end{equation}
\cond{c}{Asymptotics at infinity:} for some function $a(s)=a(s;\alpha,\omega)$ one has, as $\zeta\to\infty$,
\begin{equation}\label{MP34inf}
M^{(P34)}(\zeta)=\begin{pmatrix}1&0\\-\rmi a(s)&1\end{pmatrix}\Big(I+\frac{M_1^{(P34)}(s)}{\zeta}+\mathcal O(\zeta^{-2})\Big)\,\frac{\zeta^{-\frac14\sigma_3}}{\sqrt2}\begin{pmatrix}1&\rmi\\\rmi&1\end{pmatrix}e^{-(\frac23\zeta^{3/2}+s\zeta^{1/2})\sigma_3},
\end{equation}
where the fractional powers take their principal branches; moreover,
\begin{equation}\label{M1P34}
\big(M_1^{(P34)}(s)\big)_{12}=\rmi a(s).
\end{equation}
\cond{d}{Local behavior:} for $-\frac12<\alpha<0$ one has, as $\zeta\to0$,
\begin{equation}\label{MP340}
M^{(P34)}(\zeta)=\mathcal O(\zeta^{\alpha}),
\end{equation}
while for $\alpha\ge0$,
\begin{equation}\label{MP340b}
M^{(P34)}(\zeta)=\begin{cases}\begin{pmatrix}\mathcal O(\zeta^{\alpha})&\mathcal O(\zeta^{-\alpha})\\\mathcal O(\zeta^{\alpha})&\mathcal O(\zeta^{-\alpha})\end{pmatrix},&\zeta\in\Omega_1^P\cup\Omega_4^P,\\[10pt]
\mathcal O(\zeta^{-\alpha}),&\zeta\in\Omega_2^P\cup\Omega_3^P,\end{cases}
\end{equation}
with the sectors $\Omega_k^P$, $k=1,2,3,4$, as in Figure \ref{figP34}.
\end{conds}
\end{rhp}

For $\alpha>-\frac12$, $\omega\in\mathbb C\setminus(-\infty,0)$ and $s\in\mathbb R$, RH Problem \ref{rhpA1} admits exactly one solution \cite{IKO08,IKO09,XZ11}. The quantity $a(s)$ appearing in \eqref{MP34inf} is linked to Painlev\'e XXXIV as follows: the function
\[
v(s)=v(s;\alpha,\omega)=a'(s;\alpha,\omega)-\frac s2
\]
is a solution of the Painlev\'e XXXIV equation
\begin{equation}\label{P34eq}
v''(s)=4v(s)^2+2sv(s)+\frac{v'(s)^2-(2\alpha)^2}{2v(s)}
\end{equation}
and is pole-free on $\mathbb R$. One also has
\begin{equation}\label{vasy}
v(s;\alpha,0)=\begin{cases}\dfrac{\alpha}{\sqrt s}+\mathcal O(s^{-2}),&s\to+\infty,\\[8pt] -\dfrac s2+\mathcal O(s^{-2}),&s\to-\infty.\end{cases}
\end{equation}
Since, in addition, $a(s;\alpha,0)\to0$ as $s\to-\infty$ (see \cite{DXZ20}), integration gives
\begin{equation}\label{aint}
a(s;\alpha,0)=\int_{-\infty}^s\Big(v(\tau;\alpha,0)+\frac\tau2\Big)\rmd\tau .
\end{equation}
The parametrices of Sections \ref{sec5} and \ref{sec6} correspond to $\alpha=\frac14$ and $\omega=0$, for which \eqref{P34eq}-\eqref{aint} become \eqref{P34} and \eqref{as}.

\section*{Acknowledgments}
\addcontentsline{toc}{section}{Acknowledgments}
This work is supported by National Natural Science Foundation of China (Grant Nos. 12501336, 12371256).

\section*{Data availability}
\addcontentsline{toc}{section}{Data availability}
On behalf of all authors, the corresponding author states that there is no conflict of interest.

\end{document}